\documentclass[aps,physrev,preprint,groupedaddress,showkeys,nofootinbib,]{revtex4-2}
\usepackage{graphicx, amsfonts, amsmath, bbm, mathtools, calrsfs, orcidlink, siunitx} 

\newcommand{\orcid}[1]{\href{https://orcid.org/#1}{\includegraphics[width=8pt]{ORCID.png}}}

\begin{document}
    
\preprint{Phys. Rev. D - Accepted for publication}

\title{~\\Elliptical multipoles for gravitational lenses}

\author{Hadrien Paugnat\orcidlink{0000-0002-2603-6031} }
\email{Corresponding author: hpaugnat@astro.ucla.edu}
\affiliation{Department of Physics and Astronomy, UCLA, Los Angeles, CA 90095-1547, USA}
\author{Daniel Gilman\orcidlink{0000-0002-5116-7287}}
\altaffiliation[]{Brinson Prize Fellow}
\affiliation{Department of Astronomy \& Astrophysics, University of Chicago, Chicago, IL 60637, USA}

\preprint{\doi{10.1103/d14h-f5mn}}

\date{\today}

\begin{abstract}
Gravitational lensing galaxies are commonly modeled with elliptical density profiles, to which angular complexity is sometimes added through a multipole expansion - encoding deformations of the elliptical iso-density contours. The formalism that is widely used in current studies and software packages, however, employs perturbations that are defined with respect to a circle. In this work, we show that this popular formulation (the ‘‘circular multipoles’’) leads to perturbation patterns that depend on the axis ratio and do not agree with physical expectations (from studies of galaxy isophotal shapes) when applied to profiles that are not near-circular. We propose a more appropriate formulation, the ‘‘elliptical multipoles’’, representing deviations from ellipticity suited for any axis ratio. We solve for the lensing potentials associated with the elliptical multipoles of any order $m$, assuming that the reference profile is near-isothermal. We implement these solutions into the lens modeling package \texttt{lenstronomy}, and assess the importance of the multipole formulation by comparing flux-ratio perturbations in mock lensed systems with quadruply imaged quasars: we show that elliptical multipoles typically produce smaller flux-ratio perturbations than their circular counterparts.

\end{abstract}


\maketitle

\vspace{-1cm}
\section{Introduction} \label{sec:intro}

Gravitational lensing - a manifestation of general relativity where light traveling through distorted space-time follows bent trajectories - directly probes the underlying mass/energy distribution and its geometry, offering unique avenues to explore fundamental physics. In particular, strong lensing enables high-precision cosmological measurements which can test the predictions of the standard model of cosmology ($\Lambda$CDM). Prominent examples of tests performed with strong gravitational lensing include time-delay cosmography, to measure the Hubble constant $H_0$ independently of other late-time probes \citep[e.g.,][]{Wong2020, TDCOSMO2020}; and dark matter (DM) substructure studies, which take advantage of the fact that lensing is sensitive to all mass (both dark and luminous) to investigate the properties of DM, for instance examine alternatives to the cold dark matter (CDM) paradigm \citep[e.g.,][]{Hsueh2020, Gilman2020, Gilman2023}. While early works on galaxy lenses were limited by sample sizes and data quality, substantial progress has been made in the past decades, putting such applications within reach (for a historical perspective, see \citep{Treu2016} or \citep{Treu2022}). For instance, instead of working only with constraints from the image positions, time delays, and flux ratios of the lensed object, modern studies can exploit the full information present in the extended lensing arcs seen in high-resolution images \citep[e.g.,][]{Vegetti2012, Shajib2022, Gilman2024}. 

These high-precision measurements also require accurate models, in particular for the lensing galaxy's large-scale mass distribution (the ‘‘macromodel’’). Because of their enhanced cross-section, most known lenses are massive early-type galaxies, which can be represented, to a good approximation, by elliptical surface mass density profiles \citep[e.g.,][]{Turner1984,Chae2003, Yoo2005, Schneider2006, Auger2009}. Popular choices include the Elliptical Power Law (EPL) lens model \citep{TessoreMetcalf}, where the projected total mass density has a radial dependence $\kappa (r) \propto r^{1-\gamma}$ arising from a 3D mass distribution $\rho(r_{\rm \scriptscriptstyle 3D}) \propto r_{\rm \scriptscriptstyle 3D}^{-\gamma}$ ; and its special case for $\gamma=2$, the Singular Isothermal Ellipsoid (SIE) profile \citep{Kassiola1993, Kormann1994}. It is common practice to add an external shear component to such elliptical profiles \citep[e.g.,][]{Kormann1994_shear, Keeton1997, WittMao1997}, in order to account for perturbations introduced by nearby massive galaxies, or in general for deformations of the deflection field from other sources.

In recent years, with the increase in data quality, additional azimuthal structures in the lensing galaxy itself - beyond the external shear and ellipticity of the main deflector - have become increasingly relevant for strong lensing studies. Ignoring their effect can lead to biased values for the model parameters as they attempt to compensate for the lack of complexity in the lens model. For instance, the external shear components fitted to strong lensing data do not match the  direction and/or magnitude expected from theoretical predictions \citep{Keeton1997, Hilbert2007}, direct modeling of line-of-sight perturbers \citep{Moustakas2007,Wong2011}, or weak lensing measurements \citep{Etherington2024}. This suggests that purely elliptical models are unable to fully capture the distribution of mass in a lens.

To account for this complexity, one possible approach is to include extra angular degrees of freedom in the lens model, typically keeping the radial part unchanged while adding Fourier-type perturbations (called ‘‘multipoles’’) to the angular part of the density profile \citep[e.g.,][]{EvansWitt2003, Congdon2005, Gomer2018, Gilman2021, Gilman2023, Oh2024, Gilman2024, Cohen2024, Keeley2024}. This multipole expansion is motivated by studies of the surface photometry of early-type galaxies in the optical and infrared, where deviations from ellipticity of isophotal shapes are parametrized in a similar way - with a particular emphasis on fourth-order multipoles, yielding ‘‘boxy’’ or ‘‘disky’’ isophotes that match observations better than pure ellipses \citep{Carter1978, Bender1987, Bender1988,Hao2006,Kormendy2009, Chaware2014, Cappellari2016}. The inclusion of non-zero multipole moments can have a significant impact on the physical properties inferred from the lens models, e.g., constraints on the Hubble constant \citep{VdV2022}, detections of substructure in extended arcs \citep{Nightingale2024, O'Riordan2024}, or flux-ratio anomalies in quadruply imaged quasars \citep{Gilman2024, Oh2024, Cohen2024, Keeley2024}.

However, the usual formulation employed for strong lensing applications (that we will call ‘‘circular multipoles’’ in the rest of this paper) relies on a Fourier expansion expressed in polar coordinates - which only has a clear physical interpretation when applied to lenses with small ellipticities. Such an approach has, in fact, several shortcomings that we discuss in this paper. These flaws have already been pointed out in the context of isophote-fitting algorithms \citep{Ciambur2015, Ciambur2016}, where physically realistic perturbations can be described with coordinates more appropriate to the reference ellipses (e.g., employing the eccentric anomaly instead of the polar angle \citep{Jedrzejewski1987, Bender1988, Cappellari2016}). Despite the limitations, some recent isophotal shape studies \citep[e.g.,][]{Goullaud2018,Stacey2024, Amvrosiadis2024} continue to approximate multipoles in the low ellipticity limit, and in the context of gravitational lensing, the circular formulation is used by default.

In this work, we propose to adapt the alternate formulation with more suited coordinates - the ‘‘elliptical multipoles’’ - to a strong lensing framework. We argue that, for most cases of astrophysical interest, these provide a more realistic description of deviations from ellipticity around the Einstein radius than circular multipoles, in particular for highly flattened configurations. This comes with a challenge: the associated gravitational potentials (which were not relevant for isophote fitting) cannot be calculated as straightforwardly as in the circular case. We introduce fully analytical solutions for those potentials, and showcase their implementation in the gravitational lens
 modeling software package \texttt{lenstronomy} \footnote{\url{https://github.com/lenstronomy/lenstronomy}}\citep{lenstronomy,lenstronomy2}.

This paper is organized as follows. Section~\ref{sec:general_formalism} reviews the formalism for general isothermal models and explains how angular deviations from ellipticity can be expressed in that framework. Section~\ref{sec:multipole_comparison} presents the circular multipole perturbations, exposes the shortcomings of this parametrization, and describes the angular profile of the proposed substitute - the elliptical multipoles. Section~\ref{sec:potential_solutions} details the corresponding lensing potentials, starting with the circular multipoles and the special case of the $m=1$ order, then presenting elliptical multipole solutions for the most practical cases (the $m=1$, $m=3$ and $m=4$ orders). In Section~\ref{sec:flux_ratio_application}, we assess how differences between circular and elliptical multipole perturbations impact lens modeling analyses, considering both flux ratio predictions and imaging data reconstruction for quadruply imaged quasar systems. We summarize these results in Section~\ref{sec:conclusion} and offer perspective on future applications.

\section{Generalized isothermal models} \label{sec:general_formalism}

Let us a consider a family of lens models characterized by the following lensing potential/density pair \citep[e.g.,][]{Witt2000, Evans2001, Keeton2003}:
\begin{equation}
    \psi(r,\phi)= r F(\phi) \text{ ; } \kappa(r,\phi)= \frac{G(\phi)}{2r}
\label{eq:gen_pot_conv_pair}
\end{equation}
where $r= \sqrt{x^2+ y^2}$, $\phi \equiv \arctan (x,y)  \pmod{2\pi}$ are the usual polar coordinates\footnote{Here, we use the two-argument inverse tangent $\arctan(x, y)$, which respects the quadrant of $(x, y)$ to return the correct angle.}, and the shape functions $F(\phi), G(\phi)$ are related as a consequence of Poisson's equation $\nabla^2\psi=2\kappa$ by:

\begin{equation}
    G(\phi)=F(\phi)+F^{\prime\prime}(\phi).
    \label{eq:main_diff_eq}
\end{equation}

Such a potential/density profile is ‘‘isothermal’’: since the enclosed mass increases linearly with radius, the resulting circular velocity profile is constant. The angular dependence can be specified through the shape functions $ F(\phi)$ and $ G(\phi) $, provided that they obey Equation~(\ref{eq:main_diff_eq}). The deflection angles and second order derivatives corresponding to this potential are then given by:

\begin{equation}
\left\{\begin{array}{l}
\alpha_{x}(r,\phi)= \frac{\partial \psi}{\partial x} (r,\phi) =F(\phi)\cos\phi-F^{\prime}(\phi)\sin\phi \\ ~~ \\
\alpha_{y}(r,\phi)= \frac{\partial \psi}{\partial y}(r,\phi) = F(\phi)\sin\phi+F^{\prime}(\phi)\cos\phi \\ ~~ \\ 
\frac{\partial^2 \psi}{\partial x^2}(r,\phi)=\frac{\sin^2\phi}{r}[F(\phi)+
F^{\prime\prime}(\phi)]=\frac{\sin^2\phi}{r}G(\phi) \\ ~~ \\ 
\frac{\partial^2 \psi}{\partial y^2}(r,\phi)=\frac{\cos^2\phi}{r}[F(\phi)+ F^{\prime\prime}(\phi)]=\frac{\cos^2\phi}{r}G(\phi) \\~\\
\frac{\partial^2 \psi}{\partial x \partial y}(r,\phi)=-\frac{\sin(2\phi)}{2r}[F(\phi)+ F^{\prime\prime}(\phi)]=-\frac{\sin(2\phi)}{2r}G(\phi)
\end{array}\right.
\label{eq:pot_alpha_hessian_array}
\end{equation}
where $F^{\prime}=\frac{\partial F}{\partial \phi}$ and
$F^{\prime\prime}=\frac{\partial^2 F}{\partial \phi^2}$.

The lensing formalism is known to have many degeneracies, i.e., transformations of the physical model that leave the lensing observables (relative image positions, time delays, magnifications, ...) unchanged \citep{Falco1985, Liesenborgs2012, Schneider2014, Wagner2018, Saha2024}. One of the most simple is the so-called prismatic degeneracy \citep{Gorenstein1988}, which states that the addition of a constant deflection field ($\vec{\alpha} \mapsto \vec{\alpha} + \vec{\alpha}_0$) can be compensated by a simple translation of the source relative to the lens. As a consequence, we have the freedom to add or remove terms in $C\cdot r \cos (\phi-\phi_{c})$ (with $C$ and $\phi_{c}$ constants, i.e., independent of spatial coordinates) to the lensing potential. We will make use of this invariance to simplify equations throughout the paper.

The family of solutions described by Equations~(\ref{eq:gen_pot_conv_pair}) and (\ref{eq:main_diff_eq}) includes the Singular Isothermal Ellipsoidal (SIE) profile \citep{Kassiola1993, Kormann1994, Keeton1998}:
\begin{equation}
\begin{split}
    G_{\rm SIE}(\phi) & =\frac{\theta_E\sqrt{q}} {\sqrt{q^2\cos^2\phi+\sin^2\phi}} =  \frac{\theta_E \sqrt{\frac{1+q^2}{2q}}}{\sqrt{1 - \varepsilon \cos(2\phi)}} \\
     F_{\rm SIE}(\phi) &=\frac{\theta_E\sqrt{q}}{\sqrt{2\varepsilon(1-\varepsilon)}}\bigg[\cos(\phi)\arctan
\bigg(\frac{\sqrt{2\varepsilon}\cos\phi}{\sqrt{1-\varepsilon\cos2\phi}}\bigg)
+\sin(\phi)\arctan\bigg(\frac{\sqrt{2\varepsilon}\sin\phi}
{\sqrt{1-\varepsilon\cos2\phi}}\bigg)\bigg]
\end{split}
\end{equation}
where $\theta_E$ is the Einstein radius, $\varepsilon = \frac{1-q^2}{1+q^2}$ is the ellipticity parameter and $0<q\leq1$ is the axis ratio. $G_{\rm SIE}(\phi)$ then corresponds to the equation (in polar coordinates, relative to the center) of an ellipse $r_{\rm SIE}(\phi)$, with semi-major axis $a = \frac{\theta_E}{\sqrt{q}}$ aligned with the $x$ axis, and semi-minor axis $ b = \theta_E\sqrt{q}$ aligned with the $y$ axis (we can always choose the coordinate system to align the axes of the ellipse with the $x$ and $y$ axes). Following Equation~(\ref{eq:gen_pot_conv_pair}), the iso-convergence contours of the SIE lens are then ellipses with the same axis ratio $q$, and lengths scaled by $r^{-1}$. As a consequence, if we write instead the shape function for the convergence as $G(\phi) = G_{\rm SIE}(\phi) + G_{\rm pert}(\phi)$, then $G_{\rm pert}(\phi) = \delta r(\phi)$ encodes a self-similar deformation of the elliptical iso-$\kappa$ contours given by the SIE profile: $r_{\rm SIE}(\phi) \mapsto r_{\rm SIE}(\phi) + \delta r(\phi)$.

The calculations presented in this work are carried out under the assumption that the reference elliptical mass profile is isothermal ($\gamma=2$, where $\kappa(r)\propto r^{1-\gamma}$). Modern strong lensing studies employ more frequently the EPL profile, i.e., a generalization of the SIE to a free slope $\gamma$. Applying isothermal multipole solutions to non-isothermal profiles causes the multipole amplitude to vary with radius; however, in most astrophysical lenses, $\gamma$ does not deviate too much from $2$, so isothermal perturbations still provide a reasonable approximation. We 
discuss this further in Section~\ref{sec:conclusion}, and argue that radial dependence matters less than azimuthal dependence in the case of multipole perturbations.

\section{Multipole perturbations} \label{sec:multipole_comparison}

In this section, we describe the angular profile of the multipole perturbations (i.e., the convergence shape function $G_{\rm pert}(\phi)$), starting with the commonly used circular multipoles and their shortcomings (section~\ref{subsec:circ_mult_pbs}), then presenting the more widely applicable elliptical multipoles (section~\ref{subsec:ell_multipoles}). Figures in this section (except Figure~\ref{fig:ell_vs_circ_contours}) were generated with the lens modeling software package \texttt{lenstronomy} \citep{lenstronomy,lenstronomy2}, where we newly implemented the $m=1$ circular multipole, as well as the $m=1$, $m=3$ and $m=4$ elliptical multipoles.

\subsection{Circular multipoles}
\label{subsec:circ_mult_pbs}

First, consider a circular multipole perturbation \citep{Keeton2003, Xu2015}:
\begin{equation}
    G_m^{\rm circ}(\phi) = \delta r_m^{\rm circ} (\phi) = a_m^{\rm circ}\cos(m(\phi-\phi_m))
    \label{eq:circular_multipoles_def}
\end{equation}
where $a_m^{\rm circ}$ and $\phi_m$ are the amplitude and orientation of the $m^{\rm th}$-order perturbation with respect to the perfect ellipse. With this definition, $a_m^{\rm circ}$ is the amplitude of the perturbation for the reference $\kappa=1/2$ isocontour. Since the shape of the isodensity contours is independent of scale in Equation~(\ref{eq:gen_pot_conv_pair}), whenever we combine circular multipoles with an elliptical profile, we follow the convention used in some previous works \citep[e.g.,][]{Xu2015, Oh2024} and normalize the multipole amplitude by the semi-major axis of this $\kappa=1/2$ isocontour, i.e., we rescale the amplitude with $a_m^{\rm circ} \mapsto a_m^{\rm circ} \frac{\theta_E}{\sqrt{q}}$. We warn, however, that even with this rescaling, $a_m^{\rm circ}$ cannot be interpreted as a fractional change in semi-major axis (this is $a_m^{\rm circ}\cos(m\phi_m)$, which depends on the multipole orientation) or a fractional change in polar radius (which depends on angle $\phi$). Other equivalent conventions are sometimes used for the circular multipoles, for instance defining $G_m^{\rm circ}(\phi)$ with components along two orthogonal direction instead of $a_m^{\rm circ}$ and $\phi_m$ (see Appendix B of \citep{Oh2024}).

\begin{figure*}
    \centering
    \includegraphics[width=0.8\linewidth]{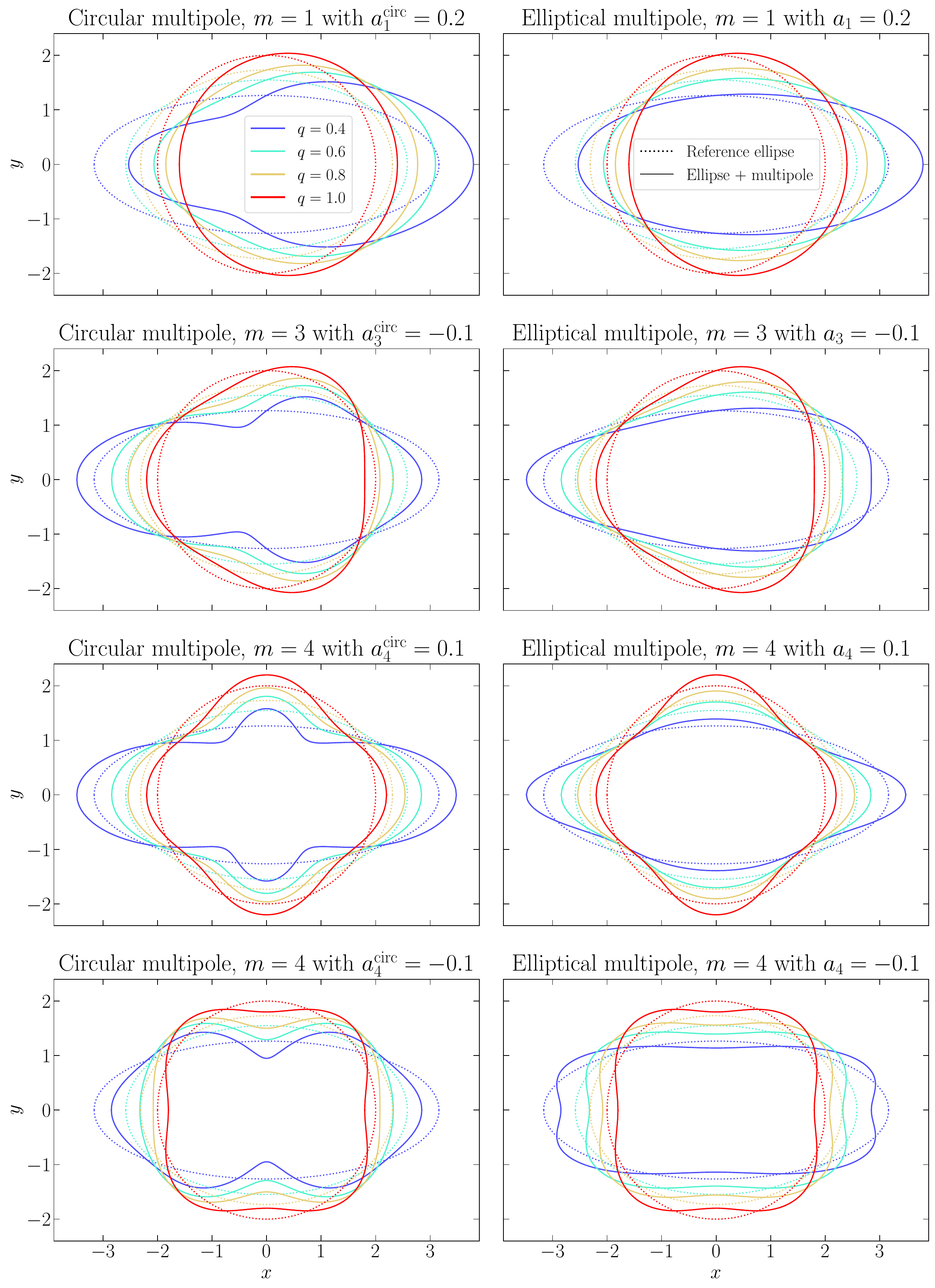}
    \vspace{-0.6cm}
    \caption{Patterns of perturbation corresponding to the $m=1$, $m=3$ and $m=4$ multipoles, in the circular (left) and elliptical (right) formulation, applied to reference ellipses with the same elliptical radius $R=2$ but varying axis ratio $q$. The multipoles are here aligned with the ellipses, i.e., $\phi_m=0$ or $\varphi_m=0$. We rescale the multipole amplitudes appropriately, but since the pattern itself does not depend on amplitude, we use values for $a_m^{\rm circ}$ and $a_m$ larger than physical expectations (of a few percent) for better visualization. The circular multipoles yield perturbation patterns that depend on the axis ratio, and that do not behave as expected for high ellipticities.}
    \label{fig:ell_vs_circ_contours}
\end{figure*}

Adding a circular multipole term of order $m$ is equivalent to adjusting the $m$-th order term in the Fourier series expansion of the original shape function \citep{EvansWitt2003} - in our case, the one associated with the SIE profile. There are, however, issues with this approach:
\begin{enumerate}
    \item For a given order $m$, the pattern of radial deviations around the ellipse depends on the axis ratio $q$ (see Figure~\ref{fig:ell_vs_circ_contours}). This is due to the fact that Equation~(\ref{eq:circular_multipoles_def}) encodes patterns of order $m$ around a circle, but gets applied to an ellipse - i.e., the perturbation does not get ‘‘squished’’ with the ellipse as one could expect.
    \item The pattern of radial deviations around the ellipse does not match the expected behaviour (derived from isophote fitting) when considering highly flattened configurations. For instance, adding a $m=4$ circular multipole term to a $q\sim0.6$ reference ellipse does not yield ‘‘boxy’’ (resp. ‘‘disky’’) contours when $a_4^{\rm circ}<0$ (resp. $a_4^{\rm circ}>0$), but ‘‘peanut’’-shaped (resp. ‘‘spinny-top’’-shaped) contours instead (see Figure~\ref{fig:ell_vs_circ_contours}). While most lens models have axis ratios $q\gtrsim 0.6$ (i.e., relatively low ellipticities) it is not rare to find systems with values close to that limit or even lower \citep[e.g.,][]{Koopmans2006,Shu2017,DINOS1}. Quadruply imaged quasars, in particular, are biased towards larger ellipticities since the inner caustics surface is larger for such systems \citep{Keeton1997, Sonnenfeld2023}.
    \item Equation~(\ref{eq:circular_multipoles_def}) does not match the reference definitions used in galaxy isophotal shape studies. Originally, \cite{Carter1978} performed a global coordinate change to map the best-fitting ellipses to circles, and only then applied a circular multipole perturbation. Other methods were developed afterwards \citep[e.g.,][]{Jedrzejewski1987, Bender1988}, which do rely on direct Fourier expansions (either of the intensity distribution or of the isophote radius), but in terms of the eccentric anomaly and not of the polar angle \citep[e.g.,][]{Peng2010, Ciambur2016, Cappellari2016}. While some studies do use the polar angle \citep[e.g.,][]{Hao2006, Chaware2014, Amvrosiadis2024}, it has already been recognized that this scheme does not accurately represent the isophotes outside of the low ellipticity regime \citep{Ciambur2015}. 
\end{enumerate}

The use of circular multipoles is not completely unsound: for lenses where the axis ratio $q$ is close to $1$, they approximately encode the desired perturbation. Even if an astrophysical lens is in a highly flattened configuration, adding several orders of circular multipole might provide the lens model with enough degrees of freedom to reproduce some observations - we test this in Section~\ref{subsec:mock lenses} by simulating mock lens systems. We show, however, that this would make interpretation of the multipole parameters more difficult, and that it might result in biased lens models, at least in the context of flux-ratio anomalies.

We emphasize that the common practice, which is to include only the circular $m=1$ and/or $m=3$ and/or $m=4$, may lead to inaccurate interpretations of
the fitted parameters $a_m^{\rm circ}$. As mentioned above, if some of the considered lenses are not close to circular, it is not possible to use $a_4^{\rm circ}$ as an indicator for boxyness/diskyness \citep[e.g.,][]{Oh2024, Etherington2024}. Similarly, depending on the axis ratio, the circular $m=1$ might not represent a profile that has been skewed in direction $\phi_1$, but rather one that has been ‘‘pinched’’ in a direction that depends on both $\phi_1$ and the orientation of the reference ellipse (see Figure~\ref{fig:convergence SIE+m=1}). This misinterpretation could explain, for instance, why high $m=1$ multipoles, fitted in light profiles when the galaxy has a companion, are sometimes not aligned with the direction of that companion \citep{Amvrosiadis2024}.

\subsection{Elliptical multipoles}
\label{subsec:ell_multipoles}

In order to overcome the flaws of the circular multipoles, we need a better parametrization - one that provides realistic, easily interpretable perturbations around elliptical countours without many additional parameters. For these ‘‘elliptical multipoles’’, we simply adopt the definition that was originally used in isophote shape studies, which has the same shape as Equation~(\ref{eq:circular_multipoles_def}) but uses different coordinates:  
\begin{equation}
    R\mapsto R + \delta R_m \text{ with } \delta R_m (\varphi) = a_m\cos(m(\varphi-\varphi_m)) \ ,
    \label{eq:elliptical_multipole_def}
\end{equation}
where $R = \sqrt{q x^2+ y^2/q}$ and $\varphi \equiv \arctan (qx,y) \pmod{2\pi}$ are the elliptical radius and the eccentric anomaly along an isocontour, respectively (note the dependence on the axis ratio). In these coordinates, the SIE profile is $\kappa_{\rm SIE}=\frac{\theta_E}{2R}$, so the reference $\kappa=1/2$ isodensity contour corresponds to $R=\theta_E$. To normalize the multipole amplitude, the appropriate rescaling is therefore $a_m\mapsto a_m\theta_E$ when combined with a reference elliptical profile. Then, $a_m$ can be consistently interpreted as a fractional change in elliptical radius for any isodensity contour.

\begin{figure}
    \centering
    \includegraphics[width=0.95\linewidth]{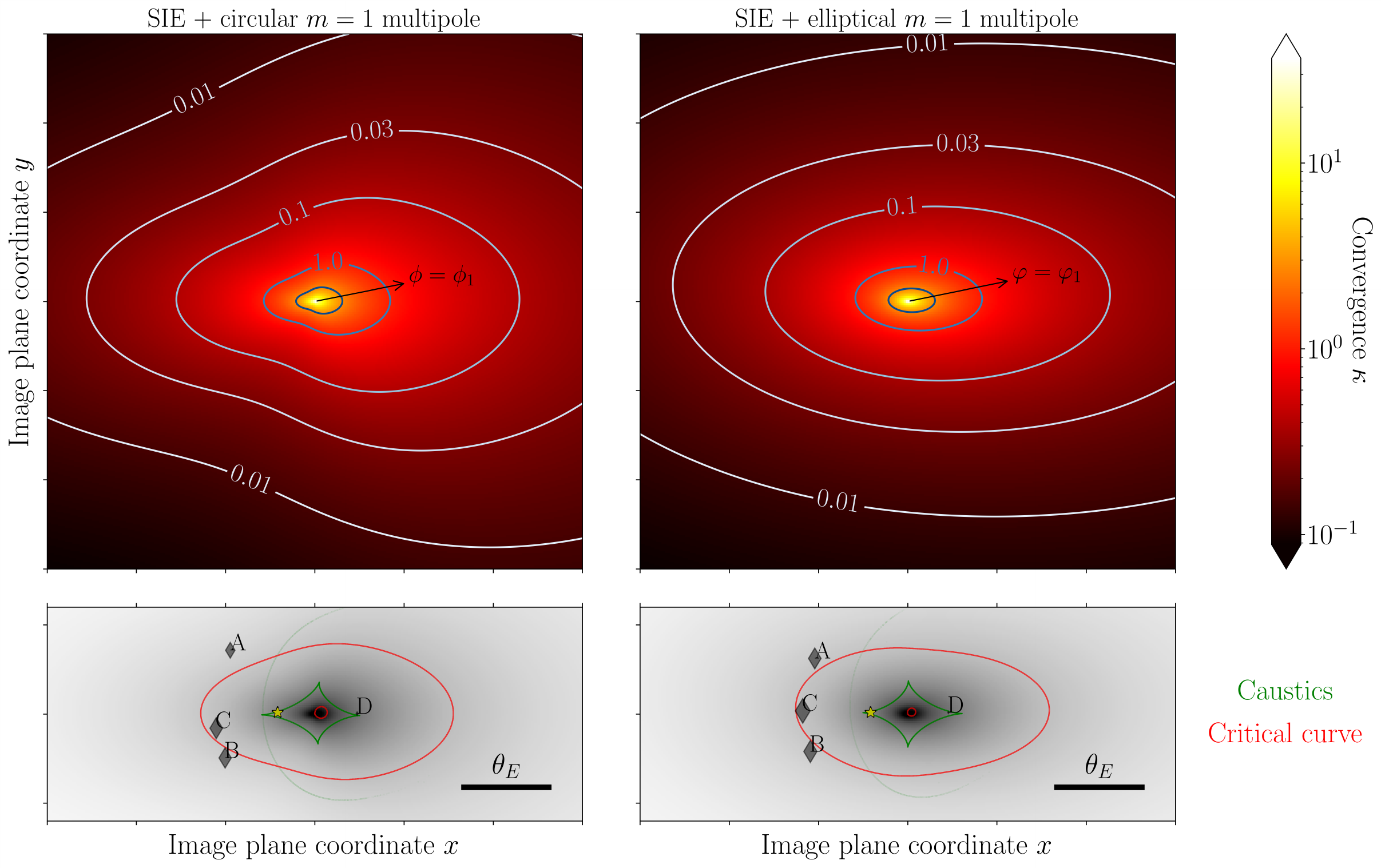}
    \caption{Top panel: Impact of a $m=1$ multipole on the convergence profile of a SIE lens model with axis ratio $q=0.5$, in the circular (left) and elliptical (right) formulations. We choose perturbations with direction $\phi_1=\frac{\pi}{12}$ (resp. $\varphi_1=\varphi(\frac{\pi}{12};q)$), and with amplitude $a_1^{\rm circ}=0.2$ (resp. $a_1=0.2$) - larger than physical expectations, for better visualization. Bottom panel: Corresponding caustics, with an example of source position (yellow star) where the difference in formulation drastically changes the image positions and magnifications. }
    \label{fig:convergence SIE+m=1}
\end{figure}

\begin{figure*}
    \centering
    \includegraphics[width=0.9\linewidth]{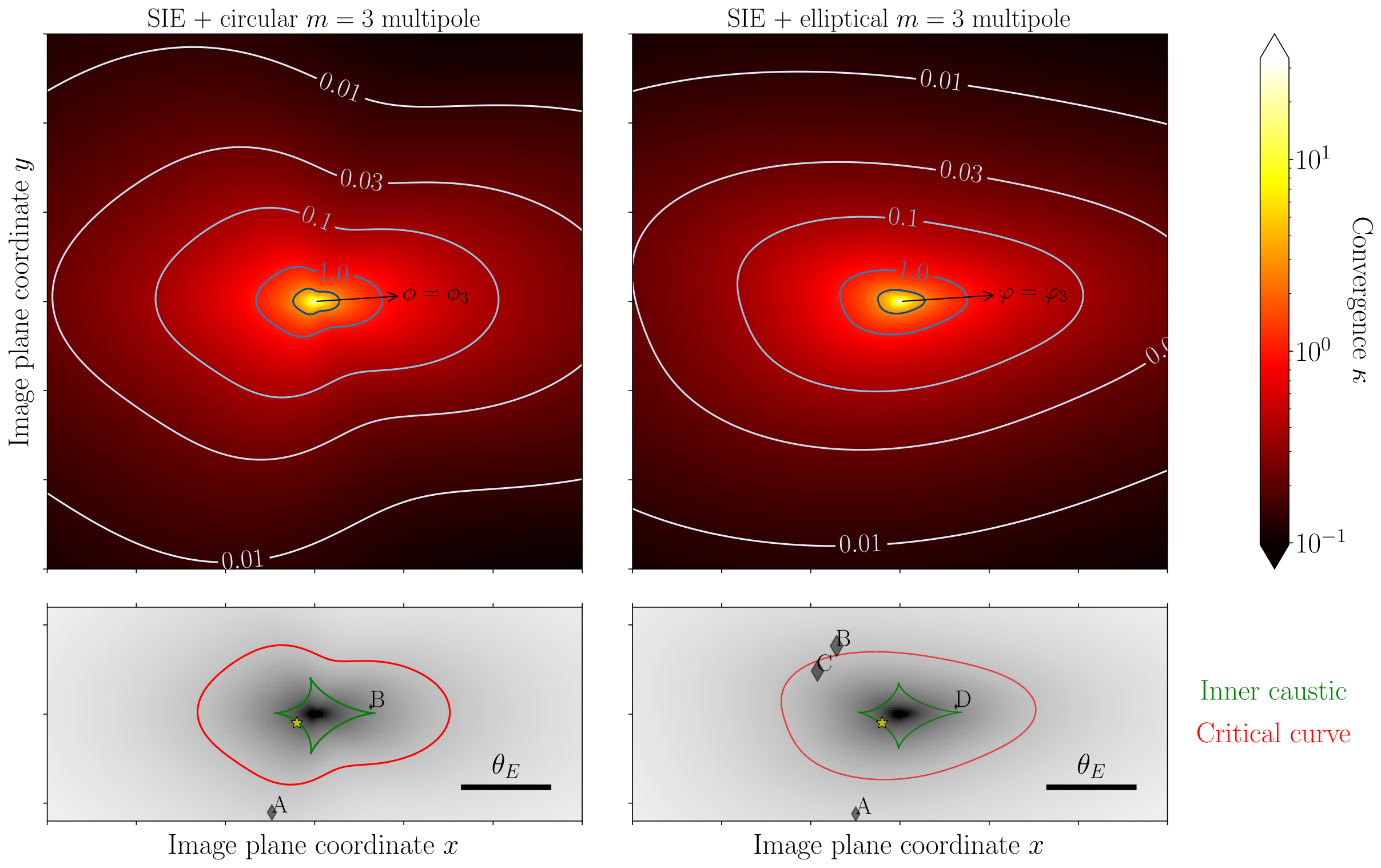}
    \caption{Same as Figure~\ref{fig:convergence SIE+m=1}, but for a $m=3$ multipole with direction $\phi_3=\frac{\pi}{36}$ (resp. $\varphi_3=\varphi(\frac{\pi}{36};q)$) and amplitude $a_3^{\rm circ}=0.075$ (resp. $a_3=0.075$).}
    \label{fig:convergence SIE+m=3}
    \vspace{0.63cm}
    \includegraphics[width=0.9\linewidth]{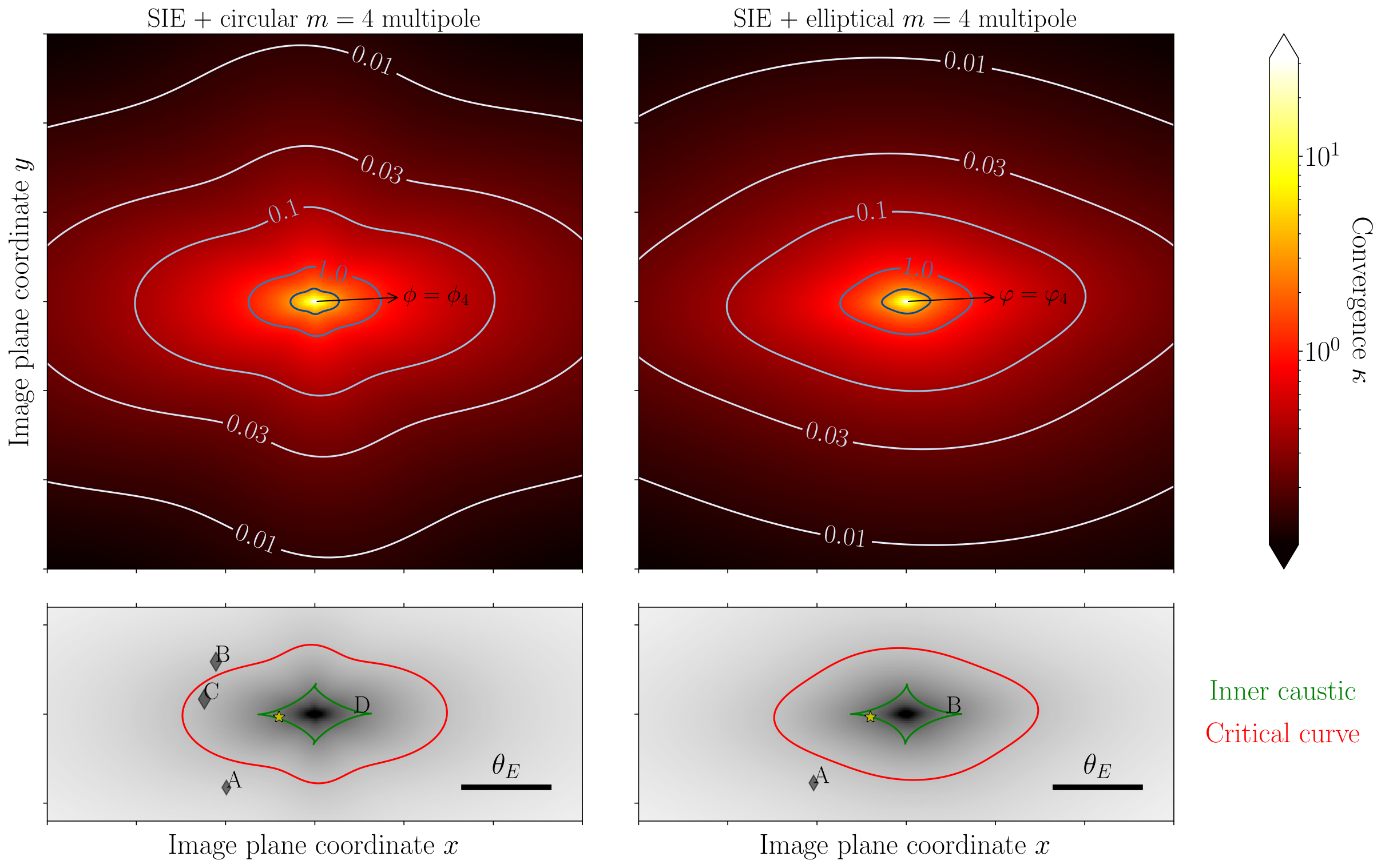}
    \caption{Same as Figure~\ref{fig:convergence SIE+m=1}, but for a $m=4$ multipole with direction $\phi_4=\frac{\pi}{48}$ (resp. $\varphi_4=\varphi(\frac{\pi}{48};q)$) and amplitude $a_4^{\rm circ}=0.05$ (resp. $a_4=0.05$).}
    \label{fig:convergence SIE+m=4}
\end{figure*}

These perturbations are to a general ellipse what the circular multipoles are to a circle: if the axis ratio is $q=1$, we recover the same definition, but this time, the pattern of perturbation is being squished along with the ellipse for $q\neq1$. The elliptical multipoles are therefore the ‘‘correct’’ formulation - in the sense that, regardless of the system's ellipticity, they actually have the expected behavior of the multipole expansion. This means that the usual interpretations of the circular multipoles \citep[e.g.,][]{O'Riordan2024} - true only if the reference contour is a circle, as we have discussed - can be extended to ellipses with any axis ratio \citep{Ciambur2015}:
\begin{itemize}
    \item the monopole ($m=0$ order) corresponds to a global rescaling - not useful in practice, since it is equivalent to a change in the Einstein radius.
    \item the dipole ($m=1$ order) encodes a skewness/lopsidedness in the convergence profile \citep[e.g.,][]{Amvrosiadis2024, Lange2024} - see Figure~\ref{fig:convergence SIE+m=1} for an illustration.
    \item the quadrupole ($m=2$ order) represents a global squeezing - not useful in practice, since the elliptical lens and external shear already have this complexity.
    \item the hexapole ($m=3$ order) and octupole ($m=4$ order) describe triangle-like and quadrangle-like deformations of the isocontours, respectively - see Figures~\ref{fig:convergence SIE+m=3} and \ref{fig:convergence SIE+m=4}. These are the most commonly considered perturbations in lensing \citep[e.g.,][]{VdV2022, Etherington2024, Oh2024, Lange2024}. From studies of galaxy isophotes, the $m=4$ is expected to have larger amplitudes than the $m=3$ (with typical fractional radial deviations along the major axis $a_m$ of the order of a percent), and to roughly align with the axis of the ellipse \citep{Bender1988, Hao2006, Nightingale2024, He2024} - in that case, the perturbation produces disky ($a_m>0$) or boxy ($a_m<0$) isophotes.
\end{itemize}

We can translate between the different coordinate systems using:
\begin{equation}
\begin{split}
    x &= r\cos{\phi} = R\cos{\varphi}/\sqrt{q} \\ 
    y &= r\sin{\phi} = R\sin{\varphi}\sqrt{q} 
\end{split}
\end{equation}
and the following equations are thus verified:
\begin{equation}
\begin{split}
    \frac{r}{R}(\phi ;q) &= \frac{r}{\sqrt{q x^2+ y^2/q}} = \frac{\sqrt{q}}{\sqrt{q^2\cos^2\phi + \sin^2\phi}} \\
    \cos{\varphi(\phi ;q)} &= \frac{q \cos \phi}{\sqrt{q^2\cos^2\phi + \sin^2\phi}} \\
    \sin{\varphi(\phi ;q)} &= \frac{\sin \phi}{\sqrt{q^2\cos^2\phi + \sin^2\phi}} 
\end{split}
\end{equation}

For a given axis ratio $q$, we can define a bijective mapping between $\varphi \in \mathbb{R}$ and $\phi \in \mathbb{R}$:
\begin{equation}
\begin{split}
    \varphi(\phi ;q) &= \arctan (q\cos\phi,\sin\phi) + \left[ \phi - \arctan (\cos\phi,\sin\phi) \right]\\
    \phi(\varphi ;q) &= \arctan (\cos\varphi,q\sin\varphi) + \left[ \varphi - \arctan (\cos\varphi,\sin\varphi) \right]
\end{split}
\label{eq:ell_angle_polar_angle_bijection}
\end{equation}
For $q>0$, $ \varphi(\phi ;q)$ is an infinitely differentiable function of $\phi$ with first derivative \begin{equation}
    \frac{\partial \varphi}{\partial \phi} = \frac{q}{q^2\cos^2\phi + \sin^2\phi}
\end{equation}
Since the mapping is bijective, a perturbation $(R,\varphi) \mapsto (R^\prime,\varphi^\prime) = (R + \delta R, \varphi)$ that is purely radial in the elliptical coordinates will also be purely radial in the usual polar coordinates ($\phi = \phi^\prime$ since $\varphi = \varphi^\prime$). Thus we can write $\delta r \cos\phi = \delta x = \delta R \cos\varphi /\sqrt{q}$, so that $\delta R/R = \delta x /x = \delta r/r $. The angular shape function corresponding to an elliptical multipole perturbation is therefore:
\begin{equation}
\begin{split}
    G_m(\phi ;q, \varphi_m) &= \delta r_m(\phi ;q) = \delta R_m(\phi ;q) \frac{r}{R}(\phi ;q)\\
            &= a_m\cos \left[ m(\varphi(\phi ;q)-\varphi_m) \right] \frac{r}{R}(\phi ;q) \\
            &= a_m \sqrt{q}\cos (m\varphi_m) G_m^{(1)}(\phi ;q) + a_m\sqrt{q}\sin (m\varphi_m) G_m^{(2)}(\phi ;q)
\end{split}
\label{eq:G_m_1_2_decomp}
\end{equation}
where we have defined \begin{equation}
\begin{split}
    G_m^{(1)}(\phi ;q) &\equiv \frac{1}{\sqrt{q}}\cos \left( m\cdot \varphi(\phi ;q) \right) \frac{r}{R}(\phi ;q) \\
            &= T_m \left( \frac{q\cos\phi}{\sqrt{q^2\cos^2\phi+\sin^2\phi}}\right) \frac{1}{\sqrt{q^2\cos^2\phi+\sin^2\phi}}\\
    \text{and } G_m^{(2)}(\phi ;q) &\equiv \frac{1}{\sqrt{q}}\sin \left( m\cdot \varphi(\phi ;q) \right) \frac{r}{R}(\phi ;q) \\
    &= U_{m-1} \left( \frac{q\cos\phi}{\sqrt{q^2\cos^2\phi+\sin^2\phi}}\right) \frac{\sin \phi}{q^2\cos^2\phi+\sin^2\phi}
\label{eq:G_m_1_2_def}
\end{split}
\end{equation}
and $T_m$ (resp. $U_m$) are the Chebyshev polynomials of the first (resp. second) kind.

Since Poisson's equation is linear, we can separately find the potentials $\psi_m^{(1)}(r,\phi ;q)$ and $ \psi_m^{(2)}(r,\phi ;q)$ associated with $\kappa_m^{(1)}$ and $\kappa_m^{(2)}$ - the convergence profiles determined by shape functions $G_m^{(1)}(\phi ;q)$ and $G_m^{(2)}(\phi ;q)$ via Equation~(\ref{eq:gen_pot_conv_pair}), respectively. We can then combine these two solutions with the adequate coefficients:
\begin{equation}
    \psi_m(r,\phi ;q) = a_m\sqrt{q}\cos (m\varphi_m) \psi_m^{(1)}(r,\phi ;q) + a_m\sqrt{q}\sin (m\varphi_m) \psi_m^{(2)}(r,\phi ;q)
    \label{eq:psi_m_1_2_decomp}
\end{equation}
Subsequently, if the potential is expressed as in Equation~(\ref{eq:gen_pot_conv_pair}), the shape function can be separated in two components $F_m^{(1)}(\phi ;q)$ and $F_m^{(2)}(\phi ;q)$, respectively solving the differential Equation (\ref{eq:main_diff_eq}) for $G_m^{(1)}(\phi ;q)$ and $G_m^{(2)}(\phi ;q)$. Then, it can be expressed as
\begin{equation}
    F_m(\phi ;q) = a_m\sqrt{q}\cos (m\varphi_m) F_m^{(1)}(\phi ;q) + a_m\sqrt{q}\sin (m\varphi_m) F_m^{(2)}(\phi ;q).
    \label{eq:F_m_1_2_decomp}
\end{equation}
We note, however, that the potential will not exactly follow the form of Equation (\ref{eq:gen_pot_conv_pair}) in the case of the circular multipole of order $m=1$ (see Section~\ref{subsec:circ_multipole_potential}), and of the elliptical multipoles of order $m$ odd (see Sections~\ref{subsec:m=1}, \ref{subsec:m=3}, and Appendix~\ref{App:m=2k+1_potential}).

The components of the convergence shape functions have the following symmetries:
\begin{equation}
    \begin{split}
        G_m^{(i)}(\phi ;q) & = G_m^{(i)}(\phi + 2\pi ;q) \text{ for } i=1,2 \\
        G_m^{(1)}(-\phi ;q) & = G_m^{(1)}(\phi ;q) \\
        G_m^{(2)}(-\phi ;q) & = - G_m^{(2)}(\phi ;q) \\
        G_m^{(1)}(\pi-\phi ;q) & = (-1)^m G_m^{(1)}(\phi ;q) \\
        G_m^{(2)}(\pi-\phi ;q) & = (-1)^{m-1} G_m^{(2)}(\phi ;q) \\
    \end{split}
    \label{eq:symmetries}
\end{equation}
so the components of the potential must, in theory, have the same symmetries. In the following, we will present some solutions to differential Equation~(\ref{eq:main_diff_eq}) where all symmetries are not necessarily enforced, in which case we write them with a tilde: $\tilde{F}_m^{(i)}(\phi ;q)$. We warn that these unsymmetrized shape functions cannot be directly plugged in Equation~(\ref{eq:gen_pot_conv_pair}), and that direct symmetrization attempts might create some unwanted discontinuities (see Sections~\ref{subsec:circ_multipole_potential}, \ref{subsec:m=1},  \ref{subsec:m=3} and Appendix~\ref{App:m=1_shape_fct_pbs}).

We also remark that the expression in Equation~(\ref{eq:ell_angle_polar_angle_bijection}) has the following property:   $\varphi \left(\phi + \frac{\pi}{2} ; 1/q \right) = \varphi \left(\phi ; q \right) + \frac{\pi}{2}$, provided that we extend this definition of $\varphi$ to values $q>1$. Similarly, we can easily show that $ \frac{r}{R}(\phi + \frac{\pi}{2} ; 1/q) =  \frac{r}{R}(\phi ;q)$. Combining these two properties, we have the following invariance for the shape function $G_m$:
\begin{equation}
    G_m\left(\phi+\frac{\pi}{2};\frac{1}{q},\varphi_m \right) = G_m\left(\phi;q,\varphi_m-\frac{\pi}{2}\right)
    \label{eq:invariance_rot_flip}
\end{equation}
This can be interpreted in the following way: rotating the multipole by $90^\circ$ with respect to the ellipse is equivalent to rotating the entire system (i.e., both the ellipse and the multipole pertubation) by $90^\circ$, then exchanging the semi-major and semi-minor axis of the ellipse. We expect the potential $\psi_m(\phi;q)$ to have the same invariance - in particular, choosing $\varphi_m = 0$ in Equation~(\ref{eq:invariance_rot_flip}) for an odd integer $m=2k+1$, then making use of the decompositions in Equations (\ref{eq:G_m_1_2_decomp}) and (\ref{eq:psi_m_1_2_decomp}), we obtain the following relation:
\begin{equation}
        \psi_{2k+1}^{(2)}(r,\phi ; q) = \frac{(-1)^{k+1}}{q} 
        \psi_{2k+1}^{(1)}\left(r,\phi+ \frac{\pi}{2} ; \frac{1}{q}\right)
    \label{eq:psi_1_psi_2_relation}
\end{equation}
Expressing this equation in cartesian coordinates $(x,y)$, taking derivatives, then transforming back to polar coordinates, we find similar relations for the deflection angles $\alpha_{x_i, m}^{(j)} = \frac{\partial \psi_m^{(j)}}{\partial x_i} $ ($j=1,2$ and $x_i=x,y$), and for the components of the Hessian:
\begin{equation}
\left\{\begin{array}{l}
\alpha_{x, 2k+1}^{(2)}(r,\phi;q)= \frac{(-1)^{k+1}}{q} \alpha_{y, 2k+1}^{(1)}(r,\phi+ \frac{\pi}{2};\frac{1}{q}) \\ ~~ \\
\alpha_{y, 2k+1}^{(2)}(r,\phi;q)= \frac{(-1)^k}{q} \alpha_{x, 2k+1}^{(1)}(r,\phi+ \frac{\pi}{2};\frac{1}{q}) \\ ~~ \\
\frac{\partial^2 \psi_{2k+1}^{(2)}}{\partial x^2}(r,\phi;q) = \frac{(-1)^{k+1}}{q} \frac{\partial^2 \psi_{2k+1}^{(1)}}{\partial y^2}(r,\phi+ \frac{\pi}{2};\frac{1}{q}) \\ ~~ \\ 
\frac{\partial^2 \psi_{2k+1}^{(2)}}{\partial y^2}(r,\phi;q) = \frac{(-1)^{k+1}}{q} \frac{\partial^2 \psi_{2k+1}^{(1)}}{\partial x^2}(r,\phi+ \frac{\pi}{2};\frac{1}{q}) \\ ~~ \\ 
\frac{\partial^2 \psi_{2k+1}^{(2)}}{\partial x \partial y}(r,\phi;q) = \frac{(-1)^k}{q} \frac{\partial^2 \psi_{2k+1}^{(1)}}{\partial x \partial y}(r,\phi+ \frac{\pi}{2};\frac{1}{q})
\end{array}\right.
\label{eq:alpha_hessian_1_2_relation}
\end{equation}
For odd order $m$, since $\psi_{m}^{(2)}$ can be deduced from $\psi_{m}^{(1)}$, it means that determining $\psi_{m}^{(1)}(r,\phi; q)$ (i.e., the potential in the $\varphi_m=0$ case) is sufficient to get the full solution.

We reiterate that the multipole framework explored in this work operates under the assumption that the reference profile is near-isothermal ($\gamma\approx2$). The multipole convergence profile can be set up with a more general radial slope, in order to match the reference EPL profile, for both the circular (see Appendix~\ref{App:isothermal_limit}) and the elliptical (see Appendix~\ref{App:isothermal_vs_free_slope}) formulations.  The calculation for the lensing potential of the elliptical multipoles, however, would become even more complex than in Section~\ref{sec:potential_solutions}, and as we show in Appendix~\ref{App:isothermal_vs_free_slope}, the isothermal approximation introduces an ‘‘error’’ in the convergence profile that only scales as $|2-\gamma|\ln\left(\frac{R}{\theta_E}\right)$. In particular, around the Einstein radius, applying the elliptical multipole solutions to non-isothermal profiles still provides a more accurate representation of the convergence than applying circular multipoles with a matched radial dependence to elliptical systems. We discuss this further in Section~\ref{sec:conclusion} and in Appendix~\ref{App:isothermal_vs_free_slope}.

\section{Lensing potential for the multipole perturbations}
\label{sec:potential_solutions}

In this section, we describe the lensing potentials associated with the different multipole perturbations discussed in Section~\ref{sec:multipole_comparison}. We start with the circular multipoles, paying particular attention to the special case of the circular $m=1$ order (section~\ref{subsec:circ_multipole_potential}). Then, we present detailed solutions for the elliptical multipoles, in the three cases that will be most practical for lensing studies \citep[e.g.,][]{VdV2022, Etherington2024, Oh2024, Gilman2024, Cohen2024, Keeley2024, Amvrosiadis2024, Lange2024, O'Riordan2024}: the $m=1$ order (section~\ref{subsec:m=1}), the $m=3$ order (section~\ref{subsec:m=3}), and the $m=4$ order (section~\ref{subsec:m=4}). Analytical expressions for the corresponding deflection angles are explicitly given in Appendix~\ref{App:m=1/3/4_alpha_explicit}.
The lensing potentials that are newly introduced in this section (and the corresponding deflections angles) were implemented in the lens modeling package \texttt{lenstronomy} \citep{lenstronomy,lenstronomy2} - used to generate Figures~\ref{fig:convergence SIE+m=1} to \ref{fig:ell_vs_circ_q58}.
Elliptical multipole lensing potentials for general $m\geq2$ can be explicitly determined at the cost of lengthy calculations: for the sake of completeness, we present them in Appendix~\ref{App:m=2k+1_potential} (for $m$ odd) and Appendix~\ref{App:m=2k_potential} (for $m$ even).

\subsection{Lensing potential for the circular multipoles}
\label{subsec:circ_multipole_potential}


The shape function for the lensing potential of the circular multipoles can be straightforwardly determined as \citep{Keeton2003, Xu2015}:
\begin{equation}
    F_m^{\rm circ}(\phi) =  \frac{a_m^{\rm circ}}{1-m^2}\cos(m(\phi-\phi_m)) \text{ for } m\neq 1 
\label{eq:shape_fct_pot_circ}
\end{equation}

Equation~(\ref{eq:shape_fct_pot_circ}) is no longer valid for $m=1$, so we need to treat this case separately. If we attempt to solve over all $\phi \in \mathbb{R}$ the differential Equation~(\ref{eq:main_diff_eq}) for $G_1^{\rm circ}(\phi)=a_1^{\rm circ}\cos(\phi-\phi_1)$, we find the following solution :
\begin{equation}
    \tilde{F}_1^{\rm circ}(\phi) = \frac{a_1^{\rm circ}}{2}(\phi-\phi_1)\sin(\phi-\phi_1) + A_1^{\rm circ}\cos(\phi-\phi_1) + B_1^{\rm circ}\sin(\phi-\phi_1)
    \label{eq:tilde_F_1_circ}
\end{equation}
where $A_1^{\rm circ}$ and $B_1^{\rm circ}$ are constants, that have no impact on the lensing observables due to the prismatic degeneracy (see Section~\ref{sec:general_formalism} or \citep{Gorenstein1988}).

The problem is that this solution does not have the expected symmetries:
like $G_1^{\rm circ}(\phi)$, the shape function for the potential should be $2\pi$-periodic, $\phi-\phi_1 \mapsto \pi -(\phi-\phi_1)$ antisymmetric and $\phi-\phi_1 \mapsto -(\phi-\phi_1)$ symmetric. If we choose $B_1^{\rm circ}=0$, the last symmetry is enforced in Equation~(\ref{eq:tilde_F_1_circ}), but not the first two, no matter how we choose $A_1^{\rm circ}$. One can attempt to symmetrize the shape function, but this can only be achieved at the expense of its differentiability: in Appendix~\ref{App:m=1_shape_fct_pbs}, we show how symmetrizing a potential of the form~(\ref{eq:gen_pot_conv_pair}) creates jump discontinuities in the deflection field. We want a potential that is solution of the Poisson equation over the entire domain $\mathbb{R}^2\setminus\{\mathbf{0}\}$ ($\mathbf{0}$ being a singular point for the SIE anyways), and not defined piecewise - in fact, for a given source/lens configuration, the lensing potential must be differentiable over a contiguous region connecting the positions of all images in order to be physically meaningful \citep{Wagner2018}. This means that there cannot be jump discontinuities in the deflection field.

This difficulty can be overcome by dropping the assumption that the potential has to be written as in Equation~(\ref{eq:gen_pot_conv_pair}), e.g., by allowing radial dependencies other than $\psi \propto r$. Ref.~\citep{Chu2013} proposes a lensing potential equivalent to the following expression:
\begin{equation}
    \psi_1^{\rm circ}(r,\phi) = \frac{a_1^{\rm circ}}{2}r\ln \left( \frac{r}{r_E} \right) \cos(\phi-\phi_1)
    \label{eq:psi_1_circ}
\end{equation}
which is twice differentiable over $\mathbb{R}^2\setminus\{\mathbf{0}\}$, solves Poisson's equation for the convergence of the $m=1$ circular multipole $\kappa_1^{\rm circ}(r,\phi) = a_1^{\rm circ}\frac{\cos(\phi-\phi_1)}{2r}$ , and has all the correct symmetries. Its derivatives are given by:

\begin{equation}
\left\{\begin{array}{l}
\alpha_{x,1}^{\rm circ} (r,\phi) = \frac{\partial \psi_1^{\rm circ}}{\partial x} =  \frac{a_1^{\rm circ}}{2} \left[ \cos\phi_1\ln \left( \frac{r}{r_E} \right) +\cos(\phi-\phi_1)\cos\phi \right] \\ ~~ \\
\alpha_{y,1}^{\rm circ} (r,\phi) = \frac{\partial \psi_1^{\rm circ}}{\partial y} = \frac{a_1^{\rm circ}}{2} \left[ \sin\phi_1\ln \left( \frac{r}{r_E} \right) +\cos(\phi-\phi_1)\sin\phi \right]  \\ ~~ \\
\frac{\partial^2 \psi_1^{\rm circ}}{\partial x^2}(r,\phi) = \frac{a_1^{\rm circ}}{2r} \left[ 2\cos\phi_1\cos\phi -\cos(\phi-\phi_1)\cos(2\phi) \right]  \\ ~~ \\
 \frac{\partial^2 \psi_1^{\rm circ}}{\partial y^2}(r,\phi) = \frac{a_1^{\rm circ}}{2r} \left[ 2\sin\phi_1\sin\phi +\cos(\phi-\phi_1)\cos(2\phi) \right] \\ ~~ \\ 
\frac{\partial^2 \psi_1^{\rm circ}}{\partial x \partial y}(r,\phi) = \frac{a_1^{\rm circ}}{2r} \left[ \sin(\phi+\phi_1) -\cos(\phi-\phi_1)\sin(2\phi) \right]  \\ ~~ \\ 
\end{array}\right.
\label{eq:derivatives_psi_1_circ}
\end{equation}
Additional arguments in favor of this potential are presented in Appendix~\ref{App:m=1_rlnr_justifications}: namely, equivalent solutions are found when taking the limit $\gamma \mapsto 2$ after generalizing circular multipole perturbations to non-isothermal profiles (Appendix~\ref{App:isothermal_limit}), and when using direct integration to calculate the deflection angles in the complex formulation of gravitational lensing (Appendix~\ref{App:m=1_complex_alpha_int}).  We present example maps of this lensing potential and deflection angles for the circular $m=1$ multipole in Figure~\ref{fig:potential_m=1}.

The normalizing radius $r_E$, introduced in Equations~(\ref{eq:psi_1_circ}) and (\ref{eq:derivatives_psi_1_circ}), modulates a term $\propto r \cos(\phi-\phi_1)$ in the lensing potential. Therefore, as a consequence of the prismatic degeneracy (see Section~\ref{sec:general_formalism} or \citep{Gorenstein1988}), it does not have any impact on the lensing observables. We choose the following convention: $r_E=\theta_E$, such that $\vec{\alpha}_1^{\rm \ circ}(\theta_E,\phi_1\pm\pi/2) =0$, i.e., at the Einstein radius, in the direction orthogonal to the multipole orientation, the additional deflection from the multipole is zero (see Figure~\ref{fig:potential_m=1}).

\subsection{The $m=1$ elliptical multipole lensing potential}
\label{subsec:m=1}

For the elliptical $m=1$ multipole, we use the decomposition of the convergence and potential introduced in Equations~(\ref{eq:G_m_1_2_decomp}) and (\ref{eq:psi_m_1_2_decomp}). We start with the case where $\varphi_1=0$, by solving the differential equation:
\begin{equation}
        F_1^{(1)\prime\prime}(\phi ;q) + F_1^{(1)}(\phi ;q) = G_1^{(1)}(\phi ;q) = \frac{ \cos \phi}{ q^2 \cos^2 \phi + \sin^2 \phi }
\end{equation}
where $^\prime$ still denotes the derivative with respect to $\phi$. We find the following solution over $\phi \in \mathbb{R}$:
\begin{multline}
    \tilde{F_1}^{(1)}(\phi ;q) = \frac{1}{1 - q^2} \Big[-\frac{q}{2}\cos(\phi) \ln \left(1 + q^2 + (q^2 - 1) \cos(2\phi)\right) + A_1^{(1)} \cos\phi  + B_1^{(1)} \sin\phi \\ - \sin(\phi) \left[q \phi - \varphi(\phi ;q) \right] \Big]
    \label{eq:f_1_1}
\end{multline}

where $A_1^{(1)}(q),B_3^{(1)}(q)$ are ‘‘constants’’ (i.e., independent of $\phi$, they can still depend on the axis ratio $q$), and we use Equation~(\ref{eq:ell_angle_polar_angle_bijection}) for the definition of $\varphi(\phi ;q)$.

Unfortunately, this has the same issues as $\tilde{F}_1^{\rm circ}(\phi)$ for the circular $m=1$ case: the expected symmetries ($2\pi$ periodicity, $\phi \mapsto -\phi$ symmetry, $\phi\mapsto \pi-\phi$ antisymmetry) are not enforced, because of the term in $(q\phi - \varphi(\phi,q))\sin\phi$ this time. Once again, trying to symmetrize the shape function directly leads to discontinuities in the deflection angle. 
To remedy this issue, we recognize that the expression $\phi - \varphi(\phi,q)$ (note the equal coefficients) is $2\pi$ periodic, and antisymmetric under $\phi \mapsto -\phi$ and $\phi\mapsto \pi-\phi$. In turn, the following ‘‘modified’’ shape function
\begin{equation}
\begin{split}
    \hat{F}_1^{(1)}(\phi;q) &\equiv \tilde{F}_1^{(1)}(\phi;q) - \frac{1-q}{1-q^2} \phi \sin\phi\\
    &= \tilde{F}_1^{(1)}(\phi;q) - \lambda_1(q)\cdot\tilde{F}_1^{\rm circ}(\phi) \Big|_{\phi_1=0, a_1^{\rm circ}=1} \text{ with } \lambda_1(q) \equiv \frac{2}{1+q} 
    \label{eq:F_1_1_hat}
\end{split}
\end{equation}
has the correct symmetries expected from the $m=1$ shape function, provided that we choose $B_1^{(1)}=0$ for the $\phi \mapsto -\phi$ symmetry. This  function has the added benefit of being infinitely differentiable with respect to $\phi$: in particular, it corresponds to a potential that has continuous deflection angles in $\mathbb{R}^2\setminus\{\mathbf{0}\} $. The convergence profile associated with this shape function is $\frac{1}{2r} [\hat{F}_1^{(1)\prime\prime}(\phi;q)+ \hat{F}_1^{(1)}(\phi;q)] = \frac{G_1^{(1)}(\phi;q)}{2r} - \lambda_1(q) \frac{\cos\phi}{2r}$, which is the profile we desire minus a $m=1$ circular multipole component. Therefore, by decomposing the potential as
\begin{equation}
    \psi_1^{(1)}(r,\phi;q) = r \hat{F}_1^{(1)}(\phi;q) + \lambda_1(q) \cdot \psi_1^{\rm circ}\left(r,\phi \right) \Big|_{\phi_1=0, a_1^{\rm circ}=1}
    \label{eq:psi_1_1}
\end{equation}
we have found a solution for the first component of the $m=1$ elliptical multipole that has (a) the correct symmetries, and (b) no discontinuities in the deflection field. Derivatives of this potential are obtained by combining (with the appropriate coefficients) Equation~(\ref{eq:pot_alpha_hessian_array}) with the shape function $\hat{F}_1^{(1)}(\phi;q)$ and Equation~(\ref{eq:derivatives_psi_1_circ}).

The constants $A_1^{(1)}$ and $B_1^{(1)}$ are a priori not constrained because of the prismatic degeneracy, but we actually need to take $B_1^{(1)}=0$ if we want to enforce the $\phi \mapsto -\phi$ symmetry. $A_1^{(1)}$ cannot be determined by symmetries alone, but we can choose its value to ensure the convergence to the circular solution in the limit $q\to 1$. Since $\lambda_1(q) \mathrel{\underset{q \to 1}{\longrightarrow}} 1$, we want $\hat{F}_1^{(1)}(\phi;q) \mathrel{\underset{q \to 1}{\longrightarrow}} 0$, or equivalently $\tilde{F}_1^{(1)}(\phi;q) \mathrel{\underset{q \to 1}{\longrightarrow}} \tilde{F}_1^{\rm circ}(\phi) \Big|_{\phi_1=0, a_1^{\rm circ}=1} = \frac{1}{2} \phi \sin\phi$. Performing a Taylor expansion of Equation~(\ref{eq:f_1_1}) in $(1-q)$, we find:
\begin{equation}
 \tilde{F_1}^{(1)}(\phi ;q) \mathrel{\underset{q \to 1}{=}} \frac{A_1^{(1)}}{1 - q^2}\cos\phi -  \frac{\ln(2)}{4(1 - q)}\cos\phi + \frac{4 + \ln(2)}{8} \cos\phi+ \frac{1}{2} \phi \sin\phi  + O[1-q] \\
\end{equation}
so, in order to converge to the circular multipole solution, we need
\begin{equation}
    A_1^{(1)}(q) = \frac{\ln 2}{4}(1+q) - \frac{1-q^2}{2} \left(1+ \frac{\ln 2}{4} \right) + f(q) \text{ with }  (1-q)f(q)\mathrel{\underset{q \to 1}{\longrightarrow}} 0
    \label{eq:A_1_1}
\end{equation}
Because of the prismatic degeneracy we can choose any function $f(q)$ without changing the lensing observables, and as long as $(1-q)f(q)\mathrel{\underset{q \to 1}{\longrightarrow}} 0$ the potential will converge to the adequate limit when $q\to1 $. We therefore choose the simplest solution by taking $f(q)=0$ in Equation~(\ref{eq:A_1_1}).

\begin{figure*}[ht!]
    \centering
    \includegraphics[width=0.85\linewidth]{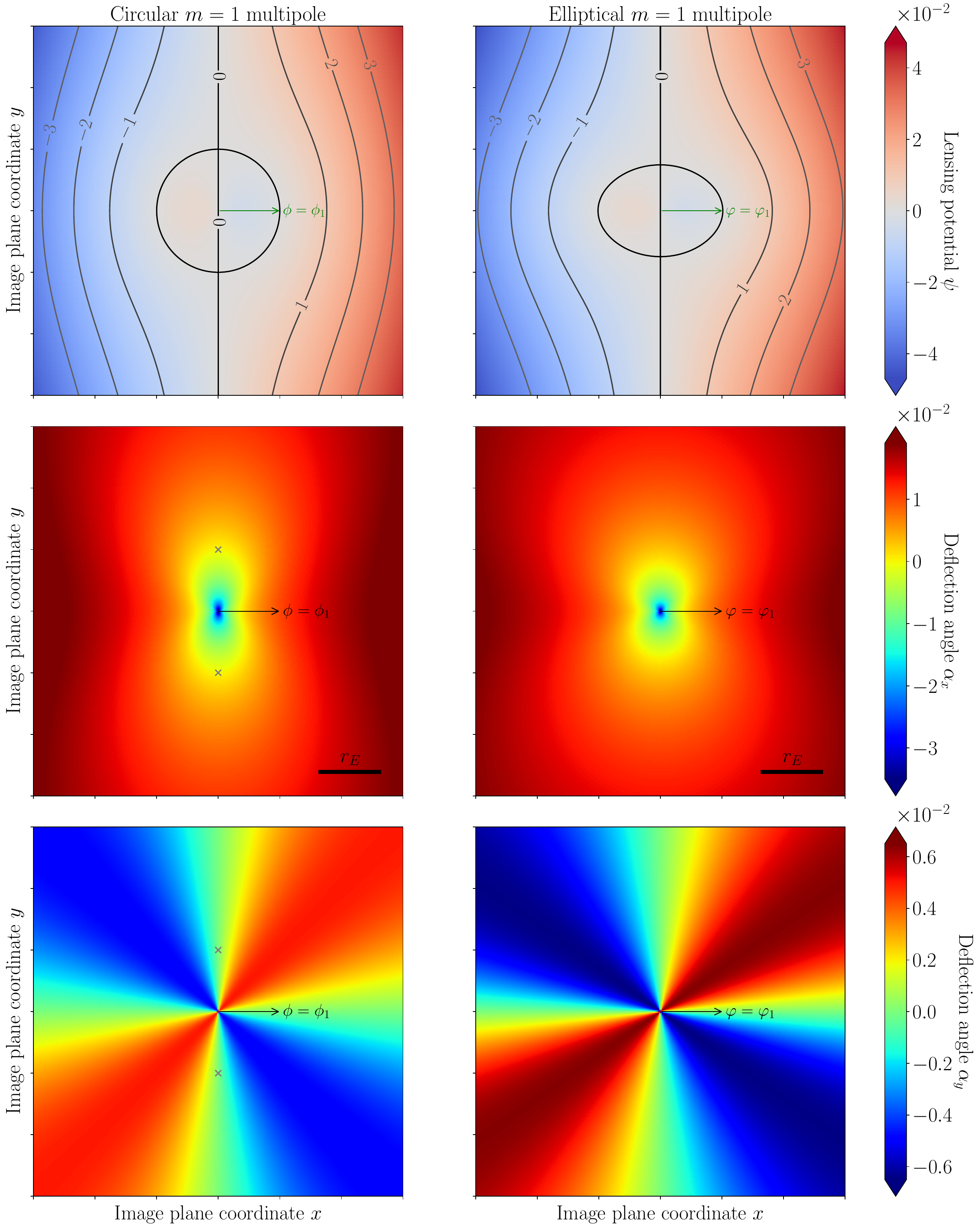}
    \caption{Example of lensing potential $\psi$ (top panel) and deflection angles $\alpha_x$ (middle panel) and $\alpha_y$ (bottom panel) associated with a $m=1$ multipole perturbation in the circular (left) and elliptical  (right) formulations. We use direction $\phi_1=0$ (resp. $\varphi_1=0$), and axis ratio $q=0.5$ for the elliptical multipole. The amplitude $a_1^{\rm circ}=0.02$ (resp.  $a_1=0.02$) is not rescaled since the multipole is not yet applied to an elliptical profile. The grey crosses are located at $r=r_E$ in the direction orthogonal to $\phi_1$, where $\alpha_x=\alpha_y=0$.}
    \label{fig:potential_m=1}
\end{figure*}

The discussion above fully describes the elliptical $m=1$ potential in the case where $\varphi_1=0$. In order to obtain the more general solution for any $\varphi_1$, we need to determine the other component $\psi_1^{(2)}$ of the decomposition presented in Equation~(\ref{eq:psi_m_1_2_decomp}). For this, we can simply exploit the invariance established earlier in Equations~(\ref{eq:psi_1_psi_2_relation}) and (\ref{eq:alpha_hessian_1_2_relation}): in particular, we have $ \psi_1^{(2)}(r,\phi;q) = -\frac{1}{q}\psi_1^{(1)}\left(r,\phi+\frac{\pi}{2};\frac{1}{q}\right)$. We note that this automatically ensures the convergence of $\psi_1^{(2)}$ to the $\phi_1 = \pi/2$ circular solution when $q\mapsto 1$. The explicit expressions for the deflection angles of the $m=1$ elliptical multipole are given in Appendix~\ref{App:m=1_alpha_explicit}. We present example maps of the lensing potential and deflection angles for the $m=1$ elliptical multipole in Figure~\ref{fig:potential_m=1}, where they are compared to the $m=1$ circular multipole solution.

\subsection{The $m=3$ elliptical multipole lensing potential}
\label{subsec:m=3}

We start by finding a solution for the $m=3$ only, $\varphi_3=0$ case, i.e., for the differential equation:
\begin{equation}
        F_3^{(1)\prime\prime}(\phi ;q) + F_3^{(1)}(\phi ;q) = G_3^{(1)}(\phi ;q) = \frac{4q^3 \cos^3 \phi}{\left( q^2 \cos^2 \phi + \sin^2 \phi \right)^2} - \frac{3q \cos \phi}{ q^2 \cos^2 \phi + \sin^2 \phi }
\end{equation}
where we have used $T_3(x) = 4x^3-3x$ in Equation~(\ref{eq:G_m_1_2_def}). The solution over all $\phi \in \mathbb{R}$ is:

\begin{equation}
\begin{split}
    \tilde{F_3}^{(1)}(\phi ;q) = \frac{1}{2 (1 - q^2)^2} \Big[& \cos(\phi) q \left(3 + q^2\right) \ln \left(1 + q^2 + (q^2 - 1) \cos(2\phi)\right) + 2 A_3^{(1)} \cos\phi  \\
    & + 2 B_3^{(1)} \sin\phi + 2 \sin(\phi) \left[q \left(3 + q^2 \right) \phi - \left(1 + 3 q^2\right) \varphi(\phi ;q) \right] \Big]
    \label{eq:f_3_1}
\end{split}
\end{equation}
where $A_3^{(1)}(q),B_3^{(1)}(q)$ are constants in the same sense as before, i.e., independent of $\phi$. We note that the expected $\phi \mapsto -\phi$ symmetry can be enforced in Equation~(\ref{eq:f_3_1}) if, and only if, $B_3^{(1)}=0$; but that this expression cannot be $2\pi$-periodic or $\phi \mapsto \pi-\phi$ antisymmetric because of the terms in $\phi$ and $\varphi(\phi ;q)$. This causes the same problem as in Sections~\ref{subsec:circ_multipole_potential} and \ref{subsec:m=1}: the shape function cannot be properly symmetrized without creating discontinuities in the deflection field. Instead, we use the same trick as in Section~\ref{subsec:m=1}, breaking the potential into a component ($\propto r$) with a shape function having the correct symmetries, and a circular $m=1$ multipole component ($\propto r\ln r$):
\begin{equation}
    \psi_3^{(1)}(r,\phi;q) = r \hat{F}_3^{(1)}(\phi;q) + \lambda_3(q) \cdot \psi_1^{\rm circ}\left(r,\phi \right) \Big|_{\phi_1=0, a_1^{\rm circ}=1}
    \label{eq:psi_3_1}
\end{equation}

where we have defined $\lambda_3(q) = -\frac{2(1-q)}{(1+q)^2}$ and 
\begin{equation}
\begin{split}
    \hat{F}_3^{(1)}(\phi;q)  &= \tilde{F_3}^{(1)}(\phi ;q) -  \frac{\lambda_3(q)}{2}\phi\sin\phi\\
    &= \tilde{F_3}^{(1)}(\phi ;q) - \lambda_3(q) \cdot\tilde{F}_1^{\rm circ}(\phi) \Big|_{\phi_1=0, a_1^{\rm circ}=1} \\
    \label{eq:F_3_1_hat}
\end{split}
\end{equation}
The additional term in $\lambda_3(q)$ equalizes the coefficients in front of $\phi$ and $\varphi(\phi;q)$ in Equation~(\ref{eq:f_3_1}): thanks to the properties of $\phi - \varphi(\phi,q)$ (see Section~\ref{subsec:m=1}), this means that $\hat{F}_3^{(1)}(\phi;q)$ is $2\pi$ periodic, symmetric under $\phi\mapsto -\phi $ and antisymmetric under $\phi\mapsto \pi-\phi $, provided that we choose $B_3^{(1)}=0$.

The value of $A_3^{(1)}(q)$ cannot be chosen from symmetries alone. Adopting the same approach as for the $m=1$ case, we look to match our solution with the circular multipole solution when $q \to 1$. In this limit, $\lambda_3(q) \to 0$, so we want $ \tilde{F_3}^{(1)}(\phi;q) \mathrel{\underset{q \to 1}{\longrightarrow}} F_3^{\rm circ}(\phi) \Big|_{\phi_3=0, a_3^{\rm circ}=1} = - \frac{1}{8} \cos(3\phi)$. Performing the Taylor expansion of Equation~(\ref{eq:f_3_1}):

\begin{equation}
\begin{split}
 \tilde{F_3}^{(1)}(\phi ;q)  \mathrel{\underset{q \to 1}{=}} \frac{A_3^{(1)}}{(1 - q^2)^2}\cos\phi +  \frac{\ln(2)\cos\phi}{2 (1 - q)^2} - \frac{(4 + \ln(2)) \cos\phi}{4 (1 - q)} + \frac{1}{2} \cos\phi \sin^2\phi + O[1-q] & \\
  \mathrel{\underset{q \to 1}{=}} \frac{A_3^{(1)}}{(1 - q^2)^2}\cos\phi + \frac{ \ln(2)\cos\phi}{2 (1 - q)^2} - \frac{ (4 + \ln(2))\cos\phi}{4 (1 - q)} + \frac{1}{8} \cos\phi - \frac{1}{8} \cos(3\phi)  + O[1-q] &
 \end{split}
\end{equation}

so, in order to converge to the $\phi_3=0$ circular multipole solution, we take:
\begin{equation}
    A_3^{(1)}(q) = - \frac{(1 + q)^2}{2} \left[\ln(2)  - 2 (1 - q) \left(1 + \frac{\ln(2)}{4}\right) + \frac{(1 - q)^2}{4}\right].
    \label{eq:A_3_1}
\end{equation}
Once again, because of the prismatic degeneracy, we could add any term $f(q)$ to $A_3^{(1)}$ as long as $f(q)(1-q)^{-2}\mathrel{\underset{q \to 1}{\longrightarrow}}0$ without changing the lensing observables, so we choose $f(q)=0$ for simplicity.

\begin{figure*}[t!]
    \centering
    \includegraphics[width=0.99\linewidth]{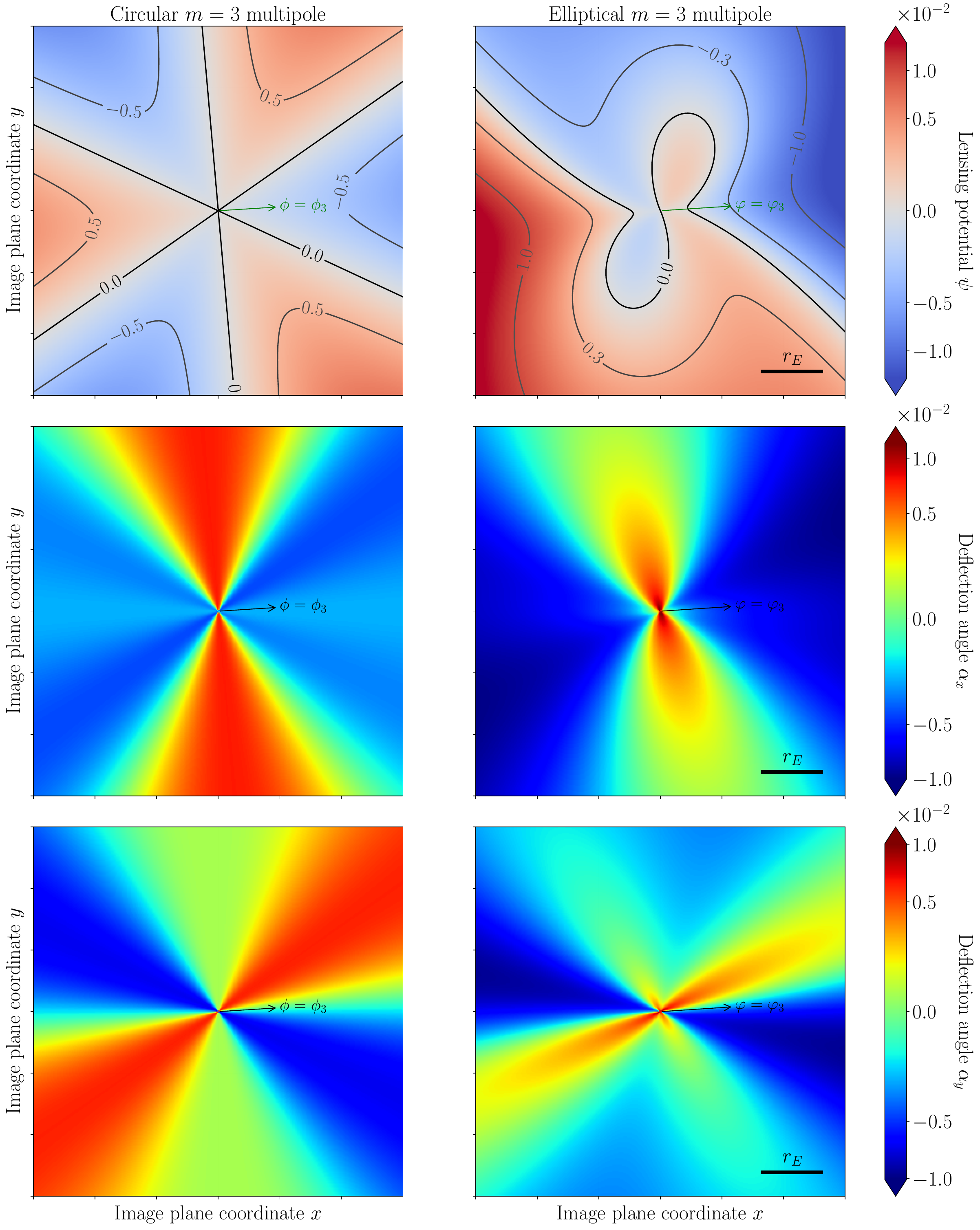}
    \caption{Same as Figure~\ref{fig:potential_m=1}, but for a $m=3$ multipole with direction $\phi_3=\frac{\pi}{36}$ (resp. $\varphi_3=\varphi(\frac{\pi}{36};q)$) and amplitude $a_3^{\rm circ}=0.02$ (resp. $a_3=0.02$). The elliptical multipole still has axis ratio $q=0.5$.}
    \label{fig:potential_m=3}
\end{figure*}

To find the general solution for the $m=3$ case, we still need to find $\psi_3^{(2)}$, i.e., solve the potential in the case where $3\varphi_3=\pi/2$. Like in Section~\ref{subsec:m=1}, we can simply exploit the invariance described by Equation~(\ref{eq:psi_1_psi_2_relation}), writing:
\begin{equation}
    \psi_3^{(2)}(r,\phi ; q) = \frac{1}{q} \psi_3^{(1)}\left(r,\phi+ \frac{\pi}{2} ; \frac{1}{q}\right)
    \label{eq:psi_3_2}
\end{equation} 
which automatically ensures that the limit $q\mapsto 1$ matches the circular solution. Explicit expressions for the deflection field of the $m=3$ elliptical multipole are given in Appendix~\ref{App:m=3_alpha_explicit}. We showcase example maps of the lensing potential and deflection angles for the $m=3$ elliptical multipole in Figure~\ref{fig:potential_m=3}, where they are compared to the $m=3$ circular multipole maps. We present an even more general solution, valid for any $m\geq 2$ odd, in Appendix~\ref{App:m=2k+1_potential}.

\subsection{The $m=4$ elliptical multipole lensing potential}
\label{subsec:m=4}

For the $m=4$ case, we give explicit solutions for $F_4^{(1)}(\phi ; q)$ and $F_4^{(2)}(\phi ; q)$, which are respectively solutions of the following differential equations:
\begin{equation}
    \begin{split}
        F_4^{(1)\prime\prime}(\phi ;q) + F_4^{(1)}(\phi ;q) &= G_4^{(1)}(\phi ;q) = \frac{1}{\sqrt{q^2 \cos^2 \phi + \sin^2 \phi }} - \frac{8q^2 \sin^2 \phi \cos^2 \phi}{ (q^2 \cos^2 \phi + \sin^2 \phi)^{5/2} }\\
        F_4^{(2)\prime\prime}(\phi ;q) + F_4^{(2)}(\phi ;q) &= G_4^{(2)}(\phi ;q) = \frac{8q^3 \sin \phi \cos^3 \phi}{ (q^2 \cos^2 \phi + \sin^2 \phi)^{5/2}} - \frac{4q \sin \phi \cos \phi}{ (q^2 \cos^2 \phi + \sin^2 \phi)^{3/2}} 
    \end{split}
\end{equation}
where we have used $T_4(x) = 8x^4-8x^2+1$ and $U_3(x) = 8x^3-4x$ in Equation~(\ref{eq:G_m_1_2_def}). The solutions over all $\phi \in \mathbb{R}$ are:

\begin{equation}
\begin{split}
F_4^{(1)}(\phi ; q) = & \frac{-4\sqrt{2}\left(1 + 4q^2 + q^4 + (q^4 - 1)\cos(2\phi)\right)}{3\left(1 - q^2\right)^2 \sqrt{1 + q^2 + (q^2 - 1)\cos(2\phi)}} \\ & +  \frac{1 + 6q^2 + q^4}{(1 - q^2)^{5/2}} \left[  A_4^{(1)} \cos\phi  + \cos(\phi) \arctan \left( \frac{\sqrt{2(1 - q^2)} \cos(\phi)}{\sqrt{1 + q^2 + (q^2 - 1) \cos(2\phi)}} \right)  \right. \\ &   \left.  + B_4^{(1)} \sin\phi +  \sin(\phi) \log \left( \frac{\sqrt{1 - q^2} \sin(\phi)}{q} + \sqrt{1 + \frac{(1 - q^2)}{q^2} \sin^2(\phi)} \right) \right]
\label{eq:F_4_1}
\end{split}
\end{equation}
\begin{equation}
\begin{split}
\text{and } F_4^{(2)}(\phi ; q) = & \ \frac{-4 \sqrt{2}\ q \sin(2\phi)}{3 (1 - q^2) \sqrt{1 + q^2 + (q^2 - 1) \cos(2\phi)}} \\ & + \frac{4q (1 + q^2)}{(1 - q^2)^{5/2}} \left[ A_4^{(2)} \cos\phi  - \sin(\phi) \arctan \left( \frac{\sqrt{2 (1 - q^2)} \cos(\phi)}{\sqrt{1 + q^2 + (q^2 - 1) \cos(2\phi)}} \right) \right. \\ & \left.   + B_4^{(2)} \sin\phi + \cos(\phi) \log \left( \frac{\sqrt{1 - q^2} \sin(\phi)}{q} + \sqrt{1 + \frac{(1 - q^2)}{q^2} \sin^2(\phi)} \right) \right]
\label{eq:F_4_2}
\end{split}
\end{equation}

We note that these expressions are already $2\pi$-periodic, and that they  have the correct symmetries:
\begin{equation}
\begin{split}
F_4^{(1)}(\phi ; q) &= F_4^{(1)}(\pi-\phi ; q)= F_4^{(1)}(-\phi ; q)   \\ 
F_4^{(2)}(\phi ;q) &=  - F_4^{(2)}(\pi-\phi ; q)= - F_4^{(2)}(-\phi ; q) 
\end{split}
\end{equation}
if, and only if, we take $A_4^{(1)}=B_4^{(1)}=A_4^{(2)}=B_4^{(2)}=0$. Furthermore, the $q\to1$ limit is already matching the circular multipole solution: we have,  for all $\phi \in \mathbb{R}$,
\begin{equation}
\begin{split}
    F_4^{(1)}(\phi;q) &\mathrel{\underset{q \to 1}{\longrightarrow}} - \frac{1}{15} \cos(4\phi) = F_4^{\rm circ}(\phi) \Big|_{\phi_4=0, \ a_4^{\rm circ}=1} \\
    \text{ and } F_4^{(2)}(\phi;q) & \mathrel{\underset{q \to 1}{\longrightarrow}} - \frac{1}{15} \sin(4\phi) = F_4^{\rm circ}(\phi) \Big|_{4\phi_4=\pi/2, \ a_4^{\rm circ}=1}.
\end{split}
\end{equation}
Subsequently, the potential is directly obtained by plugging $F_4^{(1)}(\phi;q)$ and $ F_4^{(2)}(\phi;q)$ in Equation~(\ref{eq:F_m_1_2_decomp}), then using the resulting shape function $F_4(\phi;q)$ in Equation~(\ref{eq:gen_pot_conv_pair}) - in particular, there is no need to split the potential into a $\propto r$ and a $\propto r\ln r$ component like for the $m=1$ and $m=3$ cases. The $m=4$ elliptical lensing potential is thus determined, and we report the corresponding deflection angles explicitly in Appendix~\ref{App:m=4_alpha_explicit}. In Figure~\ref{fig:potential_m=4}, we display example maps of this lensing potential and deflection angles, and compare them to the $m=4$ circular multipole maps. We present an even more general solution, valid for any $m\geq 2$ even, in Appendix~\ref{App:m=2k_potential}.

\begin{figure*}[!t]
    \centering
    \includegraphics[width=0.99\linewidth]{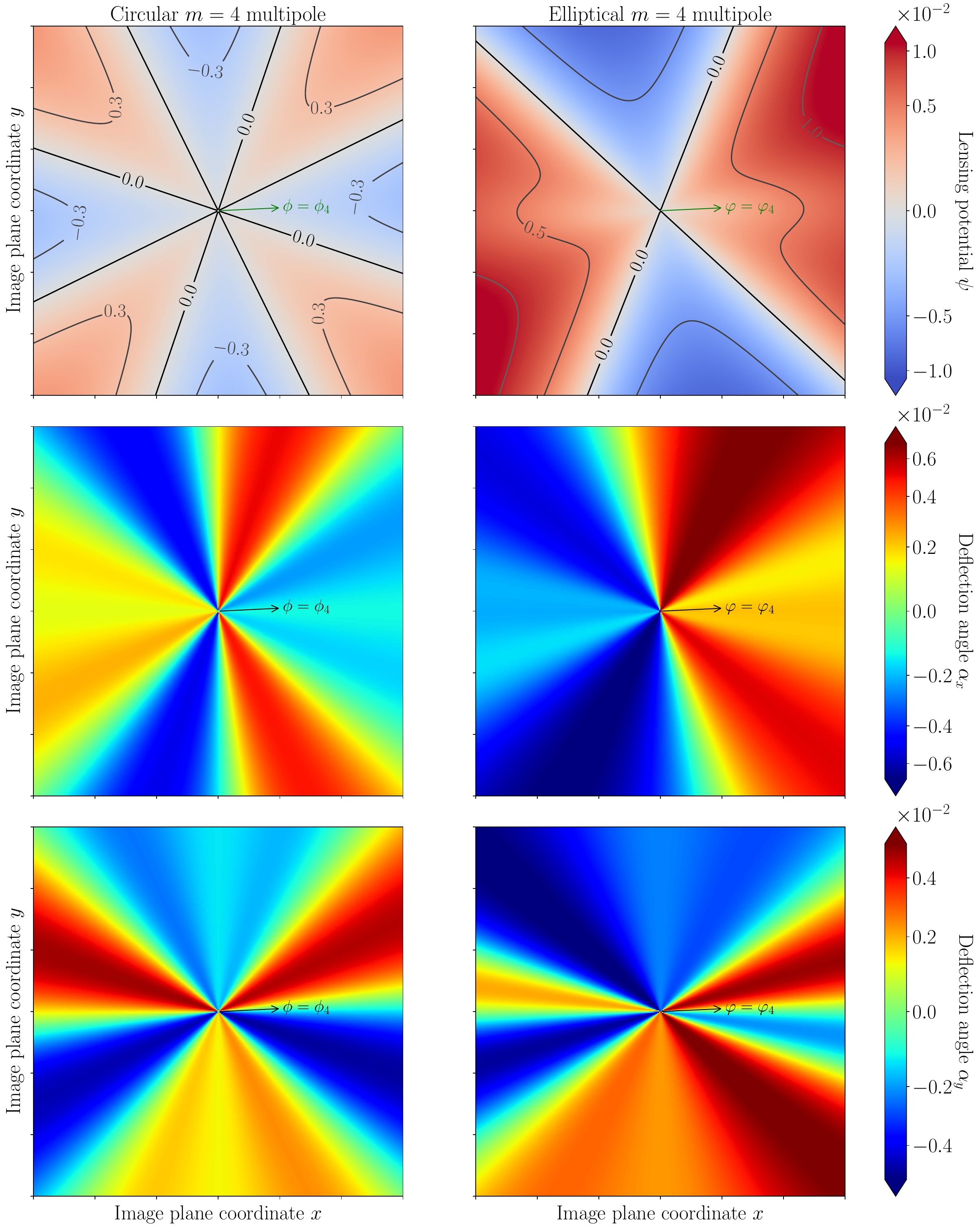}
    \caption{Same as Figure~\ref{fig:potential_m=1}, but for a $m=4$ multipole with direction $\phi_4=\frac{\pi}{48}$ (resp. $\varphi_4=\varphi(\frac{\pi}{48};q)$) and amplitude $a_4^{\rm circ}=0.02$ (resp. $a_4=0.02$). The elliptical multipole still has axis ratio $q=0.5$.}
    \label{fig:potential_m=4}
\end{figure*}

\section{Consequences for lens modeling \& flux-ratio anomalies}
\label{sec:flux_ratio_application}

Strong gravitational lensing offers a unique opportunity to directly probe the properties of DM on small scales, without relying on baryonic tracers. Strong lens systems in which a quasar becomes quadruply imaged are particularly worthy of attention: the flux ratios (i.e., the relative magnifications between lensed images) that are observed frequently disagree with the prediction made by smooth macromodels of the lensing galaxy. These ‘‘flux-ratio anomalies’’ are usually thought to be caused by the additional, localized gravitational fields associated with dark substructure (i.e., DM subhalos within the main lens) or DM halos along the line of sight. Since the abundance, mass function and internal density profiles of these halos depend on the nature of DM, population-level statistics of flux-ratio anomalies can be used to test the CDM paradigm and place constraints on various alternative DM models \citep{Mao&Schneider1998, Metcalf2001, Dalal2002, Xu2015, Hezaveh2016, Birrer2017, Gilman2019, Gilman2020, Hsueh2020, Gilman2021, Laroche2022, Zelko2022, Gilman2023, Dike2023, Gilman2024}.

These methods require careful modeling, since they implicitely assume that the macromodel is accurate enough to statistically make reliable predictions of the flux ratio in the absence of substructure. In individual systems, specific configurations of azimuthal structure can mimic the effect of DM halos on flux ratios, so sufficient model flexibility is essential to ensure that the statistics of flux-ratio anomalies are unbiased \citep{Oh2024,Cohen2024, Gilman2024, Keeley2024}. For this reason, recent studies have included (circular) multipole terms in their macromodels, primarily the $m=1$, $m=3$ and $m=4$ orders \citep[e.g.,][]{Gilman2023, Keeley2024, Lange2024, Nightingale2024}.

In this section, we assess the impact of the multipole formulation (circular vs elliptical) on flux-ratio anomaly measurements, presenting simple tests on example lens systems. Section~\ref{subsec:flux_ratio_perturbations} investigates the distribution of flux-ratio perturbations caused by multipoles under a common, physically motivated prior for the multipole amplitudes. This provides a relevant metric for studies that leverage flux-ratio anomalies to determine the relative likelihood of dark matter substructure models - since this kind of measurement aims to disentangle the flux-ratio perturbations from multipoles and those from substructure at a statistical level \citep[e.g.,][]{Gilman2024, Keeley2024}. In Section~\ref{subsec:mock lenses}, we generate mock images with a realistic elliptical multipole feature, then attempt to model the imaging data using circular multipoles, showing that some lens parameters are biased and that the flux ratios are not properly recovered.
A more definitive analysis would require to repeat the entire substructure forward-modeling procedure \citep{Gilman2019, Gilman2020, Gilman2024}, which is out of the scope of this work.

\subsection{Statistics of flux-ratio perturbations from multipoles}
\label{subsec:flux_ratio_perturbations}

\begin{figure*}[b]
    \centering
    \includegraphics[width=0.99\linewidth]{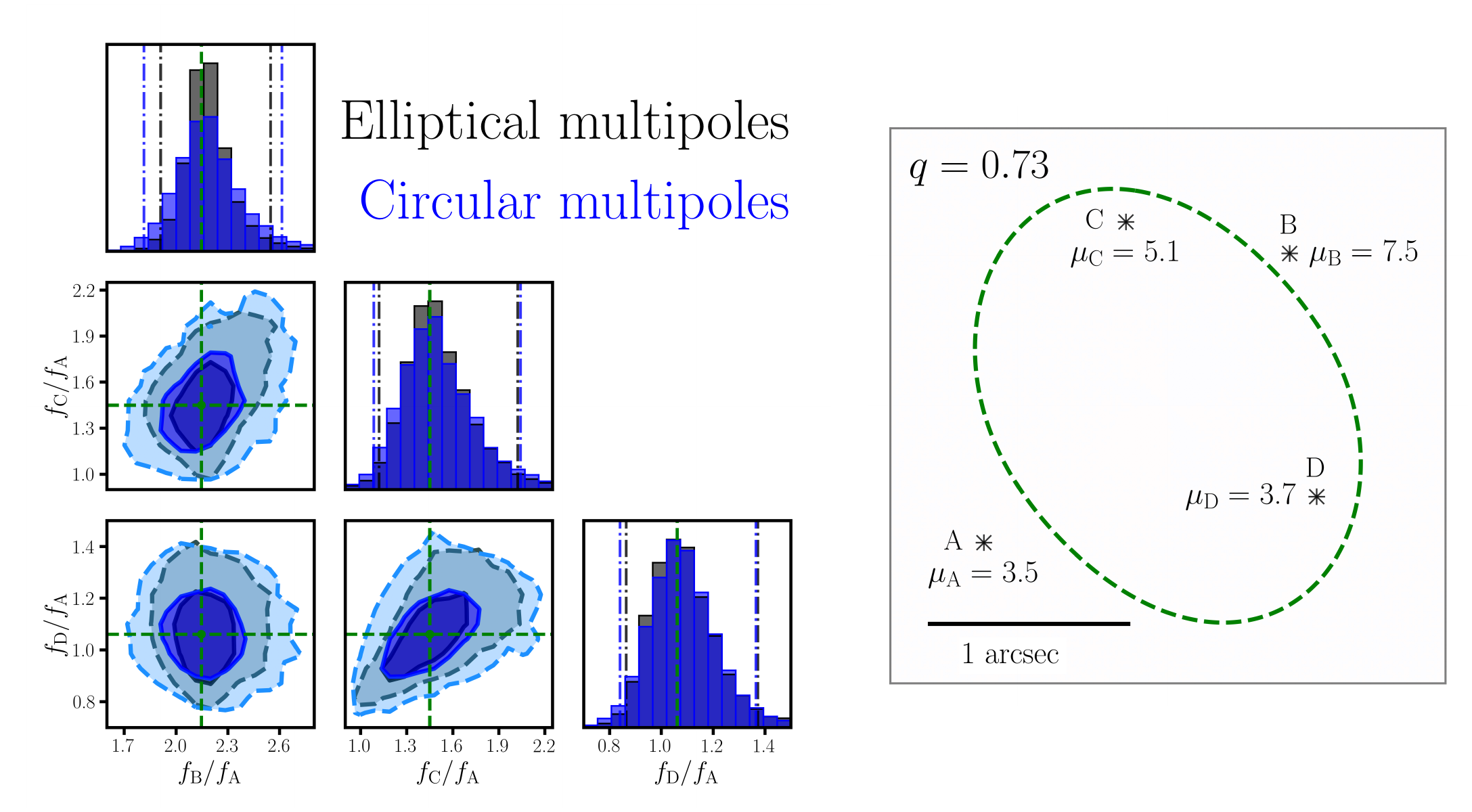}
    \caption{Distribution of flux ratios calculated from 10000 EPL + shear + $m=1$ + $m=3$ + $m=4$ lens models with axis ratio $q=0.73$, identical image positions, and randomly sampled multipole parameters, in the elliptical vs circular formulation. The 2D contours enclose 68\% and 95\% credible regions, and the vertical dosh-dotted bars in the marginal distributions represent 95\% credible intervals. The right panel shows a schematic view of the image positions and magnifications, and of the lens critical curve in the EPL+shear case (i.e., without multipoles); the corresponding flux ratios are indicated by the green crosshairs.}
    \label{fig:ell_vs_circ_q73}
\end{figure*}

\begin{figure*}[!h]
    \centering
    \includegraphics[width=0.99\linewidth]{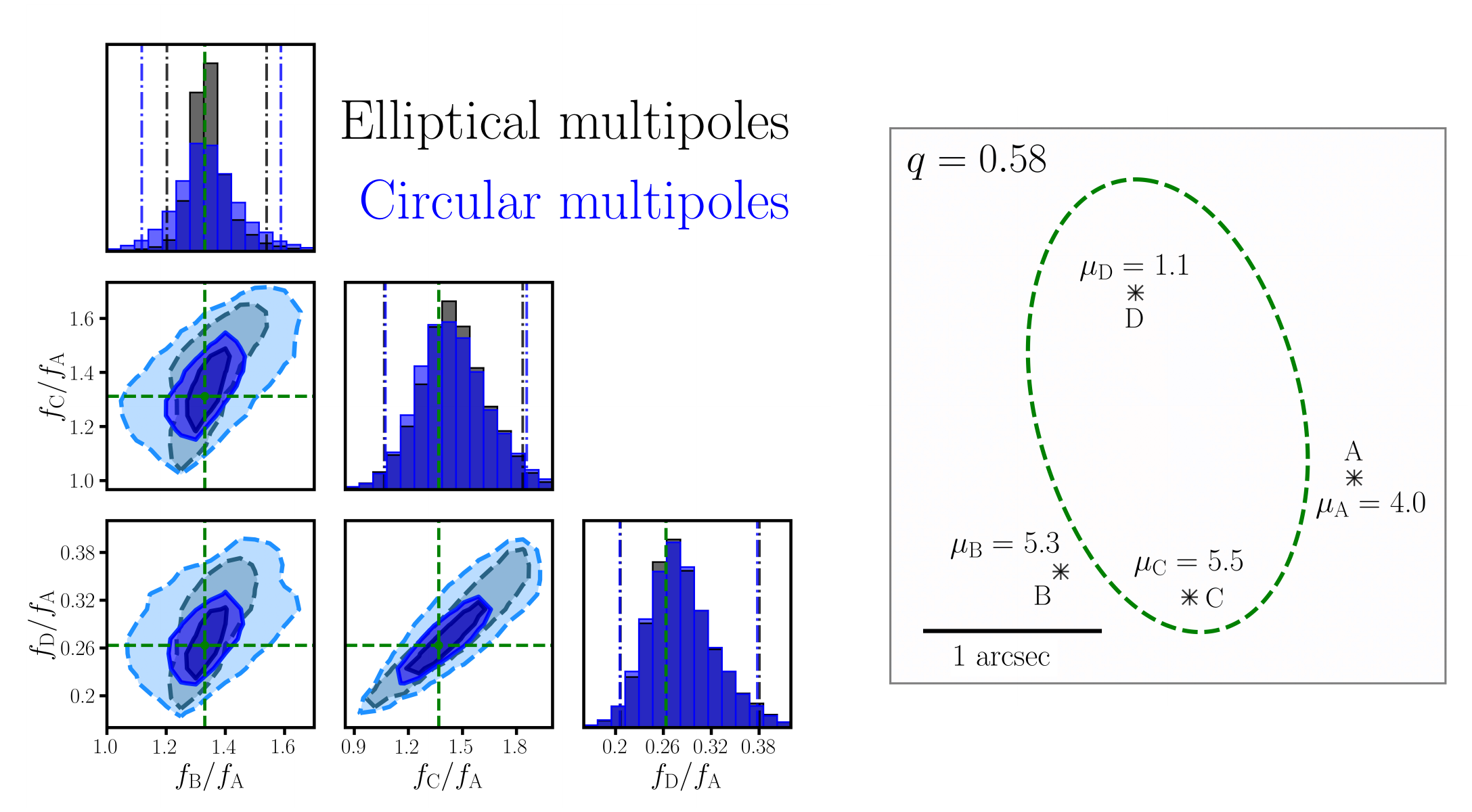}
    \caption{Same as Figure~\ref{fig:ell_vs_circ_q73}, but for lens models with axis ratio $q=0.58$.}
    \label{fig:ell_vs_circ_q58}
\end{figure*}

In our improved elliptical multipole formalism, the parameters and patterns of perturbation around the ellipse are distinct from the circular case, thus the amount of flux-ratio perturbations that can be attributed to such terms should also differ. We investigated the impact of the change in formulation by conducting the following experiment. We considered two reference mock lensed systems, generated using an EPL profile with near-isothermal slope and non-negligible ellipticities (one with $q=0.73$ and one with $q=0.58$), plus an external shear term, and a point source placed inside the inner caustic in order to produce four images. We then added $m=1$, $m=3$, and $m=4$ multipole terms with directions and amplitudes randomly sampled from physically motivated distributions. Finally, keeping these perturbations fixed, we adjusted the other degrees of freedom in the macromodel (source position and EPL + shear parameters) in order to recover the initial image positions, assuming an astrometric precision of $5$ {\rm mas} \citep{Keeley2024}. We adopted uniform priors on $[-\pi, \pi]$ for the direction of the multipoles relative to the reference ellipse ($\phi_m$ or $\varphi_m$), and Gaussian priors informed by isophotal shape studies for the amplitudes ($a_m^{\rm circ}$ or $a_m$) - with mean $0$ and standard deviation $0.005$, $0.005$, and $0.01$ for the $m=1$, $m=3$, and $m=4$ orders, respectively \citep[e.g.,][]{Amvrosiadis2024,Hao2006}. After taking $10^4$ samples of the multipole configurations, we compared the distribution of flux ratios obtained with the elliptical versus the circular formulation. These distributions, shown in Figures~\ref{fig:ell_vs_circ_q73} and \ref{fig:ell_vs_circ_q58}, and can be essentially interpreted as conditional probability distributions $\mathbf{p}(\text{flux ratios}\big| \text{image positions})$, assuming an EPL + shear + $m=1$ + $m=3$ + $m=4$ lens model (without substructure) with the chosen priors on multipole parameters. 

As Figures~\ref{fig:ell_vs_circ_q73} and \ref{fig:ell_vs_circ_q58} illustrate, the inclusion of multipoles with either formulation can generate substantial perturbations to the flux ratios predicted by the fiducial model (i.e., the one with a purely elliptical lens + external shear). The circular multipoles, however, tend to produce larger flux-ratio perturbations compared to the elliptical version (in particular, the distributions have wider wings). We can quantify this by examining, over all multipole realizations, the distribution of the strongest relative flux-ratio perturbation, which is peaked at some characteristic value
\begin{equation}
    \mu_{\rm pert} = \text{mode} \left( \max_{i\neq j} \left| \frac{(f_i/f_j) - (f_i/f_j)_{\rm EPL+shear}}{(f_i/f_j)_{\rm EPL+shear}} \right| \right)
\end{equation}
that we determined using kernel density estimation. In our examples, this value is larger for the circular multipoles (i.e., $\mu_{\rm pert}^{\rm circ} > \mu_{\rm pert}^{\rm ell}$). This is unsurprising given the fact that, for multipoles of equal amplitude applied to systems that are not near-circular, the circular perturbations yield convergence profiles displaying structure that is more localized and physically less realistic (see Section~\ref{subsec:circ_mult_pbs}) than their elliptical counterpart (see Figures~\ref{fig:convergence SIE+m=1}, \ref{fig:convergence SIE+m=3}, and \ref{fig:convergence SIE+m=4}). As expected from this interpretation, the difference becomes more apparent for highly flattened systems: in the examples shown here, the distributions differ more for the lens with $q=0.58$ (Figure~\ref{fig:ell_vs_circ_q58}) than for the lens with $q=0.73$ (Figure~\ref{fig:ell_vs_circ_q73}). In particular, circular multipoles generate flux-ratio perturbations that are typically 30\% larger ($\frac{\mu_{\rm pert}^{\rm circ}}{\mu_{\rm pert}^{\rm ell}}\approx 1.3$) for the lens with $q=0.73$, and typically 80\% larger ($\frac{\mu_{\rm pert}^{\rm circ}}{\mu_{\rm pert}^{\rm ell}}\approx 1.8$) for the lens with $q=0.58$. We warn that these numbers are merely rough indicators, since the flux-ratio perturbations also depend on the exact geometry of the lensed system.
\vspace{-0.2cm}

\subsection{Case study: fitting mock images with multipoles}
\label{subsec:mock lenses}

To further investigate the potential pitfalls of using circular multipoles to fit astrophysical lenses, we conducted another experiment: we simulated mock imaging data with the more physically realistic elliptical multipoles, then attempted to fit them with a model using the circular multipoles. We considered the same two lensed systems as in Section~\ref{subsec:flux_ratio_perturbations}, but added some diskyness to the mass profile, i.e., a fixed $m=4$ elliptical multipole component aligned with the ellipse ($\varphi_4=0$). We picked a multipole amplitude of $a_4=0.035$, which is significant but still physically realistic \citep[e.g.,][]{Bender1988,Hao2006, Ciambur2015}, especially considering that galaxies with high ellipticities ($q\lesssim 0.7$) are more likely to display strong disky features \citep{Oh2024}. Recent studies show that DM substructure analyses can use the information from extended lensed arcs to place better constraints on their lens models \citep{Oh2024, Gilman2024}, so we included both a point-like source (the quasar) and an extended source (its host galaxy), parametrized as an elliptical Sérsic profile \citep{Sersic1963, Sersic1968, Graham2005}. To generate and model the mock images, we assumed a combination of Poisson and background noise, and adopted typical Hubble Space Telescope (HST) values, with a pixel size of $0.05^{\prime\prime}$, a rms background noise of $0.06$  photons/s/pix, and an exposure time $ t_{\rm exp} = 1428$s. We employed a Gaussian point spread function with a width of $80$ mas.

We fitted these mock lenses with an EPL + shear + $m=1$ + $m=3$ + $m=4$ model with \textit{circular} multipoles for the lens mass distribution, and an elliptical Sérsic for the extended source light. The lensed quasar was modeled as four separate point sources in the image plane, whose positions and fluxes are not determined by the lens model directly - such that the image fluxes are not used as constraints, which is the usual method in flux-ratio studies \citep[e.g.,][]{Oh2024, Keeley2024} (and for lensed quasar models in general \citep[e.g.,][]{Schmidt2023}). We followed a standard \texttt{lenstronomy} fitting procedure \cite{lenstronomy,lenstronomy2}: (1) we ran a Particle Swarm Optimization (PSO) to find the maximum of the imaging likelihood while ensuring the validity of the lens equation for the quasar image positions, then (2) we estimated the posterior probability distribution of the model parameters using Markov Chain Monte-Carlo (MCMC) sampling. 

\begin{figure*}[!h]
    \centering
    \includegraphics[width=0.99\linewidth]{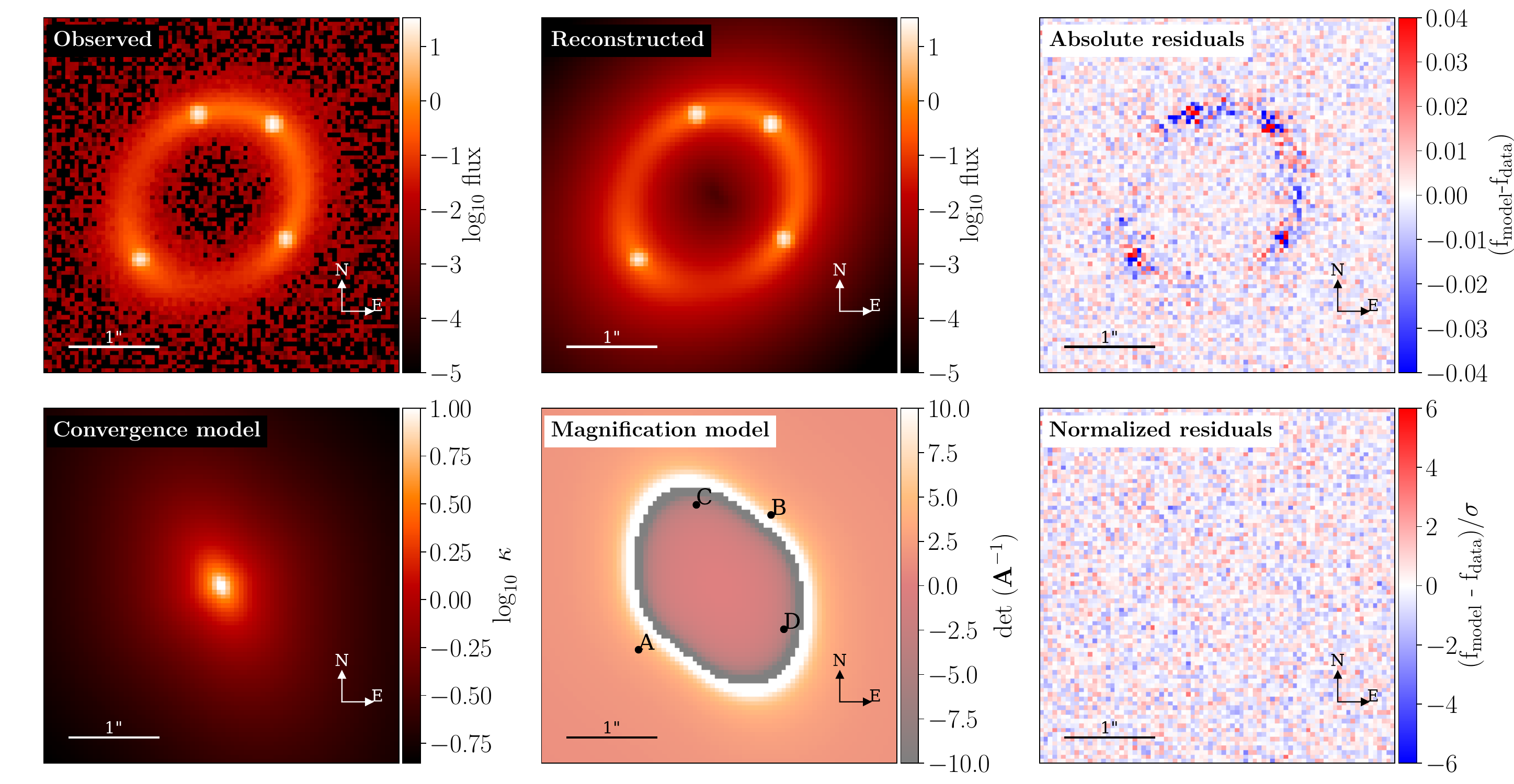}
    \vspace{-0.5cm}
    \caption{Comparison of the simulated imaging data (top left) for the $q=0.73$ mock lens, with the reconstructed image using the best-fit EPL + shear + $m=1$ + $m=3$ + $m=4$ model with circular multipoles (top middle). Also shown are the corresponding convergence (bottom left) and magnification (bottom middle) maps, with the lensed quasar positions. The right column displays the absolute (top) and normalized (bottom) residuals, after substraction of the data from the model.}
    \label{fig:q73_fit}
    \vspace{0.5cm}
    \includegraphics[width=0.99\linewidth]{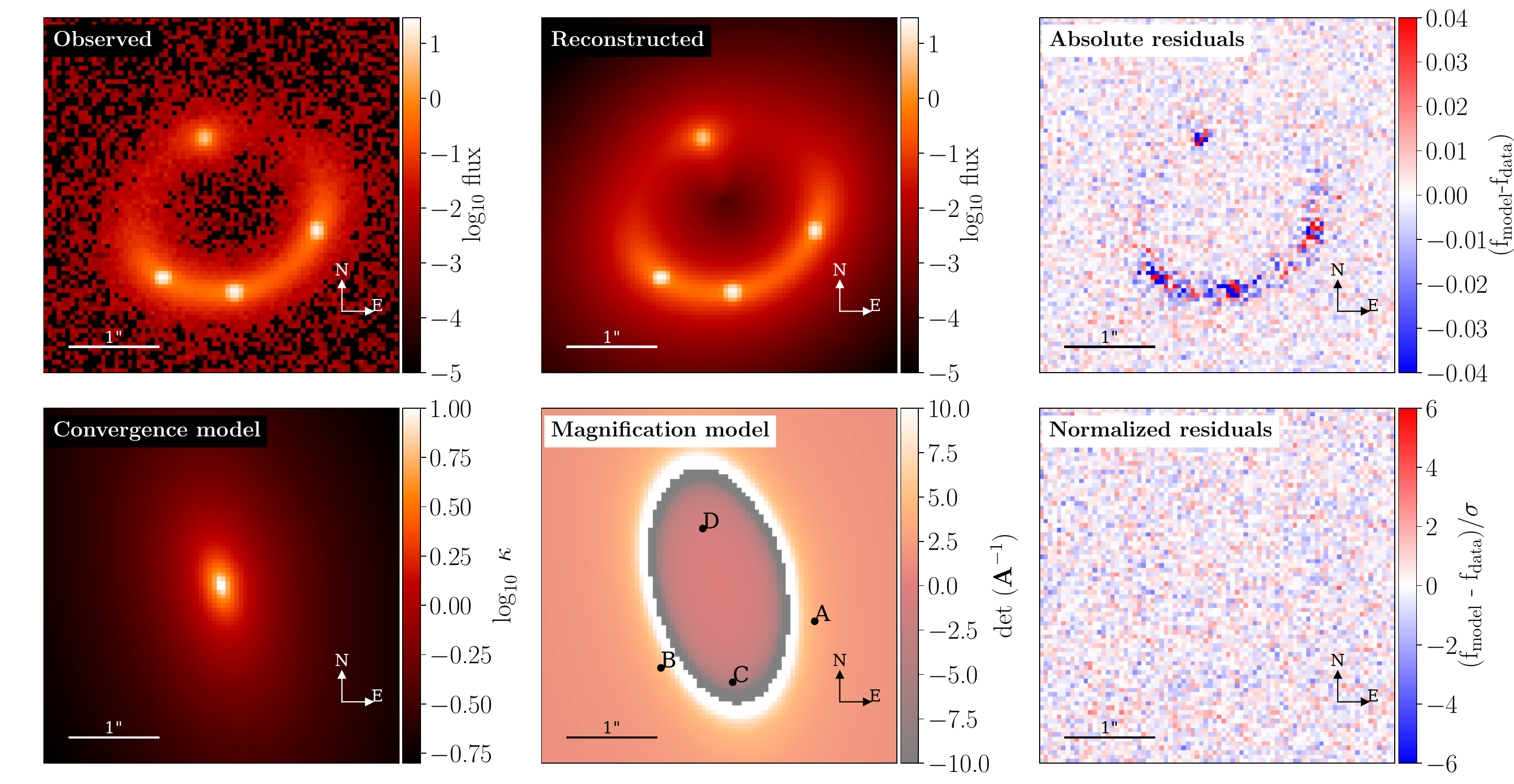}
    \vspace{-0.5cm}
    \caption{Same as Figure~\ref{fig:q73_fit}, but for the mock lens with axis ratio $q=0.58$.}
    \label{fig:q58_fit}
\end{figure*}

\begin{figure*}[!h]
    \centering
    \vspace{-0.6cm}
    \includegraphics[width=0.58\linewidth]{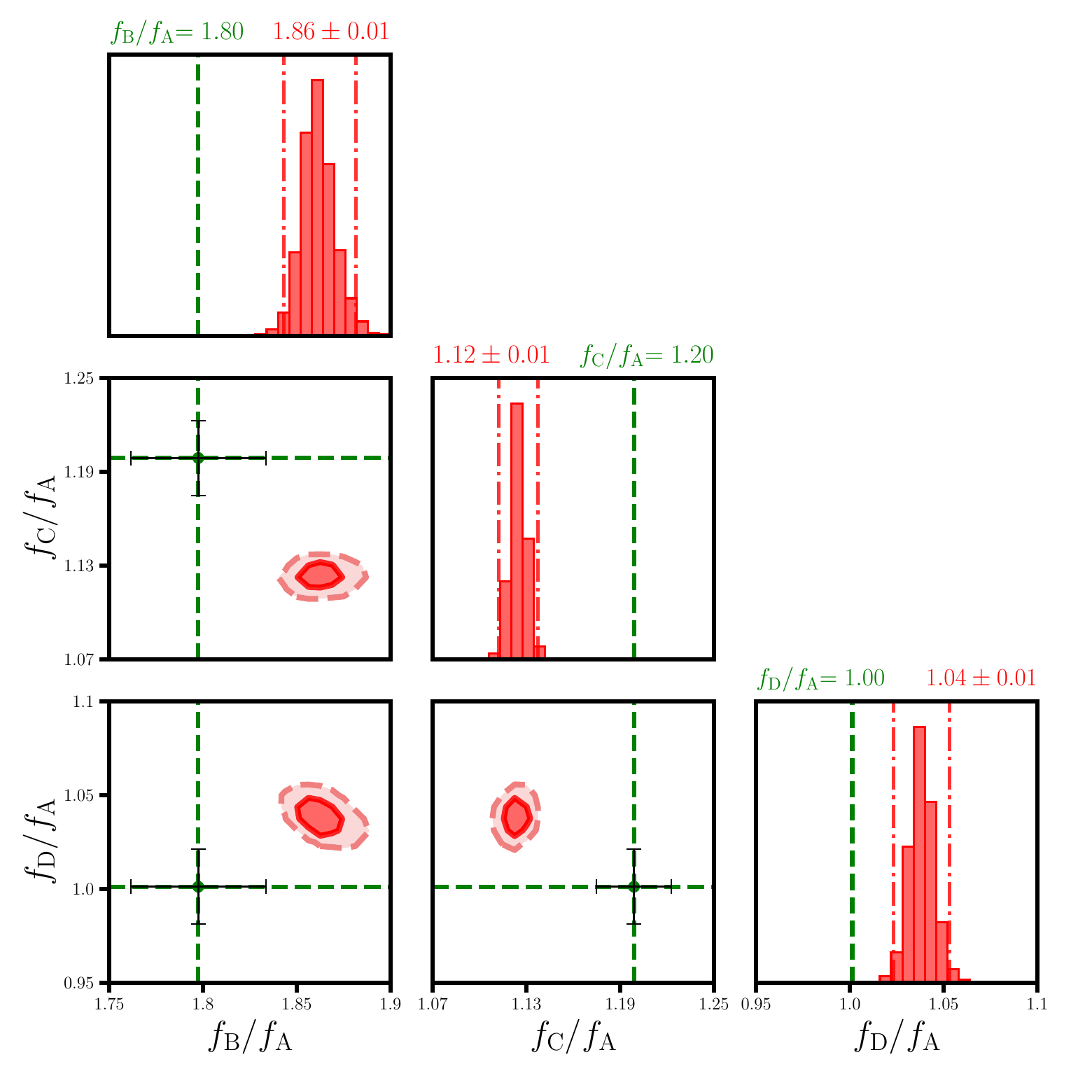}
    \vspace{-0.6cm}
    \caption{Posterior distribution for the flux ratios (in red) predicted by the EPL + shear + circular multipoles model applied to the mock image with the $q=0.73$ disky lens. The green dashed lines are the true flux-ratio values, and the black errorbars indicate typical measurement uncertainties of 2\% on the flux ratios. }
    \label{fig:q73_fitted_FRs}
    \vspace{0.6cm}
    \includegraphics[width=0.58\linewidth]{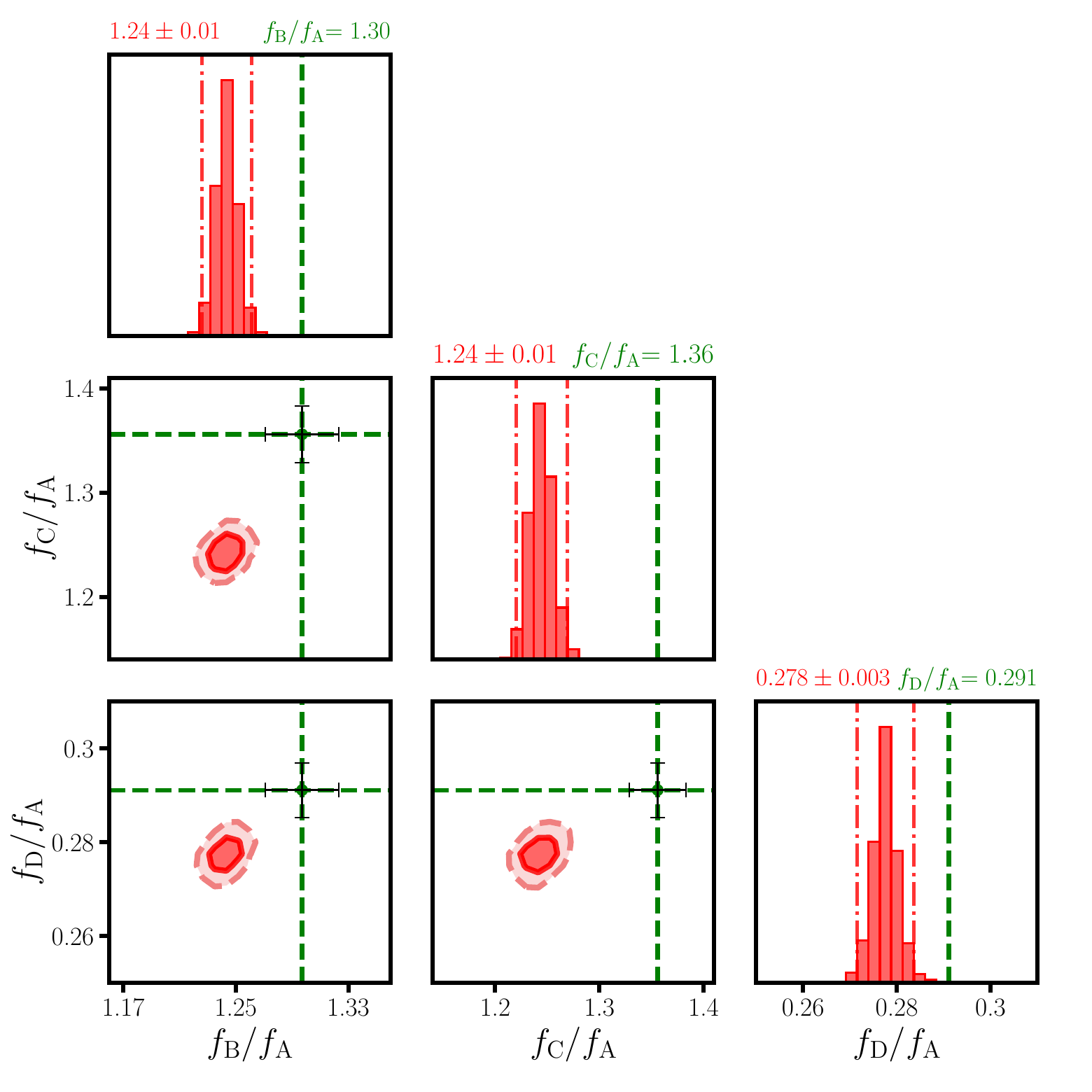}
    \vspace{-0.6cm}
    \caption{Same as Figure~\ref{fig:q73_fitted_FRs}, but for the mock lens with axis ratio $q=0.58$.}
    \label{fig:q58_fitted_FRs}
\end{figure*}

The parameter values and uncertainties inferred with this method are displayed in Table~\ref{tab:q=0.73_lens} (resp. Table~\ref{tab:q=0.58_lens}) for the lens with $q=0.73$ (resp. $q=0.58$), where they are compared to the true values used to generate the mock images. The best-fit model reconstruction is compared with the mock observations in Figure~\ref{fig:q73_fit} (resp. Figure~\ref{fig:q58_fit}), where we also plot the imaging residuals, and predicted convergence and magnification maps. Using the MCMC samples, we also determined a posterior distribution for the flux ratios predicted by the lens model at the location of the quasar images, that we display in Figure~\ref{fig:q73_fitted_FRs} (resp. Figure~\ref{fig:q58_fitted_FRs}). 

\begin{table}[!h]
\caption{\label{tab:q=0.73_lens}
True values of the non-linear parameters used to simulate the mock lens system with $q=0.73$, and values fitted with the EPL + shear + $m=1$ + $m=3$ + $m=4$ (circular multipoles) model in Section~\ref{subsec:mock lenses}.}
\begin{ruledtabular}
\begin{tabular}{lccr}
\textrm{Profile }&
\textrm{Parameter name (units)}&
\textrm{True value}&
\textrm{Fitted value}\\
\colrule
EPL & $\theta_E \ (^{\prime\prime})$ & 1.000 & $1.0000 \pm 0.003$ \\
 & $\gamma$ & 2.05 & $2.05 \pm 0.01$\\
 & $q$ & 0.730 & $0.728 \pm 0.003$ \\
 & Position angle (PA, $^\circ$) & -64.9 & $-64.2 \pm 0.2 $ \\
 & Profile center (mas) & (0,0) & $(-1.1\pm 0.3, 0.3\pm 0.3)$ \\
 \hline
External shear \footnote{The center of the shear profile is held fixed at $(0,0)$.} & Shear strength & 0.0565 & $0.0574 \pm 0.0007$\\
 & Shear PA ($^\circ$) & 67.5 & $66.4\pm 0.4$\\ 
  \hline
Multipoles (Section~\ref{subsec:mock lenses} only) \footnote{The centers of the multipoles and EPL profiles are joint.} & Type & Elliptical & Circular \footnote{To compare multipole directions, we tranform $\phi_m$ into its corresponding elliptical angle $\varphi(\phi_m;q)$.} \\
  \quad $m=1$ & $a_1$ or $a_1^{\rm circ}$ (\%) & $0.00$ & $0.00 \pm 0.01$ \\
  & $\varphi_1$ or $\varphi(\phi_1;q)$ ($^\circ$) & $0.0$ & $-8.3 \pm 2.1$\\ 
  \quad $m=3$ & $a_3$ or $a_3^{\rm circ}$ (\%) & $0.00$ & $0.28\pm 0.01$\\
 & $\varphi_3$ or $\varphi(\phi_3;q)$ ($^\circ$) & $0.0$ & $1.4\pm 0.6$ \\ 
 \quad $m=4$ & $a_4$ or $a_4^{\rm circ}$ (\%) & $3.50$ & $2.63 \pm 0.03$\\
 & $\varphi_4$ or $\varphi(\phi_3;q)$  ($^\circ$) & $0.0$ & $-2.7\pm 0.5$\\
  \hline  \hline
Quasar (point source) \footnote{
The quasar images are fitted as four separate point sources in the image plane but the lens equation is \\ \vspace{-0.2cm} enforced, so we can map them back to a single position in the source plane.} & RA$_{\rm source}$ $(^{\prime\prime})$& -0.090 & $-0.091 \pm 0.001$\\
 & DEC$_{\rm source}$ $(^{\prime\prime})$ & -0.040 & $-0.039 \pm 0.001$ \\
 \hline
 Extended source light \footnote{The amplitude parameter for light distributions can be inferred by linear minimization and therefore are \\ \vspace{-0.2cm} not sampled directly \citep{lenstronomy2}.} & Half-light radius $(^{\prime\prime})$ & $0.150$ & $0.156 \pm 0.001$ \\
 (elliptical Sérsic) & Sérsic index & 2.00 & $2.06 \pm 0.02$ \\
    & Source axis ratio & 0.806 & $0.803 \pm 0.004$ \\
 & Source PA ($^\circ$) & 34.1 &  $34.7\pm 0.7$
\end{tabular}
\end{ruledtabular}
\vspace{-0.4cm}
\end{table}

\begin{table}[!h]
\caption{\label{tab:q=0.58_lens}%
Same as Table~\ref{tab:q=0.73_lens}, but for the mock lens system with $q=0.58$.}
\begin{ruledtabular}
\begin{tabular}{lccr}
\textrm{Profile }&
\textrm{Parameter name (units)}&
\textrm{True value}&
\textrm{Fitted value}\\
\colrule
EPL & $\theta_E \ (^{\prime\prime})$ & $1.0000$& $1.0014 \pm 0.0009 $ \\
 & $\gamma$ & 2.00 & $1.99 \pm 0.01$ \\
 & $q$ & 0.576 & $0.561 \pm 0.004$\\
 & Position angle (PA, $^\circ$) & -79.1 & $-79.2 \pm 0.2$ \\
 & Profile center (mas) & (0.0,0.0) & $(1.9 \pm 0.6 , 3.4 \pm 1.0)$\\
 \hline
External shear \footnote{The center of the shear profile is held fixed at $(0,0)$.} & Shear strength & 0.022 & $0.026 \pm 0.001$ \\
 & Shear PA ($^\circ$) & 58.3 & $60.2 \pm 2.7$\\ 
  \hline
Multipoles (Section~\ref{subsec:mock lenses} only) \footnote{The centers of the multipoles and EPL profiles are joint.}  & Type & Elliptical & Circular \footnote{To compare multipole directions, we tranform $\phi_m$ into its corresponding elliptical angle $\varphi(\phi_m;q)$.} \\
  \quad $m=1$ & $a_1$ or $a_1^{\rm circ}$ (\%) & $0.00$ & $0.28 \pm 0.03$ \\
  & $\varphi_1$ or $\varphi(\phi_1;q)$ ($^\circ$) & $0.0$ & $-3.5 \pm 7.9$\\ 
  \quad $m=3$ & $a_3$ or $a_3^{\rm circ}$ (\%) & $0.00$ & $0.86 \pm 0.04$\\
 & $\varphi_3$ or $\varphi(\phi_3;q)$ ($^\circ$) & $0.0$ & $-21.5\pm 2.4$ \\ 
 \quad $m=4$ & $a_4$ or $a_4^{\rm circ}$ (\%) & $3.50$ & $0.68 \pm 0.06$\\
 & $\varphi_4$ or $\varphi(\phi_3;q)$  ($^\circ$) & $0.0$ & $22.9 \pm 2.0$\\
  \hline  \hline
Quasar (point source) \footnote{
The quasar images are fitted as four separate point sources in the image plane but the lens equation is \\ \vspace{-0.2cm} enforced, so we can map them back to a single position in the source plane.} & RA$_{\rm source}$ $(^{\prime\prime})$& 0.080 & $0.081 \pm 0.001$ \\
 & DEC$_{\rm source}$ $(^{\prime\prime})$ & -0.200 & $-0.198 \pm 0.002$\\
 \hline
 Extended source light \footnote{The amplitude parameter for light distributions can be inferred by linear minimization and therefore are \\ \vspace{-0.2cm} not sampled directly \citep{lenstronomy2}.} & Half-light radius $(^{\prime\prime})$ & $0.150$ & $0.154 \pm 0.01 $\\
 (elliptical Sérsic) & Sérsic index & 2.00 & $2.02 \pm 0.04$ \\
    & Source axis ratio & 0.806 & $0.800 \pm 0.006$\\
 & Source PA ($^\circ$) & 34.1 &  $33.7 \pm 1.0$
\end{tabular}
\end{ruledtabular}
\vspace{-0.4cm}
\end{table}

The EPL + shear + $m=1$ + $m=3$ + $m=4$ (circular multipoles) model has enough degrees of freedom to adjust itself and find a combination of parameters that can fit the imaging data down to the noise level, i.e., mimic the effect of diskyness (elliptical $m=4$) on the lensed arc (see the normalized residuals in Figures~\ref{fig:q73_fit} and \ref{fig:q58_fit}). The absolute residuals, however, show correlated structure in the extended lensed arcs. This might be of importance for direct detections of DM halos in the imaging data when parametric models are used for the main deflector \citep[e.g.][]{Sengul2022, Ballard2024, Nightingale2024}, since DM halo signatures in this context are precisely localized perturbations in the arcs. 

The parameters of the EPL+shear profile are not always recovered within the uncertainty predicted by the MCMC samples (see Tables~\ref{tab:q=0.73_lens} and \ref{tab:q=0.58_lens}), but the discrepancies are small enough that they would not impart significant bias on results that are only interested in these parameters. Unsuprisingly, however, the multipole amplitudes and directions differ significantly. In the case of the $q=0.58$ mock lens, the dominant circular multipole term is the $m=3$ (with an amplitude of $\sim 0.01$ - quite large for this order), and the $m=4$ circular multipole has a much smaller amplitude than expected. Once again, this means that the circular multipole values cannot be given their usual physical interpretation when $q \not\approx 1$.

In addition, the predicted magnifications at the quasar images yield flux ratios with statistically significant errors compared to the ground truth (see Figures~\ref{fig:q73_fitted_FRs} and \ref{fig:q58_fitted_FRs}). The largest relative flux-ratio error is $\sim 7\%$ for the lens with $q=0.73$, and $\sim 10\%$ for the lens with $q=0.58$ - much larger than the model-predicted uncertainty, but also non-negligible compared to typical measurement uncertainties (e.g., $2-10\%$ uncertainty on narrow-line flux ratio with HST \citep{Nierenberg2017, Nierenberg2020}, or $\lesssim 3\%$ precision with the more recent quasar warm dust flux ratios with the James Webb Space Telescope \citep{Nierenberg2024, Keeley2024}). Therefore, when modeling individual lenses, the use of circular multipoles might lead to false-positive detections of flux-ratio anomalies, since true azimuthal perturbations in astrophysical lenses are expected to behave more like elliptical multipoles. We note that, in practice, the multipole parameters are not strongly constrained by the imaging data when realistic sources and substructure are included \citep[e.g.,][]{Gilman2024}. In that case, the multipoles are introducing a background level of perturbation to the flux-ratio statistics, so the most appropriate test is to compare the distribution of flux-ratio perturbations - we probed this in Section~\ref{subsec:flux_ratio_perturbations}.

\section{Summary \& Discussion}
\label{sec:conclusion}

In this work, we have illustrated the shortcomings of the circular multipoles, a formalism commonly used to represent deviations from ellipticity in models of lensing galaxies. We have argued that these perturbations produce changes in azimuthal angular structure that depend on the axis ratio, and that do not correspond to physical expectations for systems that are not near-circular. We have proposed an improved formulation, the elliptical multipoles, applicable to elliptical profiles with any axis ratio. We have introduced solutions for the lensing potentials of these elliptical multipoles (as well as for the $m=1$ circular multipole), taking particular care to ensure that the expected symmetries and behavior in limiting cases were respected. We have compared the two formulations in the context of a physical application: flux-ratio anomalies of quadruply imaged quasars. Our results suggest that, at comparable amplitudes, circular multipoles might produce flux-ratio perturbations that are typically larger than the elliptical multipoles, especially for highly flattened system, likely due to the unrealistic perturbations (with important localized gradients in the surface mass density) introduced by circular multipoles outside of the near-circular regime. By simulating mock images, we have also shown on two example lens systems that several orders of circular multipoles can mimic (at a reasonable noise level) the effect of elliptical multipoles on extended arcs; however, this can bias some lens parameters (especially the multipole amplitudes/directions), and lead to discrepancies in the recovered flux ratios, that are significant compared to model-predicted and measurement uncertainties.

We remark that, in real lens systems, departures from pure elliptical profiles can manifest themselves in more than one way. The addition of multipoles (and in particular the $m=1$, $m=3$, and $m=4$ orders) is a popular choice, and one that is physically motivated by studies of galaxy isophotal shapes, but it might not fully capture this additional complexity. For instance, massive elliptical galaxies can display ellipticity gradients or isophotal twisting (i.e., a change with radius of the principal axis of the isophotes, which would trace a similar twisting in the isodensity contours) \citep[e.g.,][]{Carter1978, Madejsky_Moellenhoff_1990, Hao2006, Kormendy2009, deNicola2020, VdV2022_twisting}. Some of the more elongated lensing galaxies could also harbor an additional baryonic edge-on disk component perturbing the flux ratios, which is not exactly the same as having a disky mass distribution \citep{Hsueh2016, Gilman2017}.
Even in the context of multipole perturbations, the amplitude $a_m$ of the deviation from ellipticity is a priori different for each isocontour -  isophotal shape studies usually measure full radial profiles for the multipole amplitude \citep[e.g.,][]{Carter1978, Bender1988, Jedrzejewski1987, Ciambur2015, Goullaud2018, Amvrosiadis2024}. Therefore, the multipoles used in lensing (which assume a constant multipole amplitude relative to the semi-major axis \citep{Oh2024}) are only intended to be first-order azimuthal corrections to the elliptical mass profiles.

In that respect, the elliptical multipole expansion is not ‘‘the correct’’ parametrization for azimuthal perturbations, merely one that is more appropriate than the circular multipoles (more easily interpretable, and with patterns following expectations from isophotal shapes, independent of the axis ratio). In particular, our calculation still operates under the assumption that the reference lens mass model is an isothermal profile, i.e., it is only exact if the 3D density profile $\rho(r_{\rm 3D})\propto r_{\rm 3D}^{-\gamma}$ has a slope $\gamma=2$. Since astrophysical lenses are typically near-isothermal (e.g., the average value and dispersion in the SLACS survey are respectively $\langle \gamma \rangle = 2.08 \pm 0.03, \sigma_\gamma=0.16 \pm 0.02$ \citep{Auger2010, Shajib2021}), the formalism proposed here remains a good approximation, applicable in most realistic lens modeling setups. For this reason, the user-ready mass models implemented in \texttt{lenstronomy} combine the isothermal multipoles presented here with EPL profiles (i.e., with a free slope $\gamma$).

We note that there exists an exact generalization of the circular multipoles where the slope of the multipole is matched to the slope of the EPL profile (see \citep{EvansWitt2003, Lange2024} and Appendix~\ref{App:isothermal_limit}), available for instance in the strong lensing package \texttt{PyAutoLens} \citep{pyautolens, Nightingale2015, Nightingale2018}. We caution, however, that current implementations diverge when $\gamma \to 2$ in the case of the $m=1$ multipole (we discuss this further in Appendix~\ref{App:isothermal_limit} and propose a way to remedy this issue). With the possible exception of lenses that are both near-circular and very far from isothermal, elliptical multipoles are still arguably preferable to these ‘‘slope-matched’’ circular multipoles. We demonstrate in Appendix~\ref{App:isothermal_vs_free_slope} that applying isothermal multipoles to non-isothermal reference profiles only results in a small, logarithmic radial dependance of the multipole amplitude. As mentioned above, in real lenses, the amplitude of the multipole in the isophotes is a non-trivial function of radius, so a slope-matched multipole formulation (i.e., such that the amplitude is independent of radius) is not necessarily better suited. Furthermore, in single-plane lensing, most of the lensing information comes from radial scales $R\sim \theta_E$, so correctly describing the mass distribution around this given radius is the most crucial part - this is why the $\kappa=1/2$ isocontour is used as a reference. Hence, for multipoles (which are in the first place designed to capture perturbative azimuthal complexity), accuracy matters more in the azimuthal direction than in the radial direction. As an analogy, in most cases, an SIE would be a better lens model than a spherical power-law profile (i.e., an EPL with $q=1$) - the same reasoning applies at the level of azimuthal perturbations. We give a more quantitative criterion in Appendix~\ref{App:isothermal_vs_free_slope}, showing that, unless $|\gamma-2| \gg 1-q$, correcting for ellipticity has more impact on the convergence around the Einstein radius than matching the radial slopes. Given the complexity of the potentials presented in Section~\ref{sec:potential_solutions} and Appendix~\ref{App:gen_ell_potential} for the isothermal case, and the limited added value expected from an extension of elliptical multipoles to non-isothermal slopes, we leave such an endeavor to future work.

We have illustrated that the choice of azimuthal parametrization (circular or elliptical) for the multipole expansion can affect the modeled properties of a lensed system at levels relevant for some high-precision measurements. For instance, in the case of quadruply imaged quasars, relative to the circular multipoles, elliptical multipoles change the structure of the joint probability distribution of image flux ratios and tend to impart smaller flux-ratio perturbations in highly flattened systems (see Section~\ref{subsec:flux_ratio_perturbations}). This could impact the interpretation of Bayesian model selection between lens models that introduce perturbations from substructure, when they also include multipoles \citep[e.g.,][]{Gilman2024, Keeley2024, Lange2024}. In particular, the unphysical circular formulation might slightly overestimate the importance of multipoles, so we expect that switching to elliptical multipoles would strengthen constraints on DM models that suppress small-scale power. Further research, however, will be needed to systematically assess the impact on the inference of DM properties with flux-ratio anomalies, as well as on other science cases that rely on accurate lens models (e.g., $H_0$ inference with time-delay cosmography, or direct detection of DM halos).

\begin{acknowledgments}
We thank Tommaso Treu, Xiaolong Du, Anna Nierenberg and Simon Birrer for helpful discussions, comments and feedback. This work was supported by the National Science Foundation under grant AST-2205100, and by the Gordon and Betty Moore Foundation under grant No. 8548. DG acknowledges support for this work provided by the Brinson Foundation through a Brinson Prize Fellowship grant.
\end{acknowledgments}

\newpage


\appendix

\section{Lensing potential for the circular $m=1$ multipole}
\subsection{Problem with the $\propto r$ potential}
\label{App:m=1_shape_fct_pbs}

We have seen in Section~\ref{subsec:circ_multipole_potential} that the solution $\tilde{F}_1^{\rm circ}(\phi)$ to differential Equation~(\ref{eq:main_diff_eq}) is not $2\pi$-periodic and does not have the expected $\phi-\phi_1 \mapsto \pi-(\phi-\phi_1)$ antisymmetry. We can attempt to symmetrize the corresponding potential ‘‘manually’’, starting by choosing a principal value for the polar angle in order to impose the $2\pi$-periodicity. Because we want to keep the $\phi-\phi_1 \mapsto -(\phi-\phi_1)$ symmetry, we choose $\tilde{\phi}(\phi) = \phi_1+\arctan(\cos(\phi-\phi_1), \sin(\phi-\phi_1))$, such that $\phi-\phi_1 \in ]-\pi, \pi]$. Then, enforcing the $\phi-\phi_1 \mapsto \pi-(\phi-\phi_1)$ antisymmetry, we define the following potential:
\begin{equation}
    \tilde{\psi}_1^{\rm circ}(r, \phi) = \frac{r}{2} \left( \tilde{F}_1^{\rm circ}\left[\tilde{\phi}(\phi)\right] - \tilde{F}_1^{\rm circ}\left[\tilde{\phi}(\pi + 2\phi_1-\phi)\right] \right)
    \label{eq:psi_tilde_1_circ}
\end{equation}

\begin{figure*}[!h]
    \centering
    \includegraphics[width=0.99\linewidth]{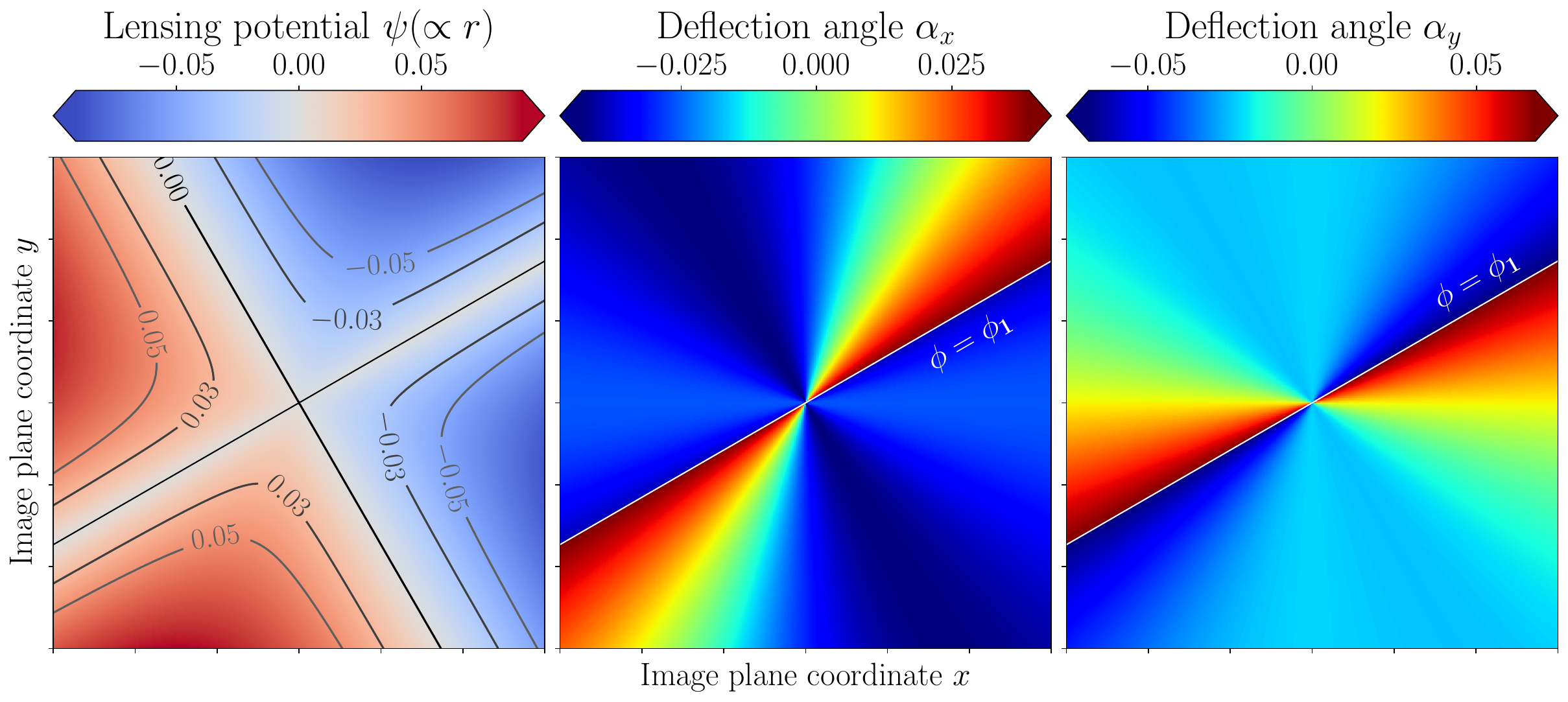}
    \caption{Illustration of the jump discontinuity in deflection angles appearing at the $\phi=\phi_1$ line when attempting to symmetrize the $\propto r$ potential for the $m=1$ circular multipole (here, we show an example with $a_1=0.1$ and $\phi_1 = \frac{\pi}{6}$).}
    \label{fig:alpha_discontinuity}
\end{figure*}

This potential has all the correct symmetries, is differentiable on $\mathbb{R}^2\setminus\{\mathbf{0}\}$, and is twice differentiable on $\mathbb{R}^2\setminus\left\{(r,\phi) \Big| \phi=\phi_1 + k\pi, k\in \mathbb{Z}\right\}$, where it verifies $\Delta \tilde{\psi}_1^{\rm circ} = 2 \kappa_1^{\rm circ}$. The problem is that its derivatives (i.e., the deflection angles) are not continuous in any point of the line defined by $\phi=\phi_1$ (see Figure~\ref{fig:alpha_discontinuity}) - this cuts the plane in two, forming a jump discontinuity. This cannot be remedied by a clever choice for $A_1^{\rm circ}$, since the corresponding term only adds a constant vector to the deflection field. Using the symmetrized lensing potential of Equation~(\ref{eq:psi_tilde_1_circ}) could therefore lead to unphysical results, as it amounts to piecewise definition, with images of the source possibly lying in disconnected regions \citep{Wagner2018}.

\subsection{Justifications for the $\propto r\ln r$ potential}
\label{App:m=1_rlnr_justifications}

\subsubsection{Isothermal limit of the non-isothermal solution}
\label{App:isothermal_limit}

The circular multipoles can easily be generalized to non-isothermal slopes \citep{EvansWitt2003, Lange2024}, i.e., convergence profiles with a radial dependence $\kappa (r) \propto r^{1-\gamma}$ where $\gamma$ is a free parameter ($1<\gamma<3$ for galaxy models, with $\gamma=2$ the isothermal case). Defining $\beta = 3-\gamma$, we can use a formalism similar to the one presented in Section~\ref{sec:general_formalism} and write the following lensing potential/density pair \citep{EvansWitt2003}:
\begin{equation}
    \psi(r,\phi) = r^\beta F(\phi)\ ; \kappa(r,\phi) = \frac{1}{2} r^{\beta-2} G(\phi)
    \label{eq:pot_density_non_isothermal}
\end{equation}
where the angular shape functions obey the differential equation $G(\phi) = \beta^2F(\phi)+F^{\prime\prime}(\phi)$ by virtue of Poisson's equation.

Circular multipole pertubations have the same shape function for the convergence $G_m^{\rm circ}(\phi)$ than in the isothermal case - see Equation~(\ref{eq:circular_multipoles_def}). The shape function for the potential, however, is different: solving the new differential equation, we find
\begin{equation}
    F_{m,\beta}^{\rm circ}(\phi) = \frac{a_m^{\rm circ}}{\beta^2 -m^2} \cos (m(\phi-\phi_m)) + A_{m,\beta}^{\rm circ} \cos(\beta(\phi-\phi_m)) + B_{m,\beta}^{\rm circ}  \sin(\beta(\phi-\phi_m))
    \label{eq:F_m_beta_circ}
\end{equation}
where $A_{m,\beta}^{\rm circ},B_{m,\beta}^{\rm circ}$ are a priori free constants, since  $\cos(\beta\phi)$ and $ \sin(\beta\phi)$ verify the homogenous equation. We can use the invariance properties of $G_m^{\rm circ}(\phi)$ to determine these constants, by requiring $F_{m,\beta}^{\rm circ}(\phi)$ to have the same symmetries. To begin with, we expect $F_{m,\beta}^{\rm circ}(\phi)$ to be $2\pi$-periodic, implying that terms in $\cos(\beta\phi)$ or $\sin(\beta\phi)$ must vanish unless $0<\beta<2$ is an integer, i.e., unless $\beta=1$. In the $\beta=1$ case, the same conclusion is reached due to invariance under $\phi \mapsto \phi + \pi/(2m)$ for $m\geq2$, and under any rotation $\phi \mapsto \phi + C$ for $m=0$. In the end, we always have $A_{m,\beta}^{\rm circ}=B_{m,\beta}^{\rm circ}=0$ except for the $m=1,\beta=1$ case, which is treated separately anyways (see Section~\ref{subsec:circ_multipole_potential}). In particular, Equation~(\ref{eq:shape_fct_pot_circ}) is naturally recovered from Equation~(\ref{eq:F_m_beta_circ}) when $\beta=1$ and $m\neq1$.

We expect that the non-isothermal potentials converge to the isothermal solutions when $\gamma \to 2$ (or equivalently $\beta \to 1$), which is trivially verified for $m\neq 1$. For the $m=1$ case, we write the $\beta\neq1$ potential in the form:
\begin{equation}
    \psi_{m=1, \beta}^{circ} (r, \phi) =  r^\beta \frac{a_1^{\rm circ} }{\beta^2 -1}   \cos (\phi-\phi_1)  + C_{1,\beta}^{\rm circ} r \cos (\phi-\phi_1) + D_{1,\beta}^{\rm circ} r \sin (\phi-\phi_1)
\end{equation}
where we have introduced two terms with free constants $C_{1,\beta}^{\rm circ},D_{1,\beta}^{\rm circ}$ in order to make the prismatic degeneracy explicit. Since $G_1^{\rm circ}(\phi)$ is invariant under $\phi-\phi_m \mapsto -(\phi-\phi_m)$, we need $D_{1,\beta}^{\rm circ}=0$ in order to have the same symmetry in the potential. $C_{1,\beta}^{\rm circ}$ cannot be constrained in a similar fashion, so we leave it free for the moment. Performing a Taylor expansion in the limit $\beta \to 1$, we find:
\begin{equation}
\begin{split}
    \psi_{m=1, \beta}^{circ} (r, \phi)  & \mathrel{\underset{\beta \to 1}{=}} r \Big(1+ (\beta-1) \ln r + O[(\beta-1)^2] \Big) \frac{a_1^{\rm circ}}{\beta^2 -1} \cos (\phi-\phi_1) +  C_{1,\beta}^{\rm circ} r \cos (\phi-\phi_1) \\
    & \mathrel{\underset{\beta \to 1}{=}} r \cos (\phi-\phi_1) \left( \frac{a_1^{\rm circ}}{\beta^2 -1}  + C_{1,\beta}^{\rm circ} \right) + \frac{a_1^{\rm circ}}{2} r\ln r \cos (\phi-\phi_1) + O[\beta -1]
\end{split}
\end{equation}
Therefore, if we want $\psi_{m=1, \beta}^{circ}$ to be convergent, we need to choose $C_{1,\beta}$ such that $\frac{a_1^{\rm circ}}{\beta^2 -1}  + C_{1,\beta}$ has a finite limit $L$ when $\beta \to 1$. Then, if we define $r_E \equiv \exp\left( - 2L/a_1^{\rm circ}\right)$, we have:
\begin{equation}
    \psi_{m=1, \beta}^{circ} (r, \phi) \mathrel{\underset{\beta \to 1}{\longrightarrow}} \frac{a_1^{\rm circ}}{2} r\ln \left( \frac{r}{r_E} \right)\cos (\phi-\phi_1),
\end{equation}
which is precisely the potential $\psi_1^{circ} (r, \phi)$ for the isothermal $m=1$ circular multipole from Equation~(\ref{eq:psi_1_circ}), equivalent to the one from Ref.~\citep{Chu2013} thanks to the prismatic degeneracy.

We warn that, in current uses \citep[e.g.,][]{Lange2024} and implementations 
of the non-isothermal circular multipoles, the term in $C_{1,\beta}$ is ignored, which would cause the multipole potential (and the deflection angles) to diverge when $\beta \to 1$. In theory, the prismatic degeneracy allows to compensate this by moving the source as much as required, but in practice this would be limited by any priors placed on the source position. To avoid biasing the lens models for near-isothermal slopes, we recommend including this term with an appropriate choice for the constant, for instance:
\begin{equation}
    C_{1,\beta} = - a_1^{\rm circ} \left( \frac{1}{\beta^2-1} + \frac{\ln r_E}{2}\right)
\end{equation}
with the same choice of normalizing radius $r_E$ as for $m=1,\beta=1$ (in our case $r_E=\theta_E$) in order to ensure the proper convergence.

\subsubsection{Direct integration in complex formulation}
\label{App:m=1_complex_alpha_int}

The gravitational lensing formalism has a complex formulation, introduced by  \citep{Bourassa1973, Bourassa1975}, in which the lens equation is written 
\begin{equation}
    z_s = z - \alpha(z)
\end{equation}
where complex quantities are used instead their 2D vector equivalent for the position in the source plane $z_s=x_s+iy_s$, the position in the image plane  $z=x+iy$, and the deflection angle $\alpha = \alpha_x + i\alpha_y$ \citep{TessoreMetcalf}. For a convergence profile $\kappa(z,\overline{z})$ (where $\overline{z}$ denotes the complex conjugate), the complex deflection angle at image position $z_0$ can be written \citep[e.g.,][]{Schramm1995}:
\begin{equation}
    \alpha(z_0) = \frac{1}{\pi} \int_{\mathbb{R}^2} \frac{\kappa(z,\overline{z})}{\overline{z}_0-\overline{z}}  dxdy.
    \label{eq:alpha_complex}
\end{equation}

Let us perform this integration with the following convergence profile:
\begin{equation}
    \kappa_1^{\rm circ}(r, \phi) =  \begin{cases}
        \frac{a_1^{\rm circ}\cos\phi}{2r} &\text{ if } r\leq r_{\rm max}  \\
        0 &\text{ otherwise.}
    \end{cases}
\end{equation}
i.e., a circular $m=1$ multipole profile with $\phi_1=0$ that has been truncated beyond some radius $r_{\rm max}$ (otherwise, the integral diverges). Changing to polar coordinates in the integral of Equation~(\ref{eq:alpha_complex}), we can write:

\begin{equation}
    \begin{split}
        \overline{\alpha_1^{\rm circ}} (z_0) &= \frac{a_1^{\rm circ}}{2\pi} \int_0^{r_{\rm max}} dr  \int_{-\pi}^{\pi} d\phi \frac{\cos\phi}{z_0 - r\cos\phi -ir\sin\phi }\\
        &= \frac{a_1^{\rm circ}}{2\pi} \left( \pi\int_0^{|z_0|}  \frac{ r }{z_0^2} dr -  \pi\int_{|z_0|}^{r_{\rm max}}  \frac{ dr}{r} \right) \\
        &= \frac{a_1^{\rm circ}}{2} \ln\left(\frac{|z_0|}{r_{\rm max}}\right)  + \frac{a_1^{\rm circ}}{4} \frac{|z_0|^2 }{z_0^2}
    \end{split}
    \label{eq:alpha_complex_calculation}
\end{equation}

where we have used the following integral:
\begin{equation}
    \int_{-\pi}^{\pi} d\phi \frac{\cos\phi}{z_0 - r\cos\phi -ir\sin\phi } = 
    \begin{cases}
        -\frac{\pi }{r} \text{ if } r \geq |z_0| \\
        \frac{r\pi }{z_0^2} \text{ if } r < |z_0|
    \end{cases}
\end{equation}
Then, taking the real/imaginary parts of the complex conjugate of Equation 
~(\ref{eq:alpha_complex_calculation}) with $z_0=re^{i\phi}$, we find the real-valued deflection angles:

\begin{equation}
    \begin{split}
        \alpha_{x,1}^{\rm circ}(r,\phi) &= \frac{a_1^{\rm circ}}{2} \left[ \ln\left(\frac{r}{r_{\rm max}} \right) + \cos^2\phi + \frac{1}{2} \right] \\
        \alpha_{y,1}^{\rm circ}(r,\phi) &= \frac{a_1^{\rm circ}}{2}  \cos\phi \sin \phi
    \end{split}
\end{equation}
which match Equation~(\ref{eq:derivatives_psi_1_circ}) with $\phi_1=0$, up to a constant $\frac{1}{2}+\ln\left( \frac{r_{\rm max}}{r_E}\right)$ for $\alpha_x$. Thanks to the prismatic degeneracy, we can add any constant to the deflection field without impacting the lensing observables, so these solutions are equivalent, even as we take $r_{\rm max}\to \infty$. The more general solution with $\phi_1$ free can be deduced from the $\phi_1=0$ case by a simple rotation of the coordinate system. 

\section{Comparison of the isothermal and slope-matched multipoles}
\label{App:isothermal_vs_free_slope}

While it might be extremely complicated to calculate the lensing potential and/or deflection field for the slope-matched elliptical multipoles (i.e., where the radial dependence of the multipole profile is the same as the underlying EPL), the convergence profile can be easily generalized. Mimicking the formalism presented in Section~\ref{App:isothermal_limit} for the extension of circular multipoles to non-isothermal slopes, we write:

\begin{equation}
    \kappa_{m,\beta} (r,\phi;q)  = \frac{1}{2} R^{\beta-2} \cdot a_m\cos[m(\varphi-\varphi_m)]
    \label{eq:slope_matched_ell_mult}
\end{equation}
where $\beta=3-\gamma$, and $R(r,\phi;q),\varphi(r,\phi;q)$ are the coordinates defined in Section~\ref{subsec:ell_multipoles}. Note that, for $\beta=1$, we recover the convergence profile from the isothermal elliptical multipoles.

Consider an EPL profile \citep{TessoreMetcalf}
\begin{equation}
    \kappa_{\rm EPL}(r, \phi) = \kappa_{\rm EPL}(R) = \frac{\beta}{2} \left( \frac{\theta_E}{R} \right)^{2-\beta} = \frac{\beta}{2} r^{\beta-2} G_{\rm SIE}(\phi)^{2-\beta}
\end{equation}
If we add the perturbation described by Equation~(\ref{eq:slope_matched_ell_mult}), the convergence profile becomes
\begin{equation}
    \kappa(r,\phi) = \frac{1}{2}R^{\beta-2} \left( \beta \theta_E^{2-\beta} + a_m \cos[m(\varphi-\varphi_m)] \right)
    \label{eq:perturbed_convergence_slope_matched}
\end{equation}
The isodensity contours are then given by:
\begin{equation}
\begin{split}
    \kappa = C \iff R_{\rm iso}(\varphi) &= \left( \frac{1}{2C} \beta \theta_E^{2-\beta} + \frac{1}{2C} a_m \cos[m(\varphi-\varphi_m)] \right)^{\frac{1}{2-\beta}} \\
    &\approx \left(\frac{\beta}{2C}\right)^{\frac{1}{2-\beta}}\theta_E \left[ 1 + \frac{1}{\beta(2-\beta)} \cdot \frac{a_m\cos[m(\varphi-\varphi_m)]}{\theta_E^{2-\beta}}\right]
    \label{eq:isocontour_slope_matched}
\end{split}
\end{equation}
where we have used the fact that the multipole describes a perturbation with respect to the EPL profile (i.e., we have $|a_m|\ll \beta \theta_E^{2-\beta} $). Thus, the profile presented in Equation~(\ref{eq:slope_matched_ell_mult}) produces the correct deformation of the isocontours (see Equation~(\ref{eq:elliptical_multipole_def})), and under the rescaling $a_m\mapsto \beta(2-\beta)\theta_E^{2-\beta}a_m$, the multipole amplitude $a_m$ is the fractional change in the elliptical radius of any isocontour (for $\beta=1$, we recover the rescaling used for the isothermal elliptical multipoles).

\begin{figure*}[hbtp]
    \centering
    \includegraphics[width=0.95\linewidth]{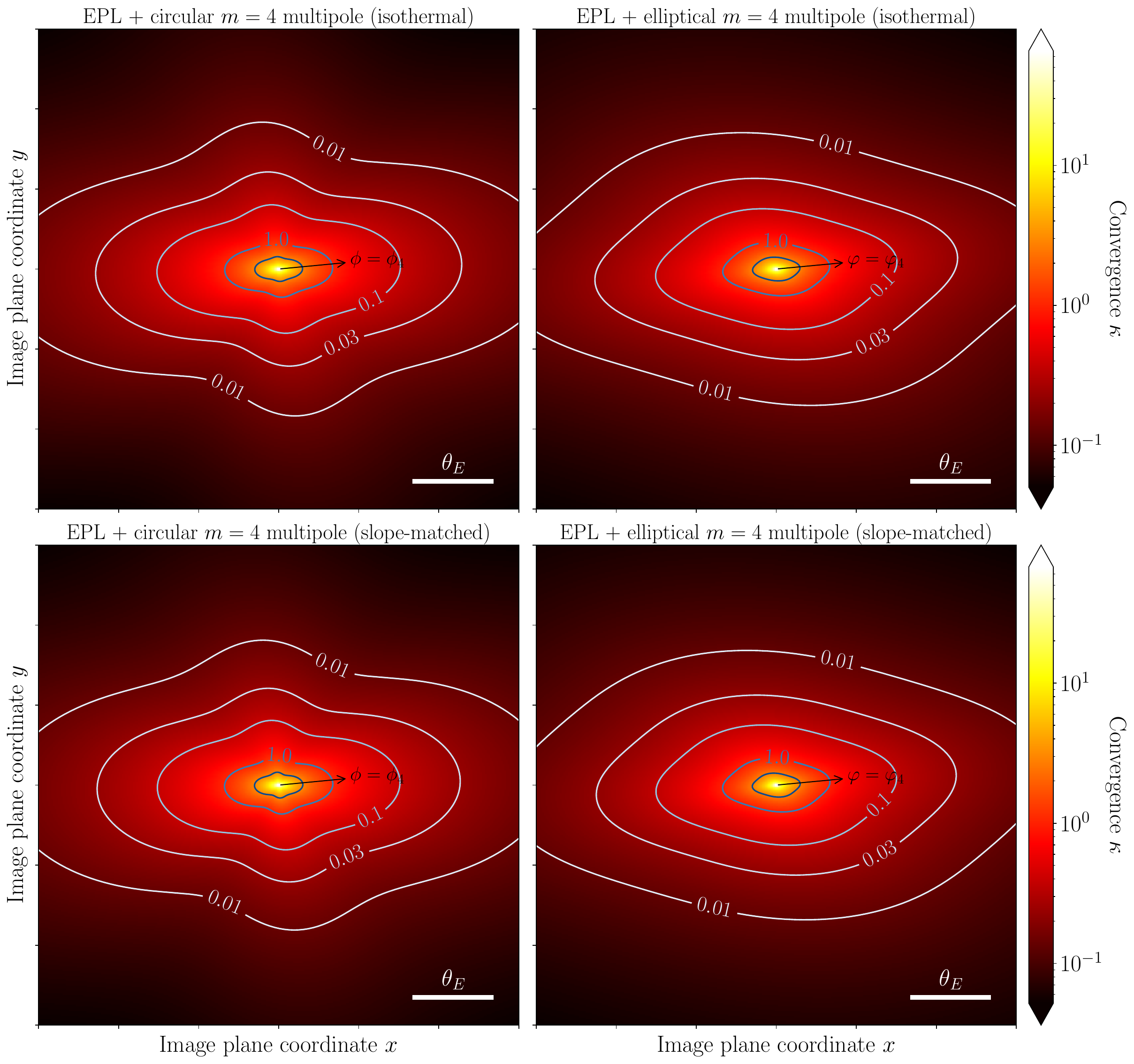}
    \caption{Impact of a $m=4$ multipole on the convergence profile of an EPL lens model with axis ratio $q=0.5$ and slope $\gamma=2.25$, in the circular (left) and elliptical (right) formulations, under the isothermal multipole approximation (top) or with a slope-matched multipole (bottom). We choose perturbations with direction $\phi_4=\frac{\pi}{24}$ (or $\varphi_4=\varphi(\frac{\pi}{24};q)$ for the elliptical multipoles), and with $a_4^{\rm circ}=0.05$ (or $a_4=0.05$) - slightly larger than physical expectations for better visualization.}
    \label{fig:isoth_freeslope_comparison}
    \vspace{1cm}
\end{figure*}

The slope-matched multipoles described by Equation~(\ref{eq:slope_matched_ell_mult}) can be related to the isothermal multipoles (after rescaling the amplitude for both) using:

\begin{equation}
    \kappa_{m,\beta=1}   = \frac{a_{m, \rm eff}(R)}{a_m} \cdot \kappa_{m, \beta}
    \label{eq:eff_mult_amplitude_def}
\end{equation}
where we have introduced an effective multipole amplitude $a_{m, \rm eff}(R) = \frac{a_m}{\beta(2-\beta)} \left( \frac{R}{\theta_E} \right)^{1-\beta}$ which is a function of (elliptical) radius. The deviation from isothermality $1-\beta$ is expected to be relatively small, so, as a first-order approximation:

\begin{equation}
    \frac{a_{m, \rm eff}(R)}{a_m} = 1 + (1-\beta)\cdot\ln\left( \frac{R}{\theta_E}\right) + O[(1-\beta)^2]
    \label{eq:effective_mult_amplitude_approx}
\end{equation}
Regardless of the slope of the EPL profile, $\kappa_{m,\beta}$ represents a multipole with constant (i.e., independent of radius) amplitude $a_m$. Therefore, as a consequence of Equations~(\ref{eq:eff_mult_amplitude_def}) and (\ref{eq:effective_mult_amplitude_approx}), if we apply the isothermal multipoles $\kappa_{m,\beta=1}$ to a non-isothermal profile, it is equivalent to introducing a logarithmic radial dependence of the multipole amplitude. These equations also show that, on scales comparable to the Einstein radius ($R\sim \theta_E$), the correction that comes from slope-matching is minor. We illustrate this in Figure~\ref{fig:isoth_freeslope_comparison}, by displaying the impact of a $m=4$ multipole on an EPL profile with a large ellipticity ($q=0.5$) and a slope significantly distinct from the isothermal case ($\gamma=2.25$), comparing the circular/elliptical and isothermal/slope-matched formulations\footnote{For the circular multipoles, to rescale the amplitude in the slope-matched case, we use a reasoning similar to Equations~(\ref{eq:perturbed_convergence_slope_matched}) and (\ref{eq:isocontour_slope_matched}), but we normalize like in the isothermal case: with respect to the semi-major axis at the $\kappa=1/2$ EPL isocontour when $\phi_m=0$, i.e., we use $a_m\mapsto \beta(2-\beta)\left(\frac{\theta_E}{\sqrt{q}}\right)^{2-\beta}$}. The isothermal multipoles provide a good approximation of the slope-matched pertubations, at least on radial scales of a few Einstein radii - which is the most important for strong lensing applications. The multipole slope only affects the radial dependence of the amplitude, not the pattern of perturbation: in particular, the slope-matched circular multipoles still display the same undesired patterns as the isothermal version (more ‘‘spinny-top’’ shaped than disky). Despite having a more easily interpretable radial dependence, the slope-matched circular multipoles yield a less realistic lens model than the (isothermal) elliptical multipoles for systems with $q\not\approx1$, since the complexity of the mass distribution near the Einstein radius is not adequately encoded.\\

To characterize the regime where isothermal elliptical multipoles are expected to be an improvement (even compared to slope-matched circular multipoles), we can perform a first order calculation to estimate and compare the differences between multipole formulations.  Making the isothermal assumption for the multipoles changes the convergence profile by:
\begin{equation}
\begin{split}
    \Delta\kappa_{\rm isoth} \equiv |\kappa_{m,\beta}-\kappa_{m, \beta=1}|&= |\kappa_{m,\beta}|\cdot\left|1-\frac{a_{m, \rm eff}(R)}{a_m}\right|\\
    &\approx \left|\kappa_{m,\beta}\right| \cdot \left|(1-\beta)\ln\left( \frac{R}{\theta_E}\right) \right|
    \label{eq:delta_kappa_isoth}
\end{split}
\end{equation}
where we have used Equations~(\ref{eq:eff_mult_amplitude_def}) and (\ref{eq:effective_mult_amplitude_approx}), assuming that $\beta-1$ is small and including the rescaling of the amplitudes. Similarly, the assumption of circularity introduces an difference:
\begin{equation}
\begin{split}
    \Delta\kappa_{\rm circ} &= |\kappa_{m,\beta}-\kappa_{m, \beta}^{\rm circ}|\\
    &= \left|\frac{\beta}{2} (2-\beta)a_m\cos[m(\varphi-\varphi_m)] \left( \frac{\theta_E}{R}\right)^{2-\beta} - \frac{\beta}{2} (2-\beta)a_m\cos[m(\phi-\phi_m)] \left( \frac{\theta_E}{r\sqrt{q}}\right)^{2-\beta}\right|\\
    &= \left|\kappa_{m,\beta}\right| \cdot \left| 1-\frac{\cos[m(\phi-\phi_m)]}{\cos[m(\varphi-\varphi_m)]} \left( \frac{R}{r\sqrt{q}}\right)^{2-\beta} \right| 
\end{split}
\end{equation}
Let us consider an example that is likely to be relevant for astrophysical applications: the $m=4$ multipole aligned with the reference ellipse (i.e., $\varphi_4=0$ or $\phi_4=0$). In that case
\begin{equation}
\begin{split}
    \Delta\kappa_{\rm circ} &= \left|\kappa_{m,\beta}\right| \cdot \left| 1-\frac{\cos(4\phi)}{\cos(4\varphi)} \left( \frac{\sqrt{q^2\cos^2(\phi)+\sin^2(\phi)}}{q}\right)^{2-\beta} \right| \\
    &\sim \left|\kappa_{m,\beta}\right| \cdot (2-\beta)(1-q)
\end{split}
\label{eq:delta_kappa_circ}
\end{equation}
where, for simplicity, we have evaluated in $\phi=\varphi=\frac{\pi}{2}$ (orthogonal to the multipole direction) and only kept the first-order term of the expansion in $(1-q)$. Comparing Equation~(\ref{eq:delta_kappa_circ}) to Equation~(\ref{eq:delta_kappa_isoth}), we find that the assumption of isothermal multipoles has less impact than the assumption of circularity if the following condition is verified:
\begin{equation}
    \left|\ln \left(\frac{R}{\theta_E}\right) \right| \lesssim \frac{(2-\beta)(1-q)}{|1-\beta|}
\end{equation}
For instance, if $\gamma=2.25$ and $q=0.8$, $\Delta\kappa_{\rm isoth} \lesssim \Delta\kappa_{\rm circ}$ for $0.4\theta_E\lesssim R \lesssim 3\theta_E$. The exact numerical factor will vary depending on the multipole order, multipole direction, and azimuthal angle, but as a rule of thumb, as long as $1-q \not \ll |1-\beta|$, the correction introduced by matching the slope of the multipoles to the EPL is less important than the correction from ellipticity on scales $R\sim \theta_E$.

\section{Explicit deflections fields for the elliptical multipoles}
\label{App:m=1/3/4_alpha_explicit}

Many strong lensing applications directly employ the deflection angles without being concerned with the lensing potential. The deflection field can be straightforwardly calculated ($\vec{\alpha}=\vec{\nabla}\psi$) from the lensing potentials presented in Section~\ref{sec:potential_solutions}, but for convenience, we report explicit analytical expressions for the deflection fields of the $m=1$, $m=3$ and $m=4$ elliptical multipoles in this section.

\subsection{The $m=1$ elliptical multipole deflection field}
\label{App:m=1_alpha_explicit}

To describe the deflection field of the elliptical $m=1$ multipole, we can start by the $\varphi_1=0$ component, taking the derivatives of the lensing potential from Equation~(\ref{eq:psi_1_1}):
\begin{equation}
\left\{\begin{array}{l}
        \alpha_{x,1}^{(1)}(r,\phi;q) = \hat{F}_1^{(1)}(\phi, q) \cos(\phi) - \hat{F}_1^{(1)\prime}(\phi, q) \sin(\phi) + \lambda_1(q) \cdot\alpha_{x,1}^{\rm circ}(r,\phi) \Big|_{\phi_1=0, a_1^{\rm circ}=1}  \\
        \alpha_{y,1}^{(1)}(r,\phi;q) = \hat{F}_1^{(1)}(\phi, q) \sin(\phi) + \hat{F}_1^{(1)\prime}(\phi, q) \cos(\phi) + \lambda_1(q) \cdot\alpha_{y,1}^{\rm circ}(r,\phi) \Big|_{\phi_1=0, a_1^{\rm circ}=1} 
\end{array}\right.
\end{equation}
where $\hat{F}_1^{(1)}$ and $\lambda_1(q)$ are given by Equation~(\ref{eq:F_1_1_hat}), $\alpha_{x,1}^{\rm circ}$ and $\alpha_{y,1}^{\rm circ}$ by Equation~(\ref{eq:derivatives_psi_1_circ}), the derivative of the modified shape function (with respect to $\phi$) is

\begin{multline}
    \hat{F}_1^{(1)\prime}(\phi, q) =  \frac{1}{(1 - q^2)} 
    \left[ \frac{q(q^2 - 1) \cos(\phi) \sin(2\phi)}{1 + q^2 + (q^2 - 1)\cos(2\phi)} + \frac{q}{2}\sin(\phi) \ln(1 + q^2 + (q^2 - 1)\cos(2\phi))  \right. \\    
    \left. - A_1^{(1)}(q)\sin\phi -  \big[\phi - \varphi(\phi; q)\big]\cos\phi - \sin(\phi) \left(1 - \frac{q}{q^2 \cos^2(\phi) + \sin^2(\phi)}\right) \right],
\end{multline}
and $A_1^{(1)}(q)$ is given by Equation~(\ref{eq:A_1_1}). Then, using Equations~(\ref{eq:psi_m_1_2_decomp}) and (\ref{eq:alpha_hessian_1_2_relation}), the deflection angles for a general orientation $\varphi_1$ can be written as:

\begin{equation}
\left\{\begin{array}{l}
        \alpha_{x,1}(r,\phi;q) = a_1\sqrt{q}\left[\alpha_{x,1}^{(1)}(r,\phi;q) \cos\varphi_1 - \frac{\sin\varphi_1}{q} \alpha_{y, 1}^{(1)}(r,\phi+ \frac{\pi}{2};\frac{1}{q})\right]  \\
        \alpha_{y,1}(r,\phi;q) = a_1\sqrt{q}\left[\alpha_{y,1}^{(1)}(r,\phi;q) \cos\varphi_1 + \frac{\sin\varphi_1}{q} \alpha_{x, 1}^{(1)}(r,\phi+ \frac{\pi}{2};\frac{1}{q})\right].
\end{array}\right.
\end{equation}

\subsection{The $m=3$ elliptical multipole deflection field}
\label{App:m=3_alpha_explicit}

For the deflection field of the elliptical $m=3$, we follow the same logic as in Section~\ref{App:m=1_alpha_explicit}, first taking the derivatives of the $\varphi_3=0$ component, i.e., $\psi_3^{(1)}$ from Equation~(\ref{eq:psi_3_1}):
\begin{equation}
\left\{\begin{array}{l}
        \alpha_{x,3}^{(1)}(r,\phi;q) = \hat{F}_3^{(1)}(\phi, q) \cos(\phi) - \hat{F}_3^{(1)\prime}(\phi, q) \sin(\phi) + \lambda_3(q) \cdot\alpha_{x,3}^{\rm circ}(r,\phi) \Big|_{\phi_3=0, a_3^{\rm circ}=1}  \\
        \alpha_{y,3}^{(1)}(r,\phi;q) = \hat{F}_3^{(1)}(\phi, q) \sin(\phi) + \hat{F}_3^{(1)\prime}(\phi, q) \cos(\phi) + \lambda_3(q) \cdot\alpha_{y,3}^{\rm circ}(r,\phi) \Big|_{\phi_3=0, a_3^{\rm circ}=1} 
\end{array}\right.
\end{equation}
where $\hat{F}_3^{(1)}$ and $\lambda_3(q)$ are given by Equation~(\ref{eq:F_3_1_hat}), $\alpha_{x,3}^{\rm circ}$ and $\alpha_{y,3}^{\rm circ}$ by Equation~(\ref{eq:derivatives_psi_1_circ}), the derivative of the modified shape function (with respect to $\phi$) is
\begin{multline}
    \hat{F}_3^{(1)\prime}(\phi, q) =  \frac{1}{(1 - q^2)^2} \left[  \frac{ q(1-q^2)(3 + q^2) \cos(\phi)\sin(2\phi)}{1 + q^2 + (q^2 - 1)\cos(2\phi)} + (1 + 3q^2) \big[\phi - \varphi(\phi; q)\big] \cos\phi \right. \\
    - \frac{q}{2}(3 + q^2)\sin(\phi) \ln(1 + q^2 + (q^2 - 1)\cos(2\phi)) - A_3^{(1)}(q)\sin\phi   \\  
    \left.  + (1 + 3q^2) \sin(\phi) \left(1 - \frac{q}{q^2 \cos^2(\phi) + \sin^2(\phi)}\right)\right],
\end{multline}
and $A_3^{(1)}(q)$ is given by Equation~(\ref{eq:A_3_1}). Then, using Equations~(\ref{eq:psi_m_1_2_decomp}) and (\ref{eq:alpha_hessian_1_2_relation}), the deflection angles for a general orientation $\varphi_3$ can be written as:

\begin{equation}
\left\{\begin{array}{l}
        \alpha_{x,3}(r,\phi;q) = a_3\sqrt{q}\left[\alpha_{x,3}^{(1)}(r,\phi;q) \cos(3\varphi_3) + \frac{\sin(3\varphi_3)}{q} \alpha_{y, 3}^{(1)}(r,\phi+ \frac{\pi}{2};\frac{1}{q})\right]  \\
        \alpha_{y,3}(r,\phi;q) = a_3\sqrt{q}\left[\alpha_{y,3}^{(1)}(r,\phi;q) \cos(3\varphi_3) - \frac{\sin(3\varphi_3)}{q} \alpha_{x, 3}^{(1)}(r,\phi+ \frac{\pi}{2};\frac{1}{q})\right].
\end{array}\right.
\end{equation}

\subsection{The $m=4$ elliptical multipole deflection field}
\label{App:m=4_alpha_explicit}

In the case of the $m=4$ elliptical multipole, the lensing potential belongs to the family of models of Equation~(\ref{eq:gen_pot_conv_pair}), so we can simply apply Equation~(\ref{eq:pot_alpha_hessian_array}) to get the deflection angles:

\begin{equation}
\left\{\begin{array}{l}
\alpha_{x,4}(r,\phi)= F_4(\phi)\cos\phi-F_4^{\prime}(\phi)\sin\phi \\ ~~ \\
\alpha_{y,4}(r,\phi)= F_4(\phi)\sin\phi+F_4^{\prime}(\phi)\cos\phi 
\end{array}\right.
\end{equation}
Following Equation~(\ref{eq:F_m_1_2_decomp}), the shape function and its derivative are decomposed as:
\begin{equation}
\begin{split}
    F_4(\phi ;q) &= a_4\sqrt{q}\cos (4\varphi_4) F_4^{(1)}(\phi ;q) + a_4\sqrt{q}\sin (4\varphi_4) F_4^{(2)}(\phi ;q)\\
    F_4^\prime(\phi ;q) &= a_4\sqrt{q}\cos (4\varphi_4) F_4^{(1)\prime}(\phi ;q) + a_4\sqrt{q}\sin (4\varphi_4) F_4^{(2)\prime}(\phi ;q)\\.
\end{split}
\end{equation}
where the expressions for $F_4^{(1)}$ and $ F_4^{(2)}$ are given Equations~(\ref{eq:F_4_1}) and (\ref{eq:F_4_2}), respectively, and their derivatives can be written as: 

\begin{multline}
F_4^{(1)\prime}(\phi ; q) =  \frac{-4 \sqrt{2} \, (1 + q^4 + (q^4 - 1)\cos(2\phi)) \sin(2\phi)}{3(1 - q^2)(1 + q^2 + (q^2 - 1)\cos(2\phi))^{3/2}} \\
 + \frac{(1 + 6q^2 + q^4)}{(1 - q^2)^{5/2}} \left[ - \sin(\phi)\arctan\left( \frac{\sqrt{2(1 - q^2)} \, \cos(\phi)}{\sqrt{1 + q^2 + (q^2 - 1)\cos(2\phi)}} \right) \right. \\ \left.
 +  \cos(\phi)\ln\left( \frac{\sqrt{1 - q^2} \, \sin(\phi)}{q} + \sqrt{1 + \frac{(1 - q^2)}{q^2} \sin^2(\phi)} \right)  \right]
\end{multline}
and 
\begin{multline}
F_4^{(2)\prime}(\phi ; q) =  
\frac{-8 \sqrt{2} \, q}{6(1 - q^2)} \left(  -\frac{(1 - q^2) \sin^2(2\phi)}{(1 + q^2 + (q^2 - 1)\cos(2\phi))^{3/2}} + \frac{2\cos(2\phi)} {\sqrt{1 + q^2 + (q^2 - 1)\cos(2\phi)}} \right) \\
- \frac{4q(1 + q^2)}{(1 - q^2)^{5/2}} \left[\cos(\phi) \arctan \left( \frac{\sqrt{2(1 - q^2)} \, \cos(\phi)}{\sqrt{1 + q^2 + (q^2 - 1)\cos(2\phi)}} \right) - \frac{\sqrt{2(1 - q^2)}}{\sqrt{1 + q^2 + (q^2 - 1)\cos(2\phi)}} \right.
\\
\left.  +\sin(\phi)  \ln\left( \frac{\sqrt{1 - q^2} \, \sin(\phi)}{q} + \sqrt{1 + \frac{(1 - q^2)}{q^2} \sin^2(\phi)} \right) \right].
\end{multline}

\section{General solutions for the lensing potential of the elliptical multipoles}
\label{App:gen_ell_potential}

In this section, we present the lensing potential solutions for the elliptical multipoles of general order $m\geq2$, separating the cases where $m$ is odd (Appendix~\ref{App:m=2k+1_potential}) and $m$ is even (Appendix~\ref{App:m=2k_potential}). This is mostly done for completeness: the solutions for the $m=3$ and $m=4$ orders were presented in Sections~\ref{subsec:m=3} and \ref{subsec:m=4} in equivalent, ‘‘simplified’’ forms, better suited for pratical applications.

\subsection{Lensing potential for elliptical multipoles with $m$ odd}
\label{App:m=2k+1_potential}

In this section, we assume that $m \geq 3$ is odd, i.e., that we can write $m=2k+1$ for $k\in \mathbb{N}^*$. We have the following explicit expression for the Chebyshev polynomials of the first kind:
\begin{equation}
    T_m(x) = \frac{m}{2} \sum_{p=0}^{\lfloor m/2 \rfloor} \left(- 1 \right)^p \frac{(m-p-1)!}{p!(m-2p)!} (2x)^{m-2p} = \frac{m}{2} \sum_{l=0}^{k} \left(- 1 \right)^{k-l} \frac{1}{k+l+1}\binom{k+l+1}{2l+1} (2x)^{2l+1}
\end{equation}
where, in the second sum, we have made the change of variables $l=k-p$. Plugging this into Equation~(\ref{eq:G_m_1_2_def}), we find that we can write the shape function as a linear combination of simpler functions:
\begin{equation}
    G_m^{(1)}(\phi;q) = \sum_{l=0}^{+\infty} g_{k l}(q) \cdot u_l(\phi;q)
    \label{eq:G_m_1_expansion_odd}
\end{equation}
where we have defined
\begin{equation}
\begin{split}
 u_l(\phi;q) & \equiv \frac{\cos^{2l+1}\phi}{(q^2\cos^2\phi +\sin^2\phi)^{l+1}} \text{ for $l\in \mathbb{R}$}\\
    \text{and } g_{k l}(q) & \equiv \begin{cases}
    (-1)^{k-l} \frac{2k+1}{2(k+l+1)} \binom{k+l+1}{2l+1} (2q)^{2l+1} \text{ if } l\leq k\\
    0 \text{ otherwise } 
\end{cases}
\end{split}
\end{equation} 

The (unsymmetrized) solution to differential Equation~(\ref{eq:main_diff_eq}) has the following form:

\begin{equation}
\begin{split}
      \tilde{F}_m^{(1)}(\phi;q) = \frac{1}{(1-q^2)^{k+1}} \Bigg[ & A_m^{(1)}(q) \cos\phi +   B_m^{(1)}(q) \sin\phi +  C_m^{(1)}(q) \mathcal{C}(\phi; q) \\ &+ D_m^{(1)}(q) \mathcal{D}(\phi; q) + \sum_{j=1}^{k-1}  \xi_{mj}^{(1)}(q) u_{j-1}(\phi; q) \Bigg]
\end{split}
\label{eq:f_m_odd_1}
\end{equation}
where the coefficients $A_m^{(1)}, B_m^{(1)}, C_m^{(1)}, D_m^{(1)},\{\xi_{mj}^{(1)}\}$ are functions of $q$ only (i.e., independent of $\phi$), and we have defined:
\begin{equation}
    \begin{split}
        \mathcal{C}(\phi; q) &= \cos(\phi)  \ln \left[ 1+q^2 + (q^2 -1) \cos(2\phi) \right] + 2\phi\sin\phi\\ 
        \mathcal{D}(\phi; q) &= \sin(\phi) \varphi(\phi;q)
    \end{split}
    \label{eq:f_m_odd_1_subfunctions}
\end{equation}
To determine the coefficients, we write the derivatives as combinations of the functions $u_l$ defined above:
\begin{equation}
    \begin{split}
        \mathcal{C}^{\prime\prime} (\phi; q)+ \mathcal{C}(\phi; q)  = &  \frac{2(1-q^2)}{(q^2\cos^2\phi +\sin^2\phi)^{2}} \left[-2(1-q^2)\cos^5\phi + (5-q^2)\cos^3\phi -3\cos\phi\right] + 4\cos\phi \\ 
        = & -2(1-3q^2)u_0 + 4q^2(1-q^2)u_1 \hfill \\
        \mathcal{D}^{\prime\prime} (\phi; q)+ \mathcal{D}(\phi; q)  = & \frac{q^3 \cos\phi}{(q^2\cos^2\phi +\sin^2\phi)^{2}} \\
        = &\ 2q^3 u_0 +2q^3(1-q^2)u_1 \\
        u_{j-1}^{\prime \prime}(\phi;q) + u_{j-1}(\phi;q) = & \Big[ (q^4-jq^2)\cos^4\phi +(j(2j+1)+q^2(2-5j))\cos^2\phi\sin^2\phi  \\
        &  +2(j-1)(2j-1)\sin^4\phi \Big] \frac{2\cos^{2j-3}\phi}{(q^2\cos^2\phi +\sin^2\phi)^{j+2}}  \\
        = &\ 2(j-1)(2j-1)u_{j-2} + \left[ 2(1-q^2)(2j(3j-2)+1)-4j(j-1)\right]u_{j-1}\\
        & + \left[ -8j^2(1-q^2) +2(1-q^2)^2 j(6j+1) \right]u_j - 4j(j+1)q^2(1-q^2)^2 u_{j+1}
    \end{split}
\end{equation}
We can then write, with the convention $\xi_{mj}^{(1)}=0$ if $j\geq k$ or $j\leq0$:
\begin{equation}
\begin{split}
    (1-&q^2)^{k+1} \left[ \tilde{F}_m^{(1)\prime \prime}(\phi;q) + \tilde{F}_m^{(1)} (\phi;q) \right] \\ = &\left[ -2(1-3q^2)C_m^{(1)}+2q^3 D_m^{(1)}\right] u_0 +\left[ 4q^2(1-q^2)C_m^{(1)} + 2q^3(1-q^2)D_m^{(1)} \right] u_1 \\
    &  
    \begin{split}
        + \sum_{j=0}^{k+1} & \left\{ -4j(j-1)q^2(1-q^2)^2 \xi_{m(j-1)}^{(1)} + \left[-8j^2(1-q^2)+2(1-q^2)^2j(6j+1) \right] \xi_{mj}^{(1)} \right. \\ + & \left. \left[2(1-q^2)(2(j+1)(3j+1)+1)-4j(j+1) \right] \xi_{m(j+1)}^{(1)} + 2(j+1)(2j+3) \xi_{m(j+2)}^{(1)}\right\} u_j
    \end{split}
    \label{eq:F_m_1_expansion}
\end{split}
\end{equation}

In order for $\tilde{F}_m^{(1)}(\phi;q)$ to be a solution of the differential Equation~(\ref{eq:main_diff_eq}), it is sufficient to match the coefficients in front of the $u_{j}$ in Equations~(\ref{eq:G_m_1_expansion_odd}) and (\ref{eq:F_m_1_expansion}). This amounts to solving a matrix–vector equation:
\begin{equation}
    M_m(q) \cdot \begin{pmatrix}
C_m^{(1)} \\
D_m^{(1)} \\
\xi_{m1}^{(1)}\\
\vdots \\
\xi_{m(k-1)}^{(1)}
\end{pmatrix}
= (1 - q^2)^{k+1} \begin{pmatrix}
g_{k0} \\
g_{k1} \\
\vdots \\
g_{kk}
\end{pmatrix}
\label{eq:matrix_eq}
\end{equation}

The $(k+1)\times (k+1)$ matrix $M_m(q)$ can be written in a block-triangular form:
\begin{equation}
    M_m(q) = \begin{pmatrix}
    M_{m}^{\rm sup}(q) & M_{m}^{\rm ur}(q) \\
    (0) & M_{m}^{\rm tri}(q)
    \end{pmatrix}
\end{equation}
where $M_{m}^{\rm sup}(q) = \begin{pmatrix}
-2(1 - 3q^2) & 2q^3 \\
4q^2(1 - q^2) & 2q^3(1 - q^2)
\end{pmatrix} $ and the two other blocks $M_{m}^{\rm ur}(q), M_{m}^{\rm tri}(q)$ are determined by:
\begin{equation}
\begin{split}
\quad M_{m}(q)[j, j+3] &= 2(j+1)(2j+3) \text{ if } 0\leq j \leq k-3, \\
\quad M_{m}(q)[j, j+2] &= 2(1 - q^2)\left(2(j+1)(3j+1)+1\right) - 4j(j+1)  \text{ if } 0\leq j \leq k-2,\\
\quad M_{m}(q)[j, j+1] &= -8j^2(1 - q^2) + 2(1 - q^2)^2 j(6j+1) \text{ if } 1 \leq j \leq k-1, \\
\quad M_{m}(q)[j, j] &= -4j(j-1)q^2(1 - q^2)^2 \text{ if } 2 \leq j \leq k,
\end{split} 
\end{equation}
all the other coefficients being zero.

As a consequence, the lower-right block $M_m^{\rm tri}$ is upper triangular and we can easily write the determinant of $M_m$:
\begin{equation}
    \begin{split}
        \det\left[M_m(q)\right] & = \det\left[M_{m}^{\rm sup}(q)\right] \times \det\left[M_{m}^{\rm tri}(q)\right] \\
        &= -4q^3(1-q^2)^2 \times \prod_{j=2}^{k} -4j(j-1)q^2(1 - q^2)^2 \\
        &=  q^{2k+1} (2(1 - q^2))^{2k} k!(k-1)! 
    \end{split}
\end{equation}
Since $\det\left[M_m(q)\right] \neq 0$ for $q \neq 1$ ( $q=1$ corresponding to the circular multipole solution), Equation~(\ref{eq:matrix_eq}) always has a solution when relevant and the coefficients $C_m^{(1)}, D_m^{(1)}, E_m^{(1)}, \{\xi_{mj}^{(1)}\}$ are fully determined.

We have therefore solved differential Equation~(\ref{eq:main_diff_eq}) for $G_m^{(1)}(\phi;q)$ - the remaining coefficients $A_m^{(1)}$ and $B_m^{(1)}$ only correspond to the terms of the homogeneous equation, and once again, they do not matter for the lensing observables due to the prismatic degeneracy (see Section~\ref{sec:general_formalism} or \citep{Gorenstein1988}). However, the solution $\tilde{F}_m^{(1)}(\phi;q)$ does not have all expected symmetries: choosing $B_m^{(1)}=0$, we enforce the $\phi \mapsto -\phi$ symmetry ; but Equation~(\ref{eq:f_m_odd_1}) cannot be made $2\pi$-periodic or $\phi \mapsto \pi-\phi$ antisymmetric, because of the terms in $\phi\sin\phi$ and $\sin(\phi)\varphi(\phi ;q)$. Direct symmetrization of the shape function leads to jump discontinuties in the deflection field, like in Sections~\ref{subsec:circ_multipole_potential}, \ref{subsec:m=1}, and \ref{subsec:m=3} ; so we use the same trick as before, dividing the lensing potential into two components:
\begin{equation}
    \psi_m^{(1)}(r,\phi;q) = r \hat{F}_m^{(1)}(\phi;q) + \lambda_m(q) \cdot \psi_1^{\rm circ}\left(r,\phi \right) \Big|_{\phi_1=0, a_1^{\rm circ}=1}
\end{equation}
where we have defined
\begin{equation}
\begin{split}
    \hat{F}_m^{(1)}(\phi;q)  &= \tilde{F}_m^{(1)}(\phi ;q) - \frac{\lambda_m(q)}{2} \phi\sin\phi \\
    &= \tilde{F}_m^{(1)}(\phi ;q) - \lambda_m(q) \cdot\tilde{F}_m^{\rm circ}(\phi) \Big|_{\phi_1=0, a_1^{\rm circ}=1}
\end{split}
\end{equation}

This modified shape function $\hat{F}_m^{(1)}(\phi;q)$ can have the correct symmetries ($2\pi$ periodic, symmetric under $\phi\mapsto -\phi $ and antisymmetric under $\phi\mapsto \pi-\phi $): we simply need to choose $\lambda_m(q)$ so that terms in $\phi$ and $\varphi(\phi,q)$ can be factorized as $\phi - \varphi(\phi,q)$ (see Section~\ref{subsec:m=1}). Examining the coefficients in Equation~(\ref{eq:f_m_odd_1}) and (\ref{eq:f_m_odd_1_subfunctions}), we obtain the following condition:
\begin{equation}
    \lambda_m(q) = \frac{2D_m^{(1)}(q)+4C_m^{(1)}(q)}{(1-q^2)^{k+1}}
\end{equation}
In practice, we find that the following expression is verified at least for $1\leq k <10$:
\begin{equation}
    \lambda_m(q) = \frac{2(q-1)^{k}}{(q+1)^{k+1}} \mathrel{\underset{q \to 1}{\longrightarrow}}0.
\end{equation}

We saw that $B_m^{(1)}(q)=0$ was imposed from the expected symmetries, but the other remaining coefficient $A_m^{(1)}(q)$ cannot be determined in such manner. Instead, using a similar approach to the $m=3$ case (see Section~\ref{subsec:m=3}), we write the Taylor expansion of $\tilde{F}_m^{(1)}(\phi;q)$ when $q \to 1$ for $\phi=0$:
\begin{equation}
     \tilde{F}_m^{(1)}(\phi=0 ;q) \mathrel{\underset{q \to 1}{=}} \frac{A_m^{(1)}}{(1 - q^2)^{k+1}} + \sum_{j=0}^{k+1}  \frac{t_{mj}}{(1 - q)^{j}}  + O[1-q]
\end{equation}
where the $t_{mj}$ can be determined with a symbolic calculator. Assuming that we indeed have $\lambda_m(q)\mathrel{\underset{q \to 1}{\longrightarrow}}0$, the potential $\psi_m^{(1)}$ will converge to the circular multipole solution in the limit $q\to 1$ if $ \tilde{F_m}^{(1)}(\phi;q) \mathrel{\underset{q \to 1}{\longrightarrow}} F_m^{\rm circ}(\phi) \Big|_{\phi_m=0, a_m^{\rm circ}=1}$. This is ensured if we choose
\begin{equation}
    A_m^{(1)}(q) = (1 - q^2)^{k+1} \left[ \frac{1}{1-m^2} - \sum_{j=0}^{k+1}  \frac{t_{mj}}{(1 - q)^{j}} \right].
\end{equation}
We could add any term $f(q)$ to $A_m^{(1)}$ as long as $f(q)(1-q)^{-(k+1)}\mathrel{\underset{q \to 1}{\longrightarrow}}0$, but thanks to the prismatic degeneracy, this would not change the lensing observables, so we can safely ignore it and choose $f(q)=0$ for simplicity.

We note that, following the method described in this section, we obtain the following coefficients: $C_3^{(1)}(q) = \frac{1}{2}q(3+q^2) $, $D_3^{(1)}(q) = -(1+3q^2)$ and $A_3^{(1)}(q)$ given by Equation~(\ref{eq:A_3_1}).
Accordingly, one can easily check that we recover the same solution for the $m=3$ lensing potential as in Section~\ref{subsec:m=3}.

\begin{figure*}[!h]
    \centering
    \includegraphics[width=0.99\linewidth]{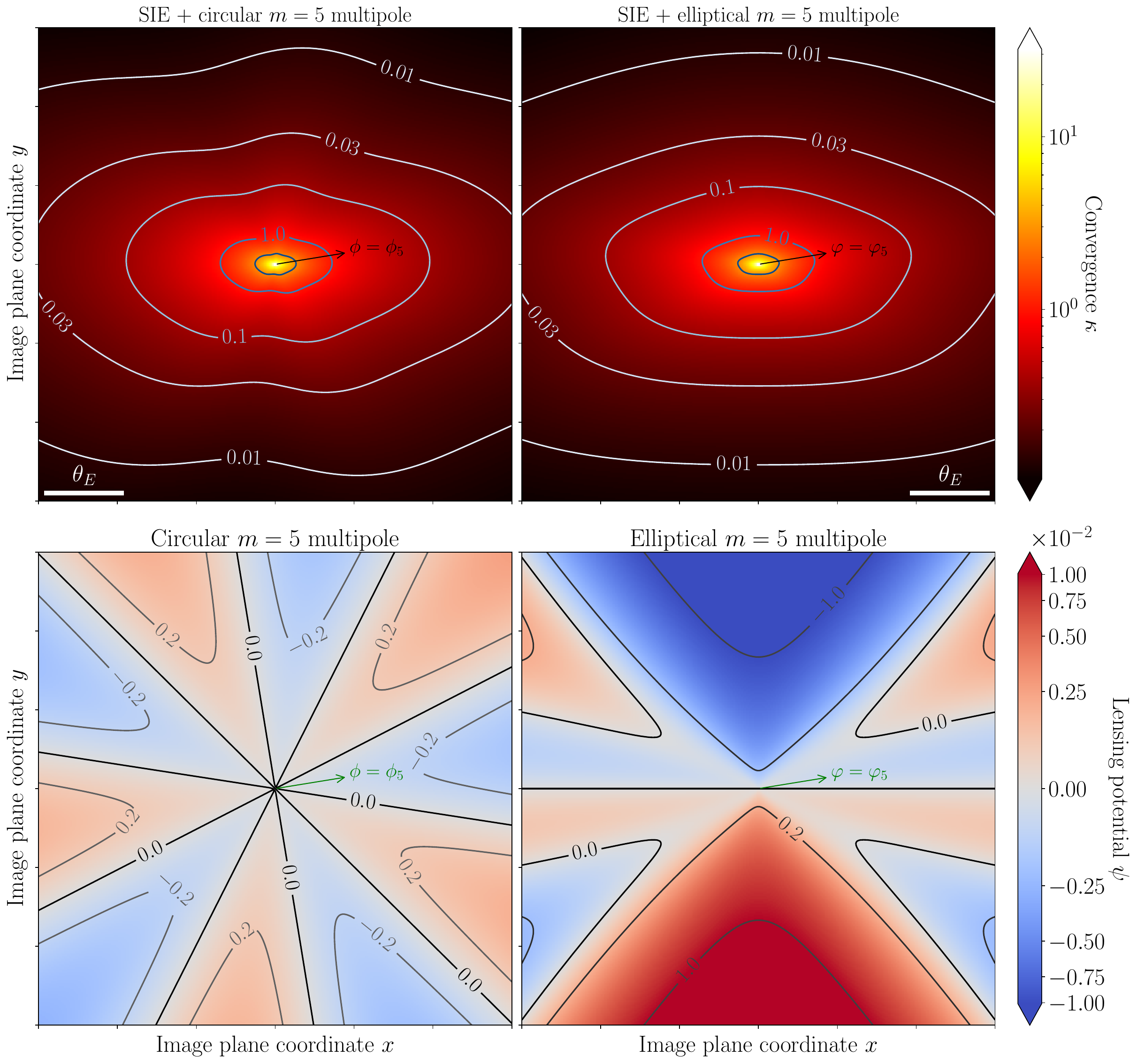}
    \caption{Top panel: Impact of a $m=5$ multipole on the convergence profile of a SIE lens model with  axis ratio $q=0.5$, in the circular (left) and elliptical (right) formulations. We choose perturbations with direction $\phi_5=\phi(\frac{\pi}{10};q)$ (resp. $\varphi_1=\frac{\pi}{10}$) and amplitude $a_1^{\rm circ}=0.027$ (resp. $a_1=0.027$). Bottom panel: lensing potentials corresponding to the $m=5$ multipole term only, without rescaling of the amplitudes.}
    \label{fig:m=5_example}
\end{figure*}

Now that $\psi_m^{(1)}(\phi;q)$ is fully formulated, the other component $\psi_m^{(2)}(\phi;q)$ can be determined without additional effort, by making use of Equation~(\ref{eq:psi_1_psi_2_relation}). This completes the description of the elliptical multipole lensing potential in the general $m=2k+1$ case. In Figure~\ref{fig:m=5_example}, we show example potential maps for the $m=5$ elliptical multipole calculated using the method described in this Appendix, as well as the effect of such a perturbation on a SIE convergence profile, in comparison with the $m=5$ circular multipole.

\subsection{Lensing potential for elliptical multipoles with $m$ even}
\label{App:m=2k_potential}

In this section, we assume that $m \geq 2$ is even, i.e., that we can write $m=2k$ for $k\in \mathbb{N}^*$. The Chebyshev polynomials of the first and second kind can be written explicitly as:
\begin{equation}
\begin{split}
    T_m(x) &= \frac{m}{2} \sum_{p=0}^{\lfloor m/2 \rfloor} \left(- 1 \right)^p \frac{(m-p-1)!}{p!(m-2p)!} (2x)^{m-2p} = k \sum_{l=0}^{k} \left(- 1 \right)^{k-l} \frac{k}{k+l} \binom{k+l}{k-l} (2x)^{2l} \\
    U_{m-1}(x) &= \sum_{p=0}^{\lfloor \frac{m-1}{2} \rfloor} \left(- 1 \right)^p \binom{m-1-p}{p}  (2x)^{m-1-2p} =  \sum_{l=0}^{k-1} \left(- 1 \right)^{k-l-1} \binom{k+l}{k-l-1}  (2x)^{2l+1}
\end{split}
\end{equation}
where we have made the change of variables $l=k-p$ in the first line, and $l=k-1-p$ in the second line. Again, plugging this into Equation~(\ref{eq:G_m_1_2_def}), we write the shape functions as a linear combination of basis functions:
\begin{equation}
\begin{split}
    G_m^{(1)}(\phi;q) &= \sum_{l=0}^{+\infty} \gamma_{k l}^{(1)}(q) \cdot u_{l-1/2}(\phi;q)\\
    G_m^{(2)}(\phi;q) &= \sum_{l=0}^{+\infty} \gamma_{k l}^{(2)}(q) \cdot w_l(\phi;q)
    \label{eq:G_m_expansion_even}
\end{split}
\end{equation}
where we use the same definition for $u_l$ as in Section~\ref{App:m=2k+1_potential} such that $u_{l-1/2}(\phi;q) =  \frac{\cos^{2l}\phi}{(q^2\cos^2\phi +\sin^2\phi)^{l+1/2}} $, and we define $w_l(\phi;q) \equiv \frac{\sin \phi \cos^{2l+1}\phi }{(q^2\cos^2\phi +\sin^2\phi)^{l+3/2}} $, as well as the coefficients
\begin{equation}
\begin{split}
    \gamma_{k l}^{(1)}(q) &\equiv \begin{cases}
    \left(- 1 \right)^{k-l} \frac{k}{k+l} \binom{k+l}{k-l} (2q)^{2l} \text{ if } l\leq k\\
    0 \text{ otherwise } 
\end{cases} \\ \text{ and } \gamma_{k l}^{(2)}(q) &\equiv \begin{cases}
    \left(- 1 \right)^{k-l-1} \binom{k+l}{k-l-1}  (2q)^{2l+1} \text{ if } l\leq k-1\\
    0 \text{ otherwise } 
\end{cases}
\end{split}
\end{equation}

The solutions to differential Equation~(\ref{eq:main_diff_eq}) have the following form:
\begin{equation}
\begin{split}
  F_m^{(1)}(\phi;q) &= \frac{1}{(1-q^2)^{k+1/2}} \Bigg[ A_m^{(1)}(q) \cos\phi +   B_m^{(1)}(q) \sin\phi +  E_m^{(1)}(q) \mathcal{E}_1(\phi; q) + \sum_{j=0}^{k-1}  \lambda_{mj}^{(1)}(q) v_{j}^{(1)}(\phi; q) \Bigg] \\
  F_m^{(2)}(\phi;q) &= \frac{1}{(1-q^2)^{k+1/2}} \Bigg[ A_m^{(2)}(q) \cos\phi +   B_m^{(2)}(q) \sin\phi +  E_m^{(2)}(q) \mathcal{E}_2(\phi; q) + \sum_{j=0}^{k-1}  \lambda_{mj}^{(2)}(q) v_{j}^{(2)}(\phi; q) \Bigg]
\label{eq:F_m_even_solution}
\end{split}
\end{equation}
where the coefficients $A_m^{(i)}, B_m^{(i)}, E_m^{(i)}, \{\lambda_{mj}^{(i)}\}$ are functions of $q$ only (i.e., independent of $\phi$), and we have defined:
\begin{equation}
    \begin{split}
        \mathcal{E}_1(\phi; q) = & +\cos(\phi) \arctan \left( \frac{\sqrt{(1 - q^2)} \cos(\phi)}{\sqrt{q^2\cos^2\phi +\sin^2\phi}} \right) \\ & +  \sin(\phi) \log \left( \frac{\sqrt{1 - q^2} \sin(\phi)}{q} + \sqrt{1 + \frac{(1 - q^2)}{q^2} \sin^2(\phi)} \right) \\ 
        \mathcal{E}_2(\phi; q) = & -\sin(\phi) \arctan \left( \frac{\sqrt{(1 - q^2)} \cos(\phi)}{\sqrt{q^2\cos^2\phi +\sin^2\phi}}  \right)  \\ & +\cos(\phi) \log \left( \frac{\sqrt{1 - q^2} \sin(\phi)}{q} + \sqrt{1 + \frac{(1 - q^2)}{q^2} \sin^2(\phi)} \right) \\
        v_{j}^{(1)}(\phi; q) = & \cos^2(\phi)u_{j-3/2}(\phi;q) = \frac{\cos^{2j}\phi}{(q^2\cos^2\phi +\sin^2\phi)^{j-1/2}} \\
        v_{j}^{(2)}(\phi; q) = & \cos(\phi) \sin(\phi) u_{j-1/2}(\phi;q) = \frac{\sin\phi \cos^{2j+1}\phi }{(q^2\cos^2\phi +\sin^2\phi)^{j+1/2}} = (q^2\cos^2\phi +\sin^2\phi)w_j(\phi; q)
    \end{split}
\end{equation}

The expressions in Equation~(\ref{eq:F_m_even_solution}) are $2\pi$-periodic, and, if (and only if) we choose $A_m^{(1)}=B_m^{(1)}=A_m^{(2)}=B_m^{(2)}=0$, they  already have the expected symmetries: $F_m^{(1)}$ (resp. $F_m^{(2)}$) is symmetric (resp. antisymmetric) under $\phi \mapsto -\phi$ and $\phi \mapsto \pi -\phi$.To determine the remaining coefficients, we follow the same approach as in Section~\ref{App:m=2k+1_potential}, and express the derivatives as linear combinations of the same basis functions used to describe $G_m^{(1)}$ and $G_m^{(2)}$:
\begin{equation}
    \begin{split}
        \mathcal{E}_1^{\prime\prime} (\phi; q)+ \mathcal{E}_1(\phi; q)  & =   \sqrt{1-q^2} \ u_{-1/2}\\ 
        v_{j}^{(1)\prime\prime}(\phi; q) + v_{j}^{(1)}(\phi; q) &=  (2j-1) \left[ 2j u_{j-3/2} +(2j -(4j+1)q^2)u_{j-1/2} - (2j+1)q^2(1-q^2) u_{j+1/2}\right] \\
       \mathcal{E}_2^{\prime\prime} (\phi; q)+ \mathcal{E}_2(\phi; q)  &=   - (1-q^2)^{3/2} \ w_0 \\ 
        v_{j}^{(2)\prime\prime}(\phi; q) + v_{j}^{(2)}(\phi; q) & =  (2j+1) \left[ 2jw_{j-1} +(2j -(4j+3)q^2)w_j - (2j+3)q^2(1-q^2) w_{j+1}\right]  \\
    \end{split}
\end{equation}
We can then express $F_m^{(i)\prime\prime}+ F_m^{(i)}$ as a function of the $u_{j-1/2}$ (for i=1) or the $w_j$ (for $i=2$), and match the coefficients to those found in Equation~(\ref{eq:G_m_expansion_even}). This results in two matrix-vector equations:

\begin{equation}
    N_m^{(1)}(q) \cdot \begin{pmatrix}
E_m^{(1)} \\
\lambda_{m0}^{(1)}\\
\vdots \\
\lambda_{m(k-1)}^{(1)}
\end{pmatrix}
= (1 - q^2)^{k+1/2} \begin{pmatrix}
\gamma_{k0}^{(1)} \\
\gamma_{k1}^{(1)} \\
\vdots \\
\gamma_{kk}^{(1)}
\end{pmatrix}
; \text{ and }     N_m^{(2)}(q) \cdot \begin{pmatrix}
E_m^{(2)} \\
\lambda_{m0}^{(2)}\\
\vdots \\
\lambda_{m(k-2)}^{(2)}
\end{pmatrix}
= (1 - q^2)^{k+1/2} \begin{pmatrix}
\gamma_{k0}^{(2)} \\
\gamma_{k1}^{(2)} \\
\vdots \\
\gamma_{k(k-1)}^{(2)}
\end{pmatrix}
\label{eq:matrix_eqs_even}
\end{equation}
where the matrices $ N_m^{(1)}(q)$ and $ N_m^{(2)}(q)$, respectively of dimension $(k+1)\times (k+1)$ and $k\times k$, with their only non-zero coefficients being:

\begin{equation}
\begin{split}
\quad N_m^{(1)}(q)[j, j] &= \begin{cases}
    \sqrt{1-q^2} \text{ if } j=0 \\
    -(2j-3)(2j-1)q^2(1-q^2) \text{ if } 1\leq j \leq k \end{cases} \\
\quad N_m^{(1)}(q)[j, j+1] &= (2j-1) \left[ 2j-(4j+1)q^2\right] \text{ if } 0 \leq j \leq k-1, \\
\quad N_m^{(1)}(q)[j, j+2] &= 2(j+1)(2j+1)  \text{ if } 0\leq j \leq k-2,\\
\quad N_m^{(2)}(q)[j, j] &= \begin{cases}
    -(1-q^2)^{3/2} \text{ if } j=0 \\
    -(2j-1)(2j+1)q^2(1-q^2) \text{ if } 1\leq j \leq k \end{cases} \\
\quad N_m^{(2)}(q)[j, j+1] &= (2j+1) \left[ 2j-(4j+3)q^2\right] \text{ if } 0 \leq j \leq k-1, \\
\quad N_m^{(2)}(q)[j, j+2] &= 2(j+1)(2j+3)  \text{ if } 0\leq j \leq k-2,\\
\end{split} 
\end{equation}

We note that $N_m^{(1)}(q)$ and
 $N_m^{(2)}(q)$ are upper triangular, and for $0<q<1$ they have only non-zero diagonal coefficients, so Equation~(\ref{eq:matrix_eqs_even}) always has a solution, and thus, all the coefficients are determined. For the $m=4$, the equations above yield:
\begin{align}
E_4^{(1)} & = 1 + 6 q^2 + q^4, & \lambda_{4,0}^{(1)} &= -\frac{8}{3}(1+2q^2)\sqrt{1-q^2}, & \lambda_{4,1}^{(1)} &= -\frac{8}{3}q^2(1-q^2)^{3/2}, \nonumber\\
E_4^{(2)} & = 4q(1+q^2), & \lambda_{4,0}^{(2)} &= -\frac{8}{3}q(1-q^2)^{3/2}, & &
\end{align}
and by plugging these coefficients into Equation~(\ref{eq:F_m_even_solution}), we recover the same shape functions as in Section~\ref{subsec:m=4}.

\begin{figure*}[!h]
    \centering
    \includegraphics[width=0.99\linewidth]{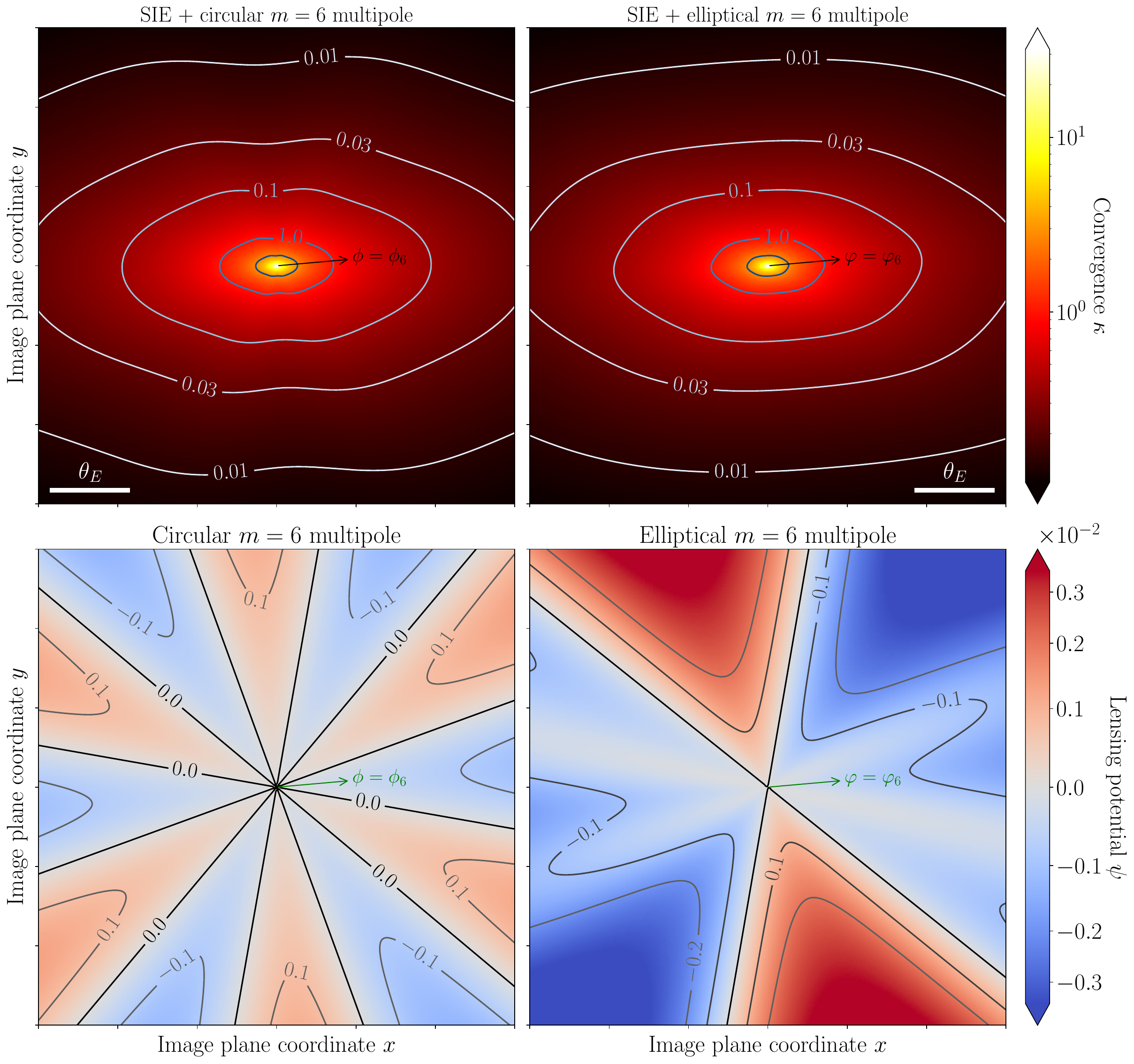}
    \caption{Same as Figure~\ref{fig:m=5_example}, but for a $m=6$ multipole with direction $\phi_6=\phi(\frac{\pi}{18};q)$ (resp. $\varphi_6=\frac{\pi}{18}$) and amplitude $a_4^{\rm circ}=0.015$ (resp. $a_4=0.015$). The axis ratio is still $q=0.5$. }
    \label{fig:m=6_example}
\end{figure*}

The lensing potential is then straightforwardly calculated: since the shape functions already have the right symmetries, there is no need to decompose the potential like in the odd $m$ case. Furthermore, we verified numerically for $1\leq k<7$ that the shape functions written in Equation~(\ref{eq:F_m_even_solution}) converged to the circular multipole solution when $q\to 1$. Thus, we simply need to plug the expressions for $F_m^{(1)}$ and $F_4^{(2)}$ in Equation~(\ref{eq:F_m_1_2_decomp}), then apply Equation~(\ref{eq:gen_pot_conv_pair}) with the resulting shape function $F_m(\phi;q)$. This completes the description of the elliptical multipole lensing potential in the general $m=2k$ case. In Figure~\ref{fig:m=6_example}, we show example potential maps for the $m=6$ elliptical multipole calculated using the method described in this Appendix, as well as the effect of such a perturbation on a SIE convergence profile, in comparison with the $m=6$ circular multipole.

\bibliographystyle{apsrev4-2}
\bibliography{biblio}{}

\end{document}